\documentclass[preprint]{aastex}

\shorttitle{The Bright SHARC Survey: The Selection Function and its Impact on
the Cluster X-ray Luminosity Function}

\begin{document}
\def\lsim{\mathrel{\rlap{\lower 4pt \hbox{\hskip 1pt $\sim$}}\raise 1pt\hbox{$<$
}}}
\def\gsim{\mathrel{\rlap{\lower 4pt \hbox{\hskip 1pt $\sim$}}\raise 1pt\hbox{$>$
}}}

\title{The Bright SHARC Survey: The Selection Function and its Impact on
the Cluster X-ray Luminosity Function}

\author{C. Adami}
\affil{IGRAP-laboratoire d'Astronomie Spatiale, Traverse du siphon, 
13012 Marseille, France \\
Department of Physics and Astronomy, Northwestern University,
Dearborn Observatory, 2131 Sheridan, 60208-2900 Evanston, USA 
}
\email{adami@lilith.astro.nwu.edu}

\author{M.P. Ulmer}
\affil{Department of Physics and Astronomy, Northwestern University,
Dearborn Observatory, 
2131 Sheridan, 60208-2900 Evanston, USA }
\email{ulmer@curie.astro.nwu.edu}

\author{A.K. Romer}
\affil{Department of Physics, Carnegie Mellon University, 5000 Forbes Avenue,
Pittsburgh, PA 15213, USA}
\email{romer@cmu.edu}

\author{R.C. Nichol}
\affil{Department of Physics, Carnegie Mellon University, 5000 Forbes Avenue,
Pittsburgh, PA 15213, USA}
\email{nichol@cmu.edu}

\author{B.P. Holden}
\affil{Department of Astronomy and Astrophysics, University of Chicago,
5640 S. Ellis Avenue, Chicago, IL 60637, USA}
\email{holden@tokyo-rose.uchicago.edu}

\author{R.A. Pildis}
\affil{Department of Physics and Astronomy, Northwestern University,
Dearborn Observatory, 
2131 Sheridan, 60208-2900 Evanston, USA}

\begin{abstract}

We present the results of a comprehensive set of simulations designed 
to quantify the selection function of the Bright SHARC survey (Romer et al. 
2000a) for distant clusters. The statistical significance of the simulations 
relied on the creation of many thousands of artificial clusters with redshifts 
and luminosities in the range $0.25<z<0.95$ and $0.5<L_X<10\times10^{44}$ erg 
s$^{-1}$ (0.5$-$2.0 keV). We created 1 standard and 19 varied distribution 
functions, each of which assumed a different set of cluster, cosmological and 
operational parameters. The parameters we varied included the values of 
$\Omega_0$, $\Omega_\Lambda$, $\beta$, core radius ($r_c$) and ellipticity 
($e$). We also investigated how non-standard surface brightness profiles (i.e 
the Navarro, Frenk \& White 1997, NFW, model); cooling flows; and the ROSAT 
pointing target, influence the selection function in the Bright SHARC survey. 
For our standard set we adopted the
parameters used during the derivation of the Bright SHARC Cluster X-ray 
Luminosity Function (CXLF, Nichol et al. 1999, N99), i.e. $\Omega_0=1$, 
$\Omega_\Lambda$=0 and an isothermal $\beta$ model with $\beta$=0.67, 
$r_c$=250 kpc and $e=0.15$. We found that certain parameters have 
a dramatic effect on our ability to detect clusters, e.g. the 
presence of a NFW profile or a strong cooling flow profile, or the value of 
$r_c$ and $\beta$. Other parameters had very little effect, e.g. the type of 
ROSAT target and the cluster ellipticity. At distant redshift
($z>0.8$), elliptical clusters are significantly easier to detect than
spherical ones in the Bright SHARC survey. 
We show also that all the tested parameters have only a small influence on the
computed luminosity of the clusters ({\it recovered luminosity} in the text)
except the presence of a strong cooling flow.
We conclude that the CXLF presented in N99 is robust (under the assumption of 
standard parameters), but
stress the importance of cluster follow-up, by {\it Chandra} and XMM, in order 
to better constrain the morphology of the distant clusters found in the Bright
SHARC and other surveys.

\end{abstract}


\section{Introduction}

Numerous authors have demonstrated that the observed evolution of
clusters of galaxies can place a strong constraint on the present day value
of the matter density of the Universe, $\rho _0 / \rho _c$=$\Omega_m$ (see 
Gunn \& Gott 1972; Press \& Schechter 1974; Lacey \& Cole 1993; Oukbir \& 
Blanchard 1992 \& 1997; Richstone, Loeb \& Turner 1992). In recent years, there 
has been considerable interest in constraining $\Omega_m$ using the observed 
abundance of clusters as a function of redshift (see Viana \& Liddle 1996 \& 
1999; Henry et al. 1997; Bahcall, Fan \& Cen 1997; Sadat et al. 1998; Reichart 
et al. 1999; Borgani et al. 1999). The effect of $\Omega_0$ (=$\Omega_m$ if
$\Omega_{\Lambda}$=0) on cluster abundances is very large. For example, the 
space density of high redshift ($z>0.3$), massive  ($T>6$ KeV) clusters in an
$\Omega_m=0.3$ universe is 100 times greater than that in an
$\Omega_m=1$ universe (e.g. Viana \& Liddle 1996, Oukbir \& Blanchard 1992, Oukbir \& Blanchard 1997, Romer et al. 2000b). Unfortunately, the various
studies performed to date have produced a wide range of results, from 
$\Omega_m=0.3\pm0.1$ (Bahcall et al. 1997) to  $\Omega_m=0.96^{+0.36}_{-0.32}$ 
(Reichart et al. 1999). This observed dispersion in $\Omega_m$ is most likely 
due to the fact that the cluster surveys currently available do not sample a 
large enough volume to include sufficient numbers of distant and
massive clusters. Nevertheless, it is still important to quantify the 
uncertainties that are applicable to the data in hand. This allows us not only 
better to understand the uncertainties in using the ROSAT database, but this 
also provides insight as how to proceed in the future.

In this paper, we discuss the selection function of one of the ROSAT 
archival surveys for distant clusters; namely, the Bright Serendipitous 
High--redshift Archival ROSAT Cluster (Bright SHARC) survey. The cluster 
catalog resulting from the Bright SHARC survey has been presented 
elsewhere (Romer et al. 2000a, R00 hereafter) as was the initial 
determination of the Bright SHARC Cluster X--ray Luminosity Function 
(CXLF; Nichol et al. 1999, N99 hereafter). Here we describe 
the results of a full set of simulations designed to determine the ability 
of the Bright SHARC survey to detect clusters of different morphologies,
luminosities and redshifts under different cosmological and observational
conditions. Earlier works such as that of Rosati et al. (1995) or Vikhlinin et 
al. (1998) set the standards of this kind of work, however, we have 
performed these simulations in a much more detailed way and have examined
the effects of different cosmologies and cluster profiles on the completeness
of the Bright SHARC survey.

An outline of the paper is as follows: Section 2 gives an
overview of the Bright SHARC survey and details the techniques used to 
perform the selection function simulations. Section 3 gives
the results of the simulations. Section 4 describes how the areal 
coverage of the survey was determined. Section 5 describes CXLF's
derived from our selection function simulations. In section 6 we
present a discussion of the results and our conclusions are in section 7.

Throughout we keep H$_0$ fixed at 50 km s$^{-1}$ Mpc$^{-1}$ but $\Omega _m$,
$\Omega _{\Lambda}$ (=$\Lambda$c$^2$/3H$_0$) and $\Omega _0$ are varied.

\section{The data and the Simulation Techniques}

\subsection{Survey Overview}

The Bright SHARC survey has been described in detail in R00, but 
we review the salient points here. The survey was comprised of 460 ROSAT PSPC
pointings. The pointings all had exposure times greater than 10 ks and 
lie at Galactic latitudes greater than $b=|20^{\circ}|$. The 460 pointings 
were directed towards nine different categories of targets; 1) normal 
stars (18 pointings), 2) white dwarfs (55 pointings), 3) cataclysmic variables
(37 pointings), 4) neutron stars and black holes (8 pointings), 5) supernova 
remnants (1 pointing), 6) normal galaxies (75 pointings), 7) AGN (150
pointings), 8) clusters of galaxies (71 pointings) and 9) other possible 
sources of diffuse X-ray emission (45 pointings). Each pointing has a unique 
six digit 
identification number, with the first digit being set by the target
category, e.g. pointing 600005 which was directed towards the galaxy
NGC 720. (For a full listing of identification numbers and pointing
targets, see Appendix A of R00.) The Bright SHARC survey took advantage of the
fact that the pointing targets covered only a small fraction of the
total field of view of the PSPC detector, leaving the rest of the field of
view (FOV) available for serendipitous cluster
detections. FOV of the ROSAT PSPC extends to a radius of $\simeq1^{\circ}$, but
in the Bright SHARC survey, we chose to use only an annular
region bounded by radii of $2.'5$ and $22.'4$. 
(Beyond $22.'4$ the point spread function degrades rapidly which makes
detections of the extended emission from clusters increasingly difficult.)

Each of the 460 pointings was run through a pipeline processing which
identified the extended sources in each field. In total 374 extended
sources were catalogued (see Appendix E of R00). These 374 sources
met the following three criteria; they were detected at a signal-to-noise 
ratio of 8 or higher, they were more than $3\sigma$ (wavelet defined, see 
R00) extended and had filling factors\footnote{The filling factor is a simple 
shape parameter designed to filter out obviously blended sources, see R00 
\S3.} of less than $f=1.3$. The Bright SHARC survey is comprised
of the brightest 94 of these 374 extended sources. These 94 all have 
count rates in excess of 0.01163 counts s$^{-1}$. Optical identifications
have been secured for all but 3 of the Bright SHARC sources, resulting
in a sample of 37 clusters ($0.3<z<0.83$: 12 clusters). 

Relevant to this paper is the method by which cluster luminosities
are derived in R00 (we refer to this as the ``R00 luminosity method''
hereafter): Once a source was identified as a cluster, 
and its redshift measured, its total count rate was determined. This
was done by laying down a metric aperture which encircles 80\% of the flux 
from an isothermal $\beta$ profile:
\begin{equation}
I(r)=\frac{I_0}{[1+(r/r_c)^2]^{3\beta -1/2}} \,\,,
\label{betamodel}
\end{equation}
at redshift $z$, where $I$ is the surface brightness at radius $r$. 
In R00 the slope and core radius were set to $\beta=0.67$ and 
$r_c=250$ kpc respectively. The effect of the ROSAT PSPC 
point spread function was taken into account by convolving with the 
appropriate off-axis PSPC PSF.
 The count rate measured inside the 80\% aperture was scaled to a
total value by dividing by $0.8$. The total count rates were then 
converted into an unabsorbed flux and a rest frame luminosity in the 
0.5$-$2.0 keV band pass using conversion factors determined using the 
XSPEC (Arnaud 1996) package. 

Also relevant to this paper is the fact that the Bright SHARC count rate 
threshold (0.01163 counts s$^{-1}$) was applied before the total derived count 
rates (see above) were known. Instead, it was applied to the wavelet count rate 
which is, on average, a factor of 2.1 times smaller than the total derived 
count rate. (Wavelet count rates were not determined in metric apertures but 
within source boundaries defined by the automated wavelet transform detection 
software, see R00 \S3.) 

We have included in the Bright SHARC survey PSPC pointings that contain 
extended on-axis targets {\it e.g.} supernova remnants (1 pointing), clusters 
of galaxies (71 pointings) and other sources of diffuse X-ray emission (45 
pointings in class 9). The supernova remnants pose no threat to the Bright 
SHARC search as only one such pointing was included. Clusters and other 
sources of diffuse emission may cause some obscuration. However, when we
simulated these pointing types separately (see \S3.7), we found that our 
success rate was 
not significantly different from the other pointing types. For the cluster
pointings included in the Bright SHARC there is an additional concern regarding
the cluster--cluster correlation function. We discuss this problem in Section 
3.7 but note here that Romer et al. 2000 examined this problem in detail for 
the Bright SHARC and concluded that at most 2 Bright SHARC clusters were 
possibly associated with the on-axis targeted cluster. Both of these 
``serendipitous'' detections were not used in our analyses {\it e.g.} N99. 
Finally, we note that other authors have shown that concerns about 
cluster--cluster correlations are unimportant in the construction of large 
serendipitous surveys of X--ray clusters (cf Ebeling et al. 1998).

\subsection{Details of the Simulation Process}

The simulation procedure is quite complex and we describe it below 
in detail, but, in essence, the basic concept is straightforward. We have 
created thousands of artificial clusters with a range of different parameters, 
placed these clusters at different positions in 
ROSAT PSPC pointings and then ran the modified pointings through 
the Bright SHARC survey pipeline processing. We used these simulations
to tell us how sensitive the survey is to each type of artificial cluster and 
also to evaluate the completeness of the SHARC survey as a function of X-ray
luminosity, shape and cosmology.

In Table 1, we outline 20 different simulation runs (see \S 2.2.1 for more
details). For each of these runs we selected a random set of
pointings from the 460 that comprise the Bright SHARC survey. In runs 1 
through 12, 200 pointings were selected. In runs 13 through 20 - where we 
concentrated on specific pointing types - between 35 and 150
pointings were selected. Into each pointing we placed an artificial cluster of 
a certain redshift and luminosity. The pointings were then processed to 
determine the percentage of times that our pipeline flagged an artificial 
cluster as an extended source. The process was repeated so that a full range of 
cluster redshifts and luminosities could be tested. Fifteen different redshifts 
($0.25<z<0.95$ with $\delta z$=0.05) and  six different luminosities ($L_{44}$ 
= 1.0, 2.0, 3.5, 5.0, 7.5 and 10.0) were used ($L_{44}$ is in units of 
$1\times$10$^{44}$ erg s$^{-1}$ in the 0.5$-$2.0 keV bandpass). Thus the total 
number of artificial clusters analyzed in each of the first 12 simulation runs 
was 200$\times$15$\times$6=18000 (more than 300000 artificial clusters in total 
when we include runs 13 through 20).

To be flagged as an extended source and to be associated with a cluster, the 
artificial clusters had to have a detection centroid not more than 5 pixels 
($1.\arcmin 25$) from the input centroid and also had to meet the three 
criteria described above (i.e. $S/N>8\sigma$, $>3\sigma$ extended, $f<1.3$).

In order to test how the survey sensitivity varied with off-axis angle, the 
artificial clusters in each run were divided among between four non-overlapping 
annuli: 9 to 28, 28 to 53, 53 to 78 and 78 to 103 ($\sim 15''$) pixels 
respectively. The exact positions of the artificial clusters within these 
annuli were selected at random.

The artificial clusters were generated as follows: First we calculated the 
expected ROSAT PSPC count rate from a cluster with a certain redshift and total 
luminosity using the R00 luminosity method in reverse. The Penn (1999) formulae 
for luminosity distance was used when $\Omega_\Lambda$ was non zero, e.g for 
run 3. Second, the count rate was converted into a number of counts using the 
correct exposure time for the chosen position on the field of view and by 
assuming a Poisson distribution in the number of counts (to take into account 
the shot-noise in the photon statistic). Third, the counts were distributed 
over a model surface brightness profile (with a random rotational orientation) 
which had been convolved with the appropriate off-axis point spread function. 
Fourth, the surface brightness image was divided up into $15''\times15''$ 
pixels and distributed so that there were only integer numbers of counts per 
pixel. Therefore, the total image was scaled so that it was in the same units 
(counts s$^{-1}$ arcmin$^{-2}$) as the pointing into which it had been placed.

We have used the L-T relation of Arnaud $\&$ Evrard (1999) to assign a 
temperature and hence a K-correction to each simulated cluster. The 
K-corrections are small ($\leq$15$\%$) for clusters temperatures in the 
range of those we have detected ([1;9]keV, see Table 2 of R00). 

The point spread convolution performed on the model
surface brightness profiles differed from that used in R00. 
In R00, the Nichol et al. (1997) empirical fit to the off-axis PSF 
degradation was used. The Nichol et al. (1997) PSF was symmetric
whereas, in reality, the PSF is not symmetric. The asymmetry was important 
to take into account in our simulations which included the ellipticity of 
the cluster profile. Here, we used, therefore,
composite images of the PSF generated by co-adding high signal 
to noise ($S/N>15\sigma$) images of point sources.
The individual point source images were accumulated in 5 pixel
wide annuli. The artificial clusters were convolved with the composite
PSF image with the closest mean off-axis angle (the mean distance between an
artificial cluster and the selected PSF: 37'').

\subsubsection{Description of the Twenty Simulation Runs}
\label{sixteen}

In Table 1 we summarize the parameters used during each of the
20 different simulation runs. Column 1 gives the run number,
column 2 the surface brightness model, column 4 the 
cluster ellipticity, column 5 the core radius, column 6 $\beta$, 
column 7 $\Omega_m$, column 8 $\Omega_\Lambda$, column 10 the number 
of pointings used and column 9 the type of pointing used. We describe
each of the 20 runs in more detail below.

\begin{itemize}
\item{Run 1: This run used a standard set of parameters
and acted as the benchmark against which the other simulations 
were compared. The parameters chosen for this run match 
closely those in the R00 luminosity method, i.e. $\Omega_m=\Omega_0=1$, 
($\Omega_\Lambda=0$) and an isothermal $\beta$ profile (equation 1.) with 
$\beta=0.67$ and $r_c=250$ kpc. In a slight departure from R00, a small
ellipticity (ellipticity: $e=0.15$) was introduced into each cluster profile. 
This value
is typical for an ensemble average of clusters and is in agreement with the
work of Wang $\&$ Ulmer (1997). This was the mean X-ray ellipticity they found
for 10 distant (0.17$\leq$z$\leq$0.54) rich clusters of galaxies.}

\item{Runs 2 and 3: In these runs we examined the effect of the 
assumed values of $\Omega_0$ and $\Omega_\Lambda$ on our selection
function. In run 2, an open model was used ($\Omega_m$=0.1,
$\Omega_\Lambda$=0). In run 3 a flat model with a positive 
cosmological constant was used ($\Omega_m$=0.4, $\Lambda$=0.6),
$\Omega_0$=1.}

\item{Runs 4 and 5: In these runs we examined the effect 
of varying the cluster ellipticity. We used
values of $e=0$ and $e=0.3$ in runs 4 and 5 respectively.}
 
\item{Runs 6, 7 and 8: In these runs we examined how different surface
brightness models affected cluster detectability. Numerical simulations, for
example by Navarro, Frenk and White (1997, NFW hereafter), have shown that 
dark matter profiles are more peaked at the cluster center than are isothermal 
$\beta$ profiles. In run 6 we used a slightly modified version of the NFW 
profile whereby we added a strong cusp to the central surface brightness to an 
isothermal $\beta$-model with $\beta$=0.67 and $r_c$=250 kpc (see Adami et al. 
1998 for more details). In run 7 and 8 we examined the effect of cooling 
flows. For this we add a Gaussian (FWHM=400 kpc) which contains 10\% of the 
total luminosity and a Gaussian (FWHM=200 kpc) which contains 50\% of the 
total luminosity to a standard isothermal $\beta$-model ($\beta$=0.67, 
$r_c$=250 kpc). The runs 7 and 8 profiles mimicked, respectively, the 
moderate and strong cooling flow surface brightness profiles given in 
Peres et al. (1998).}

\item{Runs 9 and 10: In these runs we examined the effect of varying the
value of the slope parameter $\beta$. We used values of $\beta=0.55$ and 
$\beta=0.75$ in runs 9 and 10 respectively.}

\item{Runs 11 and 12: In these runs we examined the effect 
of varying the core radius in the isothermal 
$\beta$ profile. We used core radius values of $r_c$ = 100 kpc 
and 400 kpc in runs 11 and 12 respectively.}

\item{Runs 13 to 20: In these runs we examined how the pointing target
affects cluster visibility. As stated in \S 2.1, the 460 Bright SHARC pointings
have nine different types of pointing targets. Of these nine, six types
were represented more than 35 times. In runs 13 to 20 we ran simulations
on a single pointing type (or on a single class of pointing types: point
sources versus diffuse sources). Run 13 was comprised of 55 type-2 pointings 
(white dwarves), run 14 of 37 type-3 pointings (cataclysmic variables), 
run 15 of 75 type-6 pointings (normal galaxies), run 16 of 150 type-7 
pointings (AGN), run 17 of 71 type-8 pointings (clusters), run 18 of 45 
type-9 pointings (diffuse X-ray sources), run 19 of the types 1+2+3+4 (point
sources) and run 20 of the types 5+6+7+8+9 (extended sources).}

\end{itemize}

\section{The Results}

This section describes the results presented in the Tables 3 and 4 and Figures 
1 to 11. We show only the figures describing the results for run 1 (i.e. the
standard parameter set) and those from subsequent runs which showed a 
significant departure from run 1.

\subsection{Results with the Standard Set of Parameters}

\subsubsection{Detection efficiencies and recovered luminosity}

We present the results from run 1 in Figures 1, 2 and 3 and Tables 3 and 4.
Figures 1 and 2 use grey scale shading to illustrate the 
numerical results presented in Table 3. Similarly, Figure 3 illustrates 
the results in Table 4. Each figure is comprised of 9 disks. 
Each disk represents a different redshift 
($z$=0.25 to $z$=0.95 in $z=0.1$ increments)
and is divided into 6 sectors, with each sector representing a 
different luminosity ($L_{44}$=1, 2, 3.5, 5, 7.5, 10 respectively).
Each sector is further subdivided into 4 rings to represent the 4 off-axis
annuli used in the simulations. These shaded areas represent the average
detection efficiency or the recovered luminosity (see below).

Figure 1 illustrates the percentage of artificial clusters that were
detected as extended sources when no count rate limit was
imposed (the {\it total detection efficiency} hereafter). Figure 2 illustrates 
the corresponding percentages when the Bright SHARC count rate 
limit of 0.01163 counts s$^{-1}$ is imposed (the {\it Bright SHARC 
detection efficiency} hereafter). As expected, the efficiency drops off when
a count rate threshold is imposed. It is apparent from both Figure 1
and Figure 2 that the efficiency falls off with 
decreasing luminosity
and increasing redshift. Figure 3 illustrates the percentage of the 
input luminosity recovered by the R00 luminosity method (when we used the
Bright SHARC simulations with the count rate cut: {\it recovered 
luminosity} hereafter). If the R00 method were essentially perfect, then we 
would have expected to recover close to 100\% of the input luminosity everytime 
we detected an artificial cluster as an extended source.

\subsubsection{Uncertainties of detection efficiencies and recovered 
luminosity}

In order to estimate the significance level of the statements discussed below
we have estimated the uncertainties for the detection efficiencies and 
recovered luminosities (see also Tables 3 and 4) in run 1. These uncertainties
are the statistical dispersion of the computed percentages based on 
performing the simulations 50 times for the standard parameter set per given
luminosity, redshift and location in the pointings. 

We summarize the results as follow: 
for the {\it total detection efficiency}, the mean uncertainty is 9$\pm$5$\%$.
For the {\it Bright SHARC detection efficiency}, the mean uncertainty is 
6$\pm$4$\%$. For the {\it recovered luminosity}, the mean uncertainty 
is 8$\pm$6$\%$. If we limit the analysis to the fourth bin of the
ROSAT pointings, the previous mean uncertainties are, respectively, 
11$\pm$5$\%$, 10$\pm$5$\%$ and 11$\pm$10$\%$. These values are slightly higher
due to the degraded PSF. The standard errors on the uncertainty are, however, 
not very large whatever the detection efficiency: they all have amplitudes 
similar to the mean uncertainty itself.

For redshifts lower than $\sim$0.6 and luminosities greater than 
3.5$\times$10$^{44}$, our detection efficiency versus luminosity (cf. Figs. 1 
and 2) depends on the annulus, {\it i.e.} our detection efficiency is almost 
constant for the outer annulus but increases from 40 to 100$\%$ for the 3 inner 
annulii. According to the error bars, this trend may be significant (near the 
3-$\sigma$ level) and is probably due to the larger PSF at large off-axis 
angles. A small number of SHARC point sources were probably classified as 
extended, thus slightly boosting ($\sim$5$\%$) the number of initial extended 
sources. This is a negligible effect on our results.

\subsubsection{Results}

We have used the results of the run 1 simulations to determine the lowest
detection efficiency level which can be safely used to define a selection
function. We have done this by determining the rate of spurious extended 
sources detections, i.e. the number of cases where an extended X-ray 
source, lying close to the position of an artificial cluster, is falsely associated with that cluster (the probability that this extended source is
a real cluster is low based on the number of pointings: 460 and the number of
detected clusters in R00: 37). To this end, we have carried out an additional 
simulation set for run 1 with $L_{44}=0.5$. This luminosity is so 
low that it produces a flux that is well below our detection limit for 
$z>0.30$. So, if any of the $L_{44}=0.5$ clusters were "detected" as extended 
sources, then those detections would most likely be spurious. The mean 
detection efficiency for $L_{44}=0.5$ artificial clusters is, therefore, a measure of the source confusion in the SHARC survey. Not surprisingly, we 
detected very few of the 200 $L_{44}=0.5$ artificial clusters we created; we 
measured a mean {\it Bright SHARC detection efficiency} for them of only 1\% 
and a maximum {\it Bright SHARC detection efficiency} of less than 5$\%$ (this 
value is consistent with the uncertainties estimated in 3.1.2). Given this 
confusion level, we adopted for the simulations a lower limit to the acceptable 
efficiency level of 15\% when making comparisons between run 1 and runs 2-20
(see Figures 4 to 11). This limit ensures that there are at least 3 times more 
true detections of extended sources than false detections (detections of 
point-like sources classified as extended). 

By comparing Figures 2 and 3, it is clear that the {\it recovered 
luminosity} is close to 100\% where the {\it Bright SHARC 
detection efficiency} 
was greater than 15$\%$. Averaging over all regions where the 
{\it Bright SHARC 
detection efficiency} was greater than 15$\%$, we measured a mean 
{\it recovered luminosity} of $99\pm8\%$. The {\it recovered luminosity} 
was found to vary with position on the field of view: the mean {\it recovered 
luminosities} in the 4 annuli studied were 96$\pm$13$\%$, 94$\pm$10$\%$, 
94$\pm$8$\%$ and 111$\pm$3$\%$ respectively. We attribute the 111$\pm3$\%
result to the fact that the PSF is so large in the outer annulus, that
it increased the chance of flux contamination by point sources.
The R00 luminosity method used model parameters that are very close to
those used here to generate the artificial clusters except that in run 1 the 
simulated clusters are not exactly circular ($e=0.15$).

We have also used the results from simulation run 1 to determine the
number of times artificial clusters were not detected for reasons other than
because they were too faint or not sufficiently extended.
These reasons included cases where the SHARC detection 
pipeline measured an artificial cluster centroid which was more than $1.'25$
from the input centroid, or cases where the cluster was placed
almost exactly on top of a bright point source (so that the resulting
blended source is not flagged as extended, see R00 for the definition of
a blended source). We have estimated the number of times these situations arise 
by assuming our pipeline should detect 100\% of all $L_{44}=10.0$ artificial 
clusters in the lowest redshift bin ($z=0.25$) in the first, second and third 
radial annuli. We excluded the fourth annulus from our test because of the 
degraded PSF which can cause some of these clusters to be missed for reasons
other than confusion or false positioning. We detected 97\% of these clusters. 
We conclude therefore that $\sim3\%$ of all clusters went undetected from the 
SHARC survey for these technical reasons. This is a negligible effect 
compared to the statistical uncertainties of our results.

In summary, we have shown, using simulation run 1, how the {\it Bright SHARC 
detection efficiency} falls off with decreasing luminosity and increasing 
redshift. We have also demonstrated that above efficiency levels of 15\%, we 
can be confident than significantly less than 1/3 of the detections in the 
simulations are spurious (i.e. non extended sources that were detected as 
extended). Above the 15\% efficiency level, we found that the R00 methodology 
accurately recovers the input luminosities of the clusters. Finally, we 
demonstrated that only a small percentage of the simulated clusters ($\sim 
3\%$) were not detected for technical reasons, such as bad centroid fitting or 
bright source confusion.

In the following subsections we discuss the results of the other
19 simulation runs by making comparisons with the results from 
run 1. In Figures 4 to 11, we present the percentage 
difference between the run 1 and the run $n$ {\it Bright SHARC 
detection efficiency} (or {\it total detection efficiency}) 
as a function of off-axis angle for a subset of the luminosity 
($L_{44}$=2.0,5.0 \& 10.0) and redshift ($z$=0.25, 0.35, 0.45, 
0.55, 0.65, 0.75, 0.9) values simulated in each run (the complete results
for all the redshifts are given in the tables). 
A positive percentage means that the
efficiency of detection was higher in run 1 than it was in run $n$,
and vice versa. The percentage differences were only 
calculated for those luminosity and redshift combinations where 
the run 1 detection efficiency is greater than 15\%. Regions of
parameter space where the run 1 detection efficiency is $<$15\% 
are indicated on Figures 4 to 11 by crosses on the 0\% line.

In this section, and throughout this article, we have only discussed 
simulations of extended sources with profiles similar to galaxy clusters.
However, there are other sources of extended X--ray emission -- {\it e.g.} 
supernova remnants, galaxies, low luminosity nearby galaxies, groups and 
fossil groups {\it etc.} -- and this explains the discrepancy between our 
$\geq$85$\%$ rate of extended sources being clusters (the other $\leq$15$\%$
being blends of too faint to detect on their own clusters with real ROSAT
sources) in our simulations versus
only $\sim$40$\%$ of the Bright SHARC extended sources being real clusters (see 
R00). In other words, our simulations only tell us how often a cluster of a 
given luminosity and redshift would have made it into the Bright SHARC; the 
simulations can not tell us how often we would have found other X--ray 
extended sources.

\subsection{Effect of the Cosmological Parameters}

We tested the effect of the underlying cosmological models
on our ability to detect clusters by comparing the
results of run 1 with the results of runs 2 ($\Omega_m$=0.1, 
$\Omega_\Lambda$=0 ) and 3 ($\Omega _m$=0.4, $\Lambda$=0.6).
For both the alternative cosmologies tested, we found lower values 
of the $total$ and {\it Bright SHARC detection efficiency} (the latter are 
shown in Figure 4). This was 
because, for the same redshift and total luminosity, a 
low $\Omega_m$ produced clusters which were larger 
in angular extent than did a high $\Omega_m =1$.
Despite their increased angular extent, clusters in a low $\Omega_m$ 
Universe are harder to detect than their counterparts in a $\Omega_m =1$
Universe because their surface brightness (and hence contrast against
the X-ray background) is diminished. 

For those artificial clusters detected in runs 2 and 3 we find their 
{\it recovered luminosities} to be similar to those measured in run 1. The 
mean percentage difference between the {\it recovered luminosities} in run 1 
and run 2 was 3$\pm$18$\%$. This difference was 7$\pm$30$\%$ when compared 
runs 1 and 3. The standard errors associated with these two mean percentages 
are somewhat large, but we did not detect any systematic trends with the 
redshift and/or the luminosity.

In summary, if we assume lower values of $\Omega_0$ (for a flat
or an open Universe), the luminosity estimates are as good, on average,
as when we assume $\Omega_0$=1 ($\Omega_\Lambda=0$). However, changing
the value of  $\Omega_0$ and $\Omega_\Lambda$ has a marked effect on 
the detection efficiency of the survey, especially for high redshift and/or 
low luminosity clusters. We show in Section 5 that this effect has, however,
only a small influence on the derived Cluster X-ray Luminosity Function because
of the distribution of the Bright SHARC real clusters in the (z,L$_{44}$) 
space.

\subsection{Effect of the ellipticity}

We have investigated the effect of cluster ellipticity on the detection
efficiency of the survey by comparing the results from run 1 ($e=0.15$) 
to those from runs 4 ($e=0.0$) and 5 ($e=0.3$). For a fixed cosmology, 
an elliptical cluster will have a higher central surface brightness 
than a circular cluster of the same luminosity in proportion of the surface 
difference induced by the ellipticity (if the image is contracted along the 
minor axis). In doing this, we kept the core radius constant along the major 
axis: we contracted the image along the minor axis. Instead, we could have 
elongated the image along the major axis to make clusters elliptical. However, 
doing this would lead to a lower surface brightness cluster. Such clusters 
would be similar to our low-$\beta$ and our large core radius models. 
Elliptical clusters which have longer than normal major axes would, therefore, 
not be preferentially detected at high redshift. If the majority of high 
redshift clusters were found to have extended major axes, then we could 
conclude that the detection of mostly elliptical clusters at high redshifts is 
not a selection effect.

As shown in Figure 5, the {\it total detection efficiency} is very
similar for the three runs except at the very luminous, very high redshift
end, where it appears that increased ellipticity leads to increased
detectability. We find no similar systematic trend in the {\it Bright SHARC 
detection efficiency} and so do not plot those results. We attribute this 
to the fact that very few clusters with $z>0.75$ met the Bright SHARC count 
rate threshold and, since the effect of ellipticity on the {\it total detection 
efficiency} is only seen in the $z=0.9$ bin, then there would be very little, 
if any, corresponding effect on the {\it Bright SHARC detection efficiency}.

We find no systematic trend in the {\it recovered luminosity} with ellipticity.
The mean percentage difference in {\it recovered luminosity} is 1$\pm22$\%
between runs 1 and 4  and 2$\pm$30$\%$  between runs 1 and 5. This is
good, since it shows that the R00 luminosity method, which
assumes a single ellipticity of $e=0$, did not introduce a systematic 
bias into the measured Bright SHARC cluster luminosities.

In summary, ellipticity has minimal effect on the {\it recovered luminosity}
and on the {\it Bright SHARC detection efficiency}. However, we do find a 
significant trend for increased {\it total detection efficiency} with 
ellipticity at very high redshifts. We attribute the lack of a similar trend 
in the {\it Bright SHARC detection efficiency} to the insensitivity of the 
Bright SHARC survey at very high redshifts, i.e. where the ellipticity effect 
becomes important. We have not examined with these simulations the effect of 
cluster orientation (prolate or oblate) on detection efficiency.

\subsection{Effect of the Central Surface Brightness Model}

Here we examine how the adopted surface brightness model for the artificial clusters 
affects their detectability. In run 1 a simple isothermal $\beta$-profile was 
used. We contrast this with a pseudo-NFW profile in run 6 and with a pseudo 
cooling flow profile in runs 7 and 8. Both the NFW profile and the cooling 
flow profiles are more peaked than the isothermal $\beta$-profile, and hence, 
they have higher central surface brightnesses. From figure 6 it can be 
seen that the detection efficiency for a moderate cooling flow (run 7) was 
similar to that for run 1 and there are no clear trends with redshift or 
luminosity. In other words, the presence of a moderate cooling flow does not 
have a significant impact on the detectability of a cluster. By contrast, the 
strong cooling flows (run 8: Fig. 7) and the NFW profile clusters simulated 
in run 6 were significantly (up to two times) easier to detect than the 
equivalent isothermal $\beta$-profile clusters. We attribute this to the 
surface brightness in the cluster core being significantly higher than for a 
$\beta$-profile.

We found the {\it recovered luminosities} from runs 6 and 7 to be very similar 
to those from run 1. The mean percentage difference in {\it recovered 
luminosity} is  2$\pm$17 $\%$ between runs 1 and 6 and 2$\pm$16$\%$ between 
runs 1 and 7. This demonstrates that the R00 methodology, which assumes all 
clusters have isothermal $\beta$-profiles, provides a valid approximation
for the run 6 and 7.

The {\it recovered luminosities} from run 8 are, however, underestimated 
by 27$\pm$7$\%$. Based on the uncertainty, this trend seems to be significant 
(Fig. 6b). Assuming that the R00 methodology is able to recover a valid 
luminosity for $\beta$-like profiles, we can infer that the difference comes 
mainly from the cooling-flow profile itself. If the Gaussian model peak of the 
cooling-flow is relatively bright compared to the $\beta$-profile (200 kpc 
in this case), the R00 methodology fails to recover the entire luminosity due
to the normalization technique which assumed a standard $\beta$-profile. 

However, the percentages of clusters exhibiting strong cooling flows in the 
Bright SHARC redshift range is probably low. Assuming, for example, a central 
cooling-time between 1 and 2 Gyears (e.g. Peres et al. 1998), most of the
known cooling-flows in clusters at z$\sim$0.1 have been initiated only at
redshifts lower than z$\sim$0.3 (q$_0 \leq$0.5). It is, therefore, very likely 
that we only have a small percentage of cooling-flow clusters (or at least with 
moderate cooling flows) in the redshift range of our simulations and in the 
Bright SHARC. This is confirmed by the fact that there are not many clusters 
detected at redshift higher than 0.3 with known cooling-flows (e.g. 3C 295: 
Henry $\&$ Henriksen 1986) whereas the X-ray detection of such clusters should 
be easier, theoretically, than for the non-cooling-flow clusters.  

\subsection{Effect of $\beta$}

In runs 9 and 10 we examined the effects of $\beta$ on the detectability
of clusters with isothermal $\beta$-profiles. In run 9 we use a value of
$\beta$=0.55, whereas, in run 10, we use $\beta$=0.75. We compared the results
from these runs against those from run 1 ($\beta$=0.67) in Figure 9.
From Figure 9 it is clear that clusters with higher $\beta$ values are
easier to detect (this trend is seen in both the $total$ and the 
{\it Bright SHARC 
detection efficiency}). This can be explained by the fact that
higher $\beta$ values result in more concentrated, and hence higher
surface brightness, clusters.

Lowering the value of $\beta$ to $\beta$=0.55 was found to have a 
significant effect on value of the {\it recovered luminosity} (see Figure 10). 
Where the detection is easy (low redshift + high luminosity), the 
{\it recovered luminosity} estimate is the same whatever the value of 
$\beta$. Where the detection is more difficult, however, the {\it recovered 
luminosity} of the $\beta$=0.55 tends to be low. There are two reasons for 
this. First the detection efficiency for $\beta$=0.55 clusters is significantly 
lower than that of $\beta$=0.67 clusters (see Figure 9). So that, even when the 
run 1 detection efficiency is above our minimum threshold of 15\%, the run 9 
efficiency can be much lower. (In other words well into the regime where most 
of the detections are spurious.) Second the R00 luminosity method, which 
assumes all clusters have $\beta$=0.67, breaks down when $\beta<0.67$. If the 
true $\beta$ value is smaller than $\beta$=0.67, this aperture will encircle 
less than 80$\%$ of the actual flux, which can lead to a significant
underestimate of the total luminosity. In principle, R00 could have corrected 
for the $\beta$ effect by fitting the cluster profiles and then deriving a 
flux, but the low number of counts available in the ROSAT PSPC images made such 
an approach impractical. We found that the fitted $\beta$ value was so poorly 
constrained as to lead to best fit values that could easily be very far away 
from the true value. 

\subsection{Effect of the Core Radius}

In runs 11 and 12 we examined the effects of the size of the core radius 
($r_c$) on the detectability of clusters. 
In run 11 we adopted $r_c=100$ kpc and, in run 12, we adopted $r_c=400$ kpc.
We compared the results from these runs with those from run 1 ($r_c=250$ kpc)
in Figure 11. The effect of $r_c$ turns out to be more complex than 
the effect of $\beta$. The {\it Bright SHARC 
detection efficiencies} for the $r_c=100$ kpc 
clusters was consistently lower than that of $r_c=250$ kpc clusters,
with the effect being most pronounced at lower luminosities. By contrast,
the $r_c=400$ kpc clusters could be either easier or harder to detect
than the $r_c=250$ kpc clusters, depending on the redshift and luminosity. 
They were easier to detect in
the low redshift $L_{44}=2.0$ and the intermediate redshift $L_{44}=5.0$ 
bins, but harder to detect in the $L_{44}=10.0$ and low redshift $L_{44}=5.0$  
bins. This demonstrates the competing effects of surface brightness
and extent: Clusters with higher surface brightnesses were easier for 
the wavelet pipeline to detect, but those with small angular sizes were 
less likely to be flagged as extended sources.

We found no significant difference in the {\it recovered luminosity} 
between run 11 ($r_c$=400 kpc) and run 1 ($r_c$=250 kpc). By comparison, 
we found a small, but systematic, enhancement in the {\it recovered luminosity}
for run 11 ($r_c$=100 kpc) compared to run 1 ($r_c$=250 kpc). The enhancement 
was at the $\simeq$10\% level and resulted from the fact that the R00 
luminosity method will over estimate the true luminosity if the true 
core radius is smaller than the assumed value of $r_c$=250 kpc.
Similar enhancements were found when R00 compared the Bright SHARC
luminosities to those derived by Vikhlinin et al. (1998) for the
11 clusters they had in common. (Vikhlinin et al. 1998 used the best
fit value of $r_c$ to compute cluster luminosities and, in the 
majority of cases, their $r_c$ values were smaller than 250 kpc.)

\subsection{Effect of Target Type}

In section 2.1 we mentioned that the original targets of the 460
Bright SHARC pointings are divided into 9 different categories.
It is reasonable to ask whether the target category has any influence
on the sensitivity of a particular pointing to cluster detection.
For example, pointings with targets that reside in regions of 
high local hydrogen column density would tend to have decreased 
sensitivity or diffuse extended sources can produce obscurations.

Therefore, in runs 13 to 20, we have investigated the effect of 
target type. In each of these runs we placed artificial clusters into
 pointings of only one type. We compared the results from these
runs with those from run 1 (which used pointings
with a random selection of targets). We find
very little variation in the detection efficiency between runs 13 to 20
and run 1. For both the $total$ and the {\it Bright SHARC detection 
efficiency}, we measured a mean variation of only 2$\pm$5$\%$. The 
{\it recovered luminosities} were also very similar with a mean value of 
4$\%$ without any detectable systematic trends with redshift and/or luminosity.
In order to investigate possible statistical biais due to the small number
of pointings in each of the six previous classes, we also merged all the
pointing classes into two sub-classes: point sources (run 19) and extended
sources (run 20). We found that the detection efficiencies of these runs were
similar ro run 1, and we conclude that the difference between using all kinds 
of point sources (run 19) and diffuse sources (run 20) was negligible, as we
found very little difference betweens runs 19, 20 and 1. The {\it recovered 
luminosities} were also very similar, with a mean variation of less than 5$\%$.
We can safely conclude, therefore, that the target type has no influence on 
the sensitivity of a particular pointing to serendipitous cluster detections.

We have not investigated, however, the influence of the cluster-cluster 
correlation function in any of our simulations. Clusters are known to be 
clustered (e.g. Romer et al. 1994), meaning that pointings with cluster 
targets will include, on average, a larger number of serendipitous cluster 
detections (see also Ebeling et al. 1998 for a discussion of this point). For 
example, R00 highlighted two Bright SHARC clusters that are most probably 
associated with the pointing target. The fact that we have ignored cluster 
correlations in our simulations is not a concern, however, since these two 
clusters were excluded from the Bright SHARC CXLF analyses (N99 and \S 5) 
because they were at low redshift. 

\section{Areal Coverage}

To be able to compute an  CXLF, we need, in addition to the results of
our simulations, an estimate of the area covered by the Bright SHARC
survey. R00 calculated the maximum areal coverage of the survey to be
179 deg$^2$. However, the R00 calculation did not include an uncertainty
estimate or a quantification of how the areal coverage falls off with
cluster flux. We have corrected for those shortcomings here.

The effective areal coverage of the survey is a complex function of the
cluster flux, the cluster extent and the background level.  This is because 
clusters need to meet a signal to noise threshold and an extent criterion to 
be included in the Bright SHARC cluster candidate list. So, we have computed 
the areal coverage as a function of flux for each of the 460 pointings 
individually. Since the background level and PSF varies across the FOV, we need
to beak up each pointing into several regions to ensure an accurate calculation.
We chose to break the FOV up into the same 4 annuli as we used in the 
simulations. We determined a mean noise level ($\bar N$) for an annulus by 
examining the $S/N$ values of $all$ the extended sources with a $S/N>2$ (we 
chose this low signal to noise threshold \footnote{This measure of the $\bar N$
by using $S/N>2$ does not reflect the $S/N$ of 8 used for our detection 
criterion.} to ensure we had more than 3 detections for each annulus) detected 
in that annulus (when doing this, we did not set a minimum count rate threshold 
breaking, only for this purpose, our $S/N$ criterion for Bright SHARC). We 
computed $\bar N$ as(with n the number of sources detected in a given annulus):

$\overline {(S/N)} = \sum [S_i/N_i]/n$ 

$\bar S = \sum [S_i/n]$ 

$\bar N = S/(S/N)$

From $\bar N$ we estimate the lowest signal ($S_{min}$) that would yield 
a $S/$$\bar N$$>8$ detection of an extended source in that annulus. For signals 
greater or equal to $S_{min}$, the whole\footnote{After excluding pixels which 
overlapped with other pointings and/or with the shadow of the X-ray mirror 
support, see R00 \S4.} area of the annulus can be added to the total areal
coverage of the survey. (For signals less than $S_{min}$, none of the area is 
added.) Thus, we build up an areal coverage of the whole survey, as a 
function of $S$, by summing over each annulus in each of the 460 pointings. 
It is clear that for very bright sources, all annuli in all pointings will
contribute. Thus this method yields a maximum areal coverage which
is, as expected, identical to the value calculated in R00 (179 deg$^2$).
But, as the signal decreases, more and more of the annuli drop out of the
calculation and the areal coverage falls to zero. These effects are
illustrated in Figure 12, where we have converted signal (i.e. 
count rate) into flux using the conversion tables devised for R00.

Also shown in Figure 12 is our estimate of the uncertainty in the 
areal coverage (dotted lines). We calculated this uncertainty by 
examining the distribution of $S/N$ values for the extended sources 
in each annulus. For the solid line on Figure 12 we used $\bar N$ to 
determine $S_{min}$. For the upper (lower) dotted line we used an 
estimate of $\bar N+\delta N$ ($\bar N-\delta N$). The number of extended 
sources per annulus was usually too small 
to determine a reliable standard deviation on $\bar N$. We,
therefore, chose to set $\bar N+\delta N$ to be equal to $N_{max}$ (the 
maximum noise level for that annulus) and $\bar N-\delta N$ to be equal to 
$N_{min}$. We determined that this method provided a good
estimate of the 1 sigma errors on the areal coverage. We did this as follows.  
Based on detected extended sources,  we calculated the lowest and the highest 
noise level for each pointing (without splitting the pointings in 4 annuli). 
The range of areal coverage versus flux thus derived for each pointing is so 
large as to effectively assure us that the true value of areal coverage versus 
flux level lies within this range. We then compared this with the more precise 
uncertainty estimate we derived, based on dividing the field of view of each 
pointing into 4 parts as described above. To estimate the statistical 
significance (sigma level) to assign to this uncertainty derived from the 4 
annuli approach, we compared the area covered between the dotted curves in 
Figure 12 (the area covered by the uncertainties when we split each pointing in 
four annuli) with the area covered by the maximum possible deviation to the 
areal coverage (computed without splitting the pointings in 4 annuli). This 
ratio is 0.61 which is close enough to $1\sigma$ that we take the dotted curves 
as delineating the $1\sigma$ range in the discussion that follows. 

As can be seen in Figure 12, as the flux limit is lower, the uncertainty of the 
areal coverage gets larger. This is because at decreasing flux levels, the
fluctuations due to counting statistics become relatively larger, and because
the effective coverage decreases rapidly with the flux level. In Figure 12 
we show with the thick solid vertical line, the flux at which the $1\sigma$ 
uncertainty in the areal coverage is 10\%. As can be seen from the 
thin vertical lines (Bright SHARC clusters), all the Bright SHARC clusters 
were detected at fluxes where the areal coverage error was $<10\%$. The lowest 
flux cluster used in N99 (thin solid line) has an associated areal 
coverage uncertainty of only $5\%$. Therefore, we can safely neglect the effect 
of such uncertainties when determining the Bright SHARC CXLF. However, surveys 
that go to much lower flux levels (e.g. Burke et al. 1997, Vikhlinin et al. 
1998, Rosati et al. 1998, Ebeling et al. 1999) may need to be concerned about 
uncertainties in their areal coverage at low fluxes.

\section{Influence of the Different Sets of Parameters on the CXLF}

N99 used a preliminary set of selection function simulations to
derive the Bright SHARC CXLF. This preliminary set used the same 
parameters as run 1, i.e. an isothermal $\beta$ profile with $\beta=0.67$, 
$r_c=250$ kpc, $e=0.15$, $\Omega_0=1$ and $\Omega_\Lambda=0$, but covered
a smaller range of redshifts and luminosities. Here we
redetermine the Bright SHARC CXLF using the results of run 1 in order
to test the robustness of the N99 results. We also investigate how the
CXLF changes when we use selection functions derived from the
results of runs 2 to 12. We note that we have detected a systematic trend 
affecting the accuracy of the Bright SHARC cluster luminosity measurements
only when we use a low value of the slope $\beta$ or strong cooling flow
profiles (which are not, however, very likely at high redshift). None of the 
other tested parameters have a systematic effect. This means that our 
luminosity measurements are not very dependent of the parameters tested in this 
work.

We have used the same ${\rm 1/V_a}$ methodology as N99 (adapted
where necessary for different cosmological models) to determine the Bright 
SHARC CXLF. ${\rm V_a}$ 
is defined as the available sampled volume for any given cluster in the survey
and can be computed for a cluster of luminosity $\rm L_x$ using

\begin{equation}
{\rm V_a} = \int_{z_{low}}^{z_{high}}\, \Omega({\rm L_x}, z)\,\, V(z)\, dz,
\label{va}
\end{equation}

\noindent where $z_{low}$ and $z_{high}$ are the lower and upper
bounds of the redshift shell of interest, $V(z)$ is the volume per
unit solid angle for that redshift shell and $\Omega({\rm L_x}, z)$ is
the effective area of the Bright SHARC survey. This effective area 
was calculated by multiplying the areal coverage of the survey at the 
corresponding ROSAT flux by
the appropriate {\it Bright SHARC detection efficiency}. The detection 
efficiency is a function of the cluster redshift and luminosity (see Table 3).
The effective area is also (see Figure 13 for an illustration of
how $\Omega({\rm L_x}, z)$ varies with ${\rm L_x}$ and $z$). 
Very bright clusters (e.g. $L_{44}$=10.0 clusters at $z=0.25$) will have an 
effective area equal to the maximum areal coverage of the survey (i.e. 179 
deg$^2$). Whereas very faint clusters will have an effective area that 
approaches zero. The effective area was calculated for each of the 4 annuli 
separately and then co-added, since the detection efficiency is also a function 
of off-axis angle (see Table 3). The maximum areal coverage of the inner 
annulus was 9.7 deg$^2$, 47 deg$^2$ for the second annulus, 94.3 deg$^2$ for 
the third annulus and 28 deg$^2$ for the outer annulus.

In order to compare with the earlier calculations of N99, we show in
Figure 13 the effective area computed for the standard set of
parameters. The results are similar to the preliminary simulations
presented in N99 except for the high luminosity and high redshift
clusters, where we now find a lower effective area than
thought. This is primarily due to the increased precision of the
simulations presented herein. The main consequence of this change is
in the statistical significance of any proposed deficit of high
luminosity, high redshift clusters (see N99).  We have repeated the
analysis of N99 and find that at $L_x>5\times 10^{44}$, and in the
redshift range $0.3<z<0.7$, we would expect to have detected about 2 
clusters (using the De Grandi et al. 1999 or Ebeling et al. 1997 
luminosity function) in the Bright SHARC using the effective area 
curves presented in Figure 13.

In N99, the CXLF was calculated using the 12 Bright SHARC clusters with 
$z>0.3$. In this study we used a slightly lower redshift limit ($z>0.285$) to 
increase the number of clusters in the sample. We list the detection 
efficiencies (as well as the redshift, luminosity and off-axis angle) for the 
15 $z>0.285$ Bright SHARC clusters in Table 2, of which we used 13 (see below).
Because the simulations covered only discrete values of ${\rm L_x}$ and $z$, 
we had to use linear interpolation to estimate the detection efficiency for 
these clusters. It can be seen from Table 2 that the detection efficiency 
can vary dramatically with the run number. We did not include
clusters for which the mean (over all runs 1 to 11) {\it Bright SHARC 
detection efficiency} was less than 5$\%$. We, therefore, excluded two clusters 
(RXJ1334 and RXJ1308 which were both used in N99) from our list of fifteen. The 
remaining thirteen clusters are divided into 4 luminosity bins (see Fig. 12
and 13). These bins contained 4, 5, 3 and 1
clusters respectively. The redshift ranges for these bins were slightly
different from N99; for bins 1,2 and 3 we used $0.285\leq z\leq 0.7$
(compared to $0.3\leq z\leq0.7$) and for bin 4 we used $0.285\leq z\leq 1.0$
(compared to $0.3\leq z\leq1.0$). 

Table 2, upon which our XCLF's (Figures 14 and 15) were based, has 5 clusters
below the 15$\%$ level (3 clusters we used; the two lowest efficiency clusters
were not used). These "low efficiency" clusters are not spurious as they have 
been optically verified. The uncertainty of the efficiency of detection of 
these clusters is indeed higher than the other clusters we used and we have 
factored this into the calculation of the CXLF error bars.

The results are shown in Figures 14 and 15. In these figures, we have 
overplotted the local CXLF computed by de Grandi et al. (1999) and 
Ebeling et al. (1997). In Figure 14, we also plotted the N99 CXLF and 
its uncertainty envelope. The error bars (plotted only on the run 1 
points to avoid crowding) in Figures 14 and 15 are the quadratic sum of the 
{\it Bright SHARC detection efficiency} uncertainty and of the Poisson 
statistical uncertainty due to the small number of clusters in each bin. 

Figure 14 shows the results of our CXLF calculations using the runs
1 to 3 detection efficiencies. The first point to note is that
the run 1 CXLF (open circles) is consistent with the results of 
N99. This demonstrates the robustness of the N99 result (under the 
assumption of standard parameters). In the last luminosity bin we derived a 
slightly higher number density of clusters compared to N99. This is because our 
run 1 simulation yielded a lower {\it Bright SHARC detection efficiency} in 
this bin than did the simulations used by N99. 
(When the detection efficiency is lower, the effective
area -- and hence volume -- is smaller which pushes up the number density.) 
The second point to note from Figure 14 is that the
value of $\Omega _0$ and $\Omega_\Lambda$ have a small effect on the CXLF, not 
discernable in a log-log plot. We do see a small systematic offset 
between the $\Omega _0=1$ and $\Omega _0<1$ points (because the detection 
efficiencies in runs 2 and 3 were lower than those in run 1), but 
this offset is smaller than the 1 sigma uncertainties. 

In Figure 15 we present 4 separate sets of CXLF results, comparing the
run 1 number densities to those derived from runs 4 through 12. We see, 
from the lower left panel, that ellipticity has a very small effect on 
the CXLF. However, in the other 3 panels we see some quite dramatic 
effects when we change the core radius (lower right panel), the value 
of $\beta$ (upper left panel) and surface brightness profile (upper 
right panel: we have only plotted the results using moderate cooling flows
and NFW profiles). When
clusters are more diffuse (e.g. if $\beta$ is smaller or $r_c$ is 
larger than our canonical values), the detection efficiency declines
(see \S 3.5 and \S 3.6) hence the cluster number density goes up. By contrast,
when the clusters become more concentrated (e.g. if we use a NFW profile
or a strong cooling flow instead of an isothermal $\beta$ model), then 
detection efficiency goes up (see \S 3.4) and the number density goes down.

In summary, we have shown that the N99 CXLF is robust, under the assumption
of standard parameters, despite the fact that it was derived using a 
less sophisticated set of selection function simulations than run 1. We 
find that the CXLF is not very sensitive to the values of certain parameters, 
$\Omega_0$, $\Omega_\Lambda$ and ellipticity. Other parameters have a
more significant effect on the CXLF; these are the values of $\beta$ and 
$r_c$ and the shape of the cluster surface brightness profile. If all 
clusters had NFW profiles, for example, then the N99 CXLF will 
significantly over estimate
the number density of high redshift clusters (since it was derived using
a selection function that assumed isothermal $\beta$ profiles). 
Alternatively, if all clusters had large core radii, then the N99 CXLF
will significantly under estimate the number density of clusters (since
it was derived using a selection function that assumed $r_c=250$ kpc).

\section{Discussion}

The evolution of the CXLF with redshift, especially at the bright
(i.e. high mass) end, provides - in principle - strong constrains on 
the value of $\Omega _m$ (see \S 1). If we see, for example, a much 
lower number density of high luminosity clusters at high redshift,
as opposed to the number density at low redshift, then we have strong 
support for a high $\Omega _m$ Universe (and vice versa) because 
$\Omega _{\Lambda}$ has little effect on the evolution of the CXLF out to 
at least z$\sim$1 (e.g. Holder et al. 2000). However, since
the existing cluster samples only contain very small numbers of
distant and luminous clusters, our ability to constrain 
$\Omega _m$ crucially depends on our ability to define the volume in
which those clusters were detected. For example, the Bright SHARC 
has only one cluster in the highest luminosity bin (RXJ0152). If we 
over (under) estimate the volume in which this cluster was detected 
then our number density will be too low (high) and our inferred value 
of $\Omega_0$ too high (low). 

In section 5 we attempted to determine the sensitivity of the Bright SHARC
CXLF to the assumptions that were incorporated into the selection function
simulations. We have shown that the initial choice of cosmological parameters 
has only a small influence on the CXLF. This result is fundamental because it 
means that, whatever cosmology is assumed when measuring the CXLF, we will 
still be able to probe CXLF evolution, and thus constrain the value of 
$\Omega_m$. The same cannot be said for the assumed surface brightness model; 
if all clusters follow NFW profiles or have strong cooling flows, we will 
significantly over-estimate the high $z$ cluster number density (and thus drive 
$\Omega_m$ down). By contrast, if we underestimate the value of $r_c$, or 
overestimate the value of $\beta$, we will significantly underestimate the high 
$z$ cluster number density (and thus drive $\Omega_m$ up).

Despite the fact that we have found cluster ellipticity to have minimal 
effect on our CXLF, we draw attention to the fact that, at high redshift,
highly elliptical clusters are easier to detect than circular ones in our
simulations. This may offer a partial explanation as to why the only 
$z>0.8$ X-ray cluster detected in the Bright SHARC survey (RXJ0152) is highly 
elliptical with a complex and elongated morphology (we missed probably the
faint $z>0.8$ more regular clusters). We note also that Ebeling et al. (1999)
claimed that high levels of substructure can also lead to decrease the 
probability of detection. 

With the X-ray data currently available we are not able to quantify 
just how much the various selection function assumptions might bias an  
$\Omega_m$ measurement. This is because we are not able to determine 
the distribution of $r_c$ and $\beta$ values for the Bright SHARC clusters, 
or determine what fraction of those clusters are better fit with NFW profiles 
than with isothermal $\beta$ profiles (see also Durret et al. 1994). To 
address this, we have begun a program to study cluster morphologies 
(and temperature profiles) with the new X-ray satellites XMM and 
{\it Chandra}. The follow-up of 
known clusters, such as those in the EMSS or Bright SHARC samples, will 
be one of major contributions of XMM and {\it Chandra} to cosmology. 
Both satellites have smaller fields of view than did ROSAT and so do not 
lend themselves well to serendipitous, or dedicated, cluster surveys. 
The catalogs derived from the ROSAT archive (see also Ebeling et al.
2000) will remain the pre-eminent source of high redshift X-ray clusters for 
some time to come. The recent work by Romer et al. (2000b) 
has demonstrated that in 5-10 years, XMM will have covered sufficient area to 
sufficient depth to allow new cluster catalogs to be created.
These new catalogs will contain more high redshift, high luminosity
clusters than the ROSAT and EMSS samples and so will provide much
better constraints on $\Omega_m$. These catalogs will require 
detailed simulations in order to determine their selection function. 
The work presented herein provides important guidelines as to how 
those simulations should be carried out.

Previously, several other groups (e.g. Vikhlinin et al. 1998, Burke et al.
1997, Scharf et al. 1997 or Rosati et al. 1995) have simulated 
selection functions in order to measure their CXLF's. 
Their methods were generally the same as those in this paper. All the
morphological parameters tested here were, however, not included, preventing 
them from showing the dependence of the selection function with the X-ray 
morphology of the clusters. In a future work we plan to apply our results to
the faint SHARC sample (including the Southern SHARC by Burke et al. 1997) and 
to derive the limits to the evolution of the CXLF based on the uncertainties 
we have determined.

\section{Summary and Conclusions}

The areal coverage of serendipitous surveys, such  as SHARC is, as we have
shown here, poorly determined at faint  flux limits.  For the Bright SHARC CXLF
we have demonstrated  that all the clusters in this sample are bright enough
that the error in the areal coverage is $\sim 5\%$. The uncertainty of the
completeness of the survey then,  becomes important when the areal coverage
uncertainty is so small.

We have carried out a detailed set of simulations in order to probe the 
effects of certain assumptions about distant clusters and the geometry of the
Universe on the completeness of the Bright SHARC survey.    These assumptions
have ramifications for the Cluster X-ray Luminosity Function  (CXLF) derived
for the SHARC survey and, ultimately, for the value of the  matter density
($\Omega_m$) derived from CXLF evolution. Under the assumption  of a standard
set of morphological parameters, we find that the Bright SHARC CXLF as
determined by N99 (using a less sophisticated set of selection function 
simulations) is robust. The new Bright SHARC CXLF presented in Figure 14
agrees  (within the 1 sigma envelope) with earlier estimates of the SHARC CXLF
(Nichol  et al. 1997, N99, Burke et al. 1997) and shows no statistical evidence
for  strong evolution at any luminosity out to z=0.7. At present, we are unable
to make any definitive statement about a possible deficit of high redshift,
high  luminosity clusters in the Bright SHARC (see N99) since we are still
hampered  by a combination of small number statistics, uncertainties in the
local CXLF,  incompleteness in our optical and X--ray follow-up as well as the
systematic  uncertainties in the Bright SHARC selection due to the unknown
surface  brightness profiles of distant clusters. 

We find that certain assumptions have little effect on the detection efficiency
of the survey: the target of the X-ray pointing, the moderate (e$\leq$0.3)
ellipticity of the cluster or the presence of moderate cooling flows.  None of 
these assumptions has a significant impact on our ability to measure cluster 
luminosities using the simple method adopted in R00. Other assumptions, 
specifically those associated with the cluster X-ray morphology (e.g. sharply 
peaked surface brightness profile or high ellipticity), can have a significant 
impact on cluster detectability (and the inferred completeness of the survey) 
and the derived CXLF.  We have shown, then, that even with an increased number
of detected clusters, the results based on attempting to measure CXLF
evolution to constrain $\Omega_m$ will remain highly uncertain.  Follow-up 
studies by {\it Chandra} and XMM that measure the X-ray morphology of distant 
clusters will, therefore, play a key role in the measurement of $\Omega_m$.

\acknowledgments

CA thanks the staff of the Dearborn Observatory for hospitality 
during his postdoctoral fellowship. The authors thank the Institut Gassendi 
pour la Recherche Astronomique en Provence for numerical support. The authors
thank the referee Harald Ebeling for detailed comments which have lead to a
greatly improved paper. This project was funded in part by NASA grants 
NAG5-2432 and NAG5-6548.

\clearpage

\figcaption[f1.ps]{{\it total detection efficiency} (see text for the
definition) for run 1 (standard parameter set). Each disk represents the 
results for a different redshift bin (we note that not all redshifts used in 
the simulations are represented). We have split each disk into 4 radial bins 
and 6 sectors, each of the 24 regions represents the results from a different 
luminosity ($L_{44}$= 1, 2, 3.5, 5, 7.5, 10) and off-axis annulus combination. 
The uncertainties on the detection levels are 9$\pm$5$\%$, therefore, less
than the color variation of the color bar between 0 and 20$\%$. \label{fig1}}

\figcaption[f2.ps]{As Fig. 1, but for the {\it Bright SHARC detection 
efficiency} (see text for definition) instead of the {\it total detection 
efficiency}. The uncertainties on the detection levels are 9$\pm$5$\%$, 
therefore, less than the color variation of the color bar between 0 and 
20$\%$. \label{fig2}}

\figcaption[f3.ps]{As Fig. 1, but here grayscale levels show
the {\it recovered luminosity} (see text for definition). The uncertainties on 
the {\it recovered luminosity} are 8$\pm$6$\%$, therefore, less than the color 
variation of the color bar between 25 and 55$\%$. \label{fig3}}

\figcaption[f4.ps]{Percentage difference (inner y-axes) between the {\it 
Bright SHARC detection efficiency} for run 1 ($\Omega _0$=1, $\Omega 
_\Lambda$=0) and 
run 2 ($\Omega _0$=0.1, $\Omega _\Lambda$=0, circles) and run 3 ($\Omega_0$
=0.4, $\Lambda$=0.6, filled squares). The outer x and y-axes give the 
cluster redshift and luminosity respectively. The inner x-axes are the 
distances from the ROSAT PSPC pointing center in ($15''$) pixels. The crosses 
depict redshift and luminosity combinations where the detection efficiency for 
run 1 was $<15\%$. The 1-$\sigma$ error bars are only plotted on the circles.
The error bars are approximatively the same size for the squares as for the 
open circles and are not shown for the squares in the figures. \label{fig4}}

\figcaption[f5.ps]{As Fig. 4, but here we test the effects of ellipticity on the
{\it total detection efficiency}. The circles represent 
the percentage difference between run 1 ($e=0.15$) and run 5 ($e=0.3$), 
whereas the filled squares represent the percentage difference between 
run 1 and run 4 ($e=0$). The 1-$\sigma$ error bars are only plotted on the 
circles. \label{fig5}}

\figcaption[f6.ps]{As Fig. 4, but here we test the effects of surface 
brightness profile on the {\it Bright SHARC detection efficiency}. The circles 
represent the detection difference between run 1 (isothermal 
$\beta$ profiles) and run 7 (moderate cooling flow profiles), whereas the 
filled squares represent the percentage difference between run 1 and run 6 
(NFW profiles). The 1-$\sigma$ error bars are only plotted on the circles.
\label{fig6}}

\figcaption[f7.ps]{As Fig. 6. Circles represent the {\it Bright 
SHARC detection efficiency} percentage difference between run 1 (isothermal 
$\beta$ profiles) and run 8 (strong cooling flow profiles), whereas the filled 
squares represent the percentage difference between run 1 and run 6 (NFW 
profiles). The 1-$\sigma$ error bars are only plotted on the circles.
 \label{fig7}}

\figcaption[f8.ps]{Percentage difference (inner 
y-axes) between the {\it Bright SHARC recovered luminosity} for run 1 
($\beta$=0.67) and run 8 (strong cooling-flow profile) and for run 1 
($\beta$=0.67) and run 6 (NFW profile, squares). The 1-$\sigma$ error bars 
are only plotted on the circles. \label{fig8}}

\figcaption[f9.ps]{As Fig. 4, but here we test the effects of varying $\beta$ 
on the {\it Bright SHARC 
detection efficiency}. The 
circles represent the percentage difference between run 1 ($\beta$=0.67) 
and run 9 ($\beta$=0.55), whereas the filled squares represent 
the percentage difference between run 1 and run 10 ($\beta$=0.75). 
The 1-$\sigma$ error bars are only plotted on the circles.
\label{fig9}}

\figcaption[f10.ps]{Percentage difference (inner y-axes) between the {\it Bright SHARC 
recovered luminosity} for run 1 ($\beta$=0.67) and run 9 ($\beta$=0.55, 
circles) and for run 1 ($\beta$=0.67) and run 10 ($\beta$=0.75, squares). The 
1-$\sigma$ error bars are only plotted on the circles. 
\label{fig10}}

\figcaption[f11.ps]{As Fig. 4, but here we test the effects of 
varying the core radius on the {\it Bright SHARC detection efficiency}. The 
circles represent the percentage difference between run 1 ($r_c$=250 kpc) 
and run 12 ($r_c$=400 kpc), whereas the filled squares represent 
the percentage difference between run 1 and run 11 ($r_c$=100 kpc). 
The 1-$\sigma$ error bars are only plotted on the circles.
\label{fig11}}

\figcaption[f12.ps]{The solid curve shows how the Bright SHARC areal coverage varies with 
extended source flux (where f$_{-13}$ is the 0.5-2.0 keV flux in units of 
$10^{-13}$ erg s$^{-1}$ cm$^{-2}$) or count rate. The two 
dotted curves depict the estimated 1-$\sigma$ errors. The thick vertical solid 
line is drawn where the 1-$\sigma$ errors are $\pm10\%$. The fluxes of each of 
the 13 Bright SHARC distant clusters (z$\geq$0.285) are plotted as thin 
vertical lines. The faintest cluster is depicted by the thin vertical solid 
line (here the error is $\pm5\%$). \label{fig12}}

\figcaption[f13.ps]{Effective area of the Bright sample as a function of 
cluster luminosity and redshift. The six curves represent the six different
input luminosities (in units of 10$^{44}$ ergs.s$^{-1}$). 
\label{fig13}}

\figcaption[f14.ps]{The Bright SHARC CXLF as a function of assumed 
cosmological model.
The x-axis gives the log of the cluster luminosity (erg s$^{-1}$) 
in the [0.5-2.0 keV] band. The y-axis gives the log of cluster number
density in units of Mpc$^{-3}$ ($10^{44}$ erg s$^{-1}$)$^{-1}$. 
The error bars are shown for the run 1 ($\Omega _0$=1) points (circles).
We plot also the error envelope from N99 (two dashed lines). 
The two solid lines are the envelope of the local CXLF from De Grandi et al. 
(1999) and Ebeling et al. (1997). \label{fig14}}

\figcaption[f15.ps]{The Bright SHARC CXLF as a function of cluster morphology: 
(lower left) ellipticity ; (lower right) core radius ; (upper left) slope 
$\beta$ ; (upper right) surface brightness profile. 
The x-axis gives the log of the cluster luminosity (erg s$^{-1}$) 
in the [0.5-2.0 keV] band. The y-axis gives the log of cluster number
density in units of Mpc$^{-3}$ ($10^{44}$ erg s$^{-1}$)$^{-1}$. 
The error bars are shown for the run 1 points (circles).
The two solid lines are the envelope of the local CXLF from De Grandi et al. 
(1999) and Ebeling et al. (1997).
\label{fig15}}

\clearpage

\begin{deluxetable}{cccccccc}
\footnotesize
\tablecaption{The parameters used for each of the 16 different simulation runs:
run number, surface brightness profile, ellipticity, core radius, 
profile slope, $\Omega _0$, $\Lambda$, number of pointings and
pointing type. 
\label{tbl-1}}
\tablewidth{0pt}
\tablehead{
\colhead{run} & \colhead{profile} & \colhead{$e$} & \colhead{$r_c$ (kpc)} & 
\colhead{$\beta$} & \colhead{$\Omega _m$} & \colhead{$\Lambda$} & 
 \colhead{ptg. type}
}
\startdata
1 & Isothermal & 0.15 & 250 & 0.67 & 1. & 0.    & all \\ 
2 & Isothermal & 0.15 & 250 & 0.67 & 0.1 & 0.   & all \\
3 & Isothermal & 0.15 & 250 & 0.67 & 0.4 & 0.6  &all \\
4 & Isothermal & 0.   & 250 & 0.67 & 1. & 0.   & all \\
5 & Isothermal & 0.30 & 250 & 0.67 & 1. & 0.   & all \\
6 & NFW        & 0.15 & 250 & 0.67 & 1. & 0.   & all \\ 
7 & moderate cool. flow & 0.15 & 250 & 0.67 & 1. & 0. & all \\ 
8 & strong cool. flow & 0.15 & 250 & 0.67 & 1. & 0. & all \\ 
9 & Isothermal & 0.15 & 250 & 0.55 & 1. & 0.  & all \\ 
10 & Isothermal & 0.15 & 250 & 0.75 & 1. & 0.  & all \\ 
11 & Isothermal & 0.15 & 100 & 0.67 & 1. & 0.   & all \\
12 & Isothermal & 0.15 & 400 & 0.67 & 1. & 0.   & all \\
13 & Isothermal & 0.15 & 250 & 0.67 & 1. & 0.   & type 2 \\ 
14 & Isothermal & 0.15 & 250 & 0.67 & 1. & 0.   & type 3 \\ 
15 & Isothermal & 0.15 & 250 & 0.67 & 1. & 0.   & type 6 \\ 
16 & Isothermal & 0.15 & 250 & 0.67 & 1. & 0.  & type 7 \\ 
17 & Isothermal & 0.15 & 250 & 0.67 & 1. & 0.   & type 8 \\ 
18 & Isothermal & 0.15 & 250 & 0.67 & 1. & 0.   & type 9 \\ 
19 & Isothermal & 0.15 & 250 & 0.67 & 1. & 0.   & point sources \\ 
20 & Isothermal & 0.15 & 250 & 0.67 & 1. & 0.   & diffuse sources \\ 
\enddata
\end{deluxetable}

\clearpage

\begin{deluxetable}{ccccccccccccccc}
\footnotesize
\tablecaption{Percentage of {\it Bright SHARC detection efficiency} 
for the 15 Bright SHARC clusters (noted Cl.) with $z>0.285$ for the run 1 
to 12. The redshift z, luminosity L$_{44}$ and Offaxis angle (in pixels) are 
from R00. The two clusters labelled by a ``*'' have been discarded from the 
CXLF analysis because their mean {\it Bright SHARC detection efficiency}, over 
simulation runs 1 to 12, was lower than 5$\%$. We note than when an individual 
success rate (one of the numbers in this table) was less than 5$\%$, we have 
used the value 5$\%$ (see text \S 3.1) as the efficiency to compute the CXLF's.
\label{tbl-2}}
\tablewidth{0pt}
\tablehead{
\colhead{Cl.} & \colhead{z} & \colhead{L$_{44}$} & \colhead{Off} & 
\colhead{1} & \colhead{2} & \colhead{3} & \colhead{4} & 
\colhead{5} & \colhead{6} & \colhead{7} & \colhead{8} & 
\colhead{9} & \colhead{11} & \colhead{12}
}
\startdata
0152 & 0.83 & 8.26 & 57 & 11.5 &  4.5 &  5.9 & 15.2 &  7.4 & 12.6 & 14.4 & 
73.1 & 15.7 &  4.1 & 26.6 \\
0221 & 0.45 & 2.87 & 74 & 25.9 & 10.2 &  8.7 & 21.6 &  8.1 & 19.2 & 22.9 & 
63.3 & 20.2 &  0.6 & 38.3 \\
0256 & 0.36 & 4.16 & 84 & 62.9 & 61.0 & 60.0 & 25.9 & 34.0 & 59.6 & 62.8 & 
75.4 & 56.2 & 20.7 & 61.5 \\
0318 & 0.37 & 3.28 & 43 & 70.4 & 62.2 & 55.7 & 60.2 & 34.2 & 71.8 & 70.8 & 
88.9 & 66.0 &  2.7 & 85.4 \\
0426 & 0.38 & 3.20 & 23 & 46.3 & 41.9 & 41.0 & 25.4 & 22.3 & 47.0 & 48.9 & 
69.4 & 38.1 &  4.2 & 53.6 \\
1120 & 0.60 & 5.01 & 48 & 19.6 &  3.5 &  3.1 & 19.0 &  3.2 & 13.6 & 13.9 & 
89.7 & 15.6 &  0.7 & 30.2 \\
1211 & 0.34 & 1.59 & 62 & 14.4 &  7.5 &  6.5 & 28.0 &  3.2 & 15.3 & 13.5 & 
54.1 & 15.5 &  0.2 &30.8 \\
1241 & 0.39 & 3.10 & 77 & 45.8 & 36.2 & 34.7 & 32.1 & 19.3 & 45.5 & 46.0 & 
71.6 & 37.3 &  2.1 & 57.4 \\
\enddata
\end{deluxetable}

\clearpage

\begin{deluxetable}{cccccccccccc}
\footnotesize
\tablecaption{Detection efficiency percentages for run 1 (standard 
parameter set)\label{tbl-3}}
\tablewidth{0pt}
\tablehead{
\multicolumn{6}{c}{{\it total detection efficiency}$^a$} & \multicolumn{6}{c}{
{\it Bright SHARC detection efficiency}$^a$} \\
\cline{1-6} \cline{7-12} \\
\colhead{z} & \colhead{$L_{44}$} & \colhead{1$^{st}$} & \colhead{2$^{nd}$} & 
\colhead{3$^{rd}$} & \colhead{4$^{th}$} &
\colhead{$z$} & \colhead{$L_{44}$} & \colhead{1$^{st}$} & \colhead{2$^{nd}$} & 
\colhead{3$^{rd}$} & \colhead{4$^{th}$}
}
\startdata
0.25 & 1.0 &  40$\pm$11 &  42$\pm$12 &  30$\pm$17 &  25$\pm$8 & 0.25 & 1.0 &   2$\pm$ 5 &   7$\pm$13 &   7$\pm$11 &  17$\pm$9 \\
0.30 & 1.0 &  27$\pm$7 &  33$\pm$16 &  28$\pm$18 &  20$\pm$19 & 0.30 & 1.0 &   0$\pm$2 &   1$\pm$7 &   2$\pm$12 &  13$\pm$12 \\
0.35 & 1.0 &  12$\pm$10 &  16$\pm$10 &  17$\pm$7 &   5$\pm$9 & 0.35 & 1.0 &   
0$\pm$3 &   0$\pm$5 &   0$\pm$2 &   0$\pm$4 \\
0.40 & 1.0 &   4$\pm$13 &  14$\pm$17 &   5$\pm$7 &   5$\pm$6 & 0.40 & 1.0 &   0$\pm$2 &   1$\pm$2 &   0$\pm$2 &   2$\pm$4 \\
0.45 & 1.0 &   3$\pm$7 &   5$\pm$11 &   3$\pm$8 &   1$\pm$6 & 0.45 & 1.0 &   0$\pm$2 &   0$\pm$2 &   0$\pm$2 &   0$\pm$4 \\
0.25 & 2.0 &  74$\pm$9 &  72$\pm$9 &  73$\pm$13 &  56$\pm$7 & 0.25 & 2.0 &  73$\pm$9 &  72$\pm$9 &  73$\pm$13 &  56$\pm$17 \\
0.30 & 2.0 &  68$\pm$12 &  76$\pm$11 &  58$\pm$11 &  54$\pm$8 & 0.30 & 2.0 &  49$\pm$14 &  68$\pm$12 &  51$\pm$11 &  54$\pm$8 \\
0.35 & 2.0 &  60$\pm$12 &  71$\pm$8 &  42$\pm$21 &  40$\pm$11 & 0.35 & 2.0 &  14$\pm$12 &  23$\pm$11 &  16$\pm$20 &  36$\pm$11 \\
\enddata
\tablenotetext{a}{The percentage 
{\it total detection efficiency} (left columns) and the 
{\it Bright SHARC detection efficiency} (right columns). The columns 1 and 7 give
the cluster redshift, columns 2 and 8 the cluster luminosities,
columns 3 and 9 the percentage detection efficiency 
in the inner annulus, columns 4 and 10 the percentage detection efficiency
in the second annulus, etc.}
\end{deluxetable}

\clearpage

\begin{deluxetable}{cccccccccccc}
\footnotesize
\tablecaption{Recovered luminosity percentages for run 1 (standard 
parameter set) \label{tbl-4}}
\tablewidth{0pt}
\tablehead{
\multicolumn{6}{c}{total sample$^a$} & \multicolumn{6}{c}{Bright 
SHARC sample$^a$} \\
\cline{1-6} \cline{7-12} \\
\colhead{z} & \colhead{$L_{44}$} & \colhead{1$^{st}$} & \colhead{2$^{nd}$} & 
\colhead{3$^{rd}$} & \colhead{4$^{th}$} &
\colhead{$z$} & \colhead{$L_{44}$} & \colhead{1$^{st}$} & \colhead{2$^{nd}$} & 
\colhead{3$^{rd}$} & \colhead{4$^{th}$}
}
\startdata
0.25 & 1.0 & 109$\pm$23 & 100$\pm$22 & 104$\pm$8 & 115$\pm$24 & 0.25 & 1.0 & 130$\pm$8 & 125$\pm$8 & 110$\pm$8 & 119$\pm$24 \\
0.30 & 1.0 & 108$\pm$27 &  94$\pm$20 & 102$\pm$26 & 124$\pm$13 & 0.30 & 1.0 &   0 & 173$\pm$8 & 176$\pm$8 & 132$\pm$8 \\
0.35 & 1.0 & 112$\pm$36 & 107$\pm$14 & 106$\pm$25 & 116$\pm$13 & 0.35 & 1.0 &   0 &   0 &   0 &   0 \\
0.40 & 1.0 & 141$\pm$83 & 121$\pm$23 & 130$\pm$13 & 184$\pm$13 & 0.40 & 1.0 &   0 & 200$\pm$8 &   0 & 999 \\
0.45 & 1.0 &  90$\pm$16 & 149$\pm$19 & 156$\pm$13 & 128$\pm$13 & 0.45 & 1.0 &   0 &   0 &   0 &   0 \\
0.25 & 2.0 & 106$\pm$13 & 103$\pm$4 & 102$\pm$11 & 116$\pm$6 & 0.25 & 2.0 & 106$\pm$13 & 103$\pm$4 & 102$\pm$11 & 116$\pm$6 \\
0.30 & 2.0 &  97$\pm$16 &  94$\pm$13 &  99$\pm$9 & 116$\pm$11 & 0.30 & 2.0 &  99$\pm$16 &  96$\pm$13 &  99$\pm$9 & 116$\pm$11 \\
0.35 & 2.0 &  99$\pm$22 &  95$\pm$14 &  97$\pm$20 & 109$\pm$10 & 0.35 & 2.0 & 124$\pm$28 & 107$\pm$4 & 105$\pm$8 & 111$\pm$10 \\
\enddata
\tablenotetext{a}
{The {\it recovered luminosity} (in percent) when no count rate limit is 
imposed 
(left columns) and when the Bright SHARC count rate limit is
imposed (right columns). The columns 1 and 7 give
the cluster redshift, columns 2 and 8 the cluster luminosities,
columns 3 and 9 the percentage detection efficiency 
in the inner annulus, columns 4 and 10 the percentage detection efficiency
in the second annulus, etc. Incidences of 999 indicate places where 
we recovered more than 200$\%$ of the input luminosity, i.e. spurious
detections.}

\end{deluxetable}

\clearpage

\begin{deluxetable}{cccccccccccc}
\footnotesize
\tablecaption{Detection efficiency percentage for the NFW set of 
parameters\label{tbl-3}}
\tablewidth{0pt}
\tablehead{
\multicolumn{6}{c}{{\it total detection efficiency}} & \multicolumn{6}{c}{
{\it Bright SHARC detection efficiency}} \\
\cline{1-6} \cline{7-12} \\
\colhead{z} & \colhead{$L_{44}$} & \colhead{1st$^a$} & \colhead{2nd$^a$} & 
\colhead{3rd$^a$} & \colhead{4th$^a$} &
\colhead{z} & \colhead{$L_{44}$} & \colhead{1st$^a$} & \colhead{2nd$^a$} & 
\colhead{3rd$^a$} & \colhead{4th$^a$}
}
\startdata
0.25 & 1.0 &  52 &  56 &  45 &  46 & 0.25 & 1.0 &  52 &  56 &  45 &  46 \\
0.30 & 1.0 &  54 &  48 &  39 &  54 & 0.30 & 1.0 &  53 &  48 &  39 &  54 \\
0.35 & 1.0 &  47 &  42 &  37 &  41 & 0.35 & 1.0 &  24 &  33 &  30 &  41 \\
0.40 & 1.0 &  34 &  27 &  30 &  22 & 0.40 & 1.0 &   3 &   5 &   7 &  14 \\
0.45 & 1.0 &  13 &  20 &  11 &  11 & 0.45 & 1.0 &   2 &   2 &   0 &   1 \\
0.25 & 2.0 &  60 &  78 &  67 &  62 & 0.25 & 2.0 &  60 &  78 &  67 &  62 \\
0.30 & 2.0 &  70 &  77 &  68 &  65 & 0.30 & 2.0 &  70 &  77 &  68 &  65 \\
0.35 & 2.0 &  67 &  70 &  69 &  55 & 0.35 & 2.0 &  66 &  70 &  69 &  55 \\
\enddata

\end{deluxetable}

\clearpage

\begin{deluxetable}{cccccccccccc}
\footnotesize
\tablecaption{Recovered luminosity percentage for the NFW set of 
parameters\label{tbl-3}}
\tablewidth{0pt}
\tablehead{
\multicolumn{6}{c}{total sample} & \multicolumn{6}{c}{Bright 
SHARC sample} \\
\cline{1-6} \cline{7-12} \\
\colhead{z} & \colhead{$L_{44}$} & \colhead{1st$^a$} & \colhead{2nd$^a$} & 
\colhead{3rd$^a$} & \colhead{4th$^a$} &
\colhead{z} & \colhead{$L_{44}$} & \colhead{1st$^a$} & \colhead{2nd$^a$} & 
\colhead{3rd$^a$} & \colhead{4th$^a$}
}
\startdata
0.25 & 1.0 & 102 & 101 & 100 & 113 & 0.25 & 1.0 & 102 & 101 & 100 & 113 \\
0.30 & 1.0 &  95 &  96 &  98 & 110 & 0.30 & 1.0 &  95 &  96 &  98 & 110 \\
0.35 & 1.0 &  96 &  93 &  96 & 108 & 0.35 & 1.0 &  96 &  96 &  96 & 108 \\
0.40 & 1.0 &  94 &  99 &  91 & 109 & 0.40 & 1.0 &  98 & 130 & 107 & 113 \\
0.45 & 1.0 & 123 & 102 &  93 & 110 & 0.45 & 1.0 & 168 & 168 &   0 & 125 \\
0.25 & 2.0 &  99 & 100 & 100 & 110 & 0.25 & 2.0 &  99 & 100 & 100 & 110 \\
0.30 & 2.0 &  95 &  95 &  96 & 111 & 0.30 & 2.0 &  95 &  95 &  96 & 111 \\
0.35 & 2.0 &  94 &  92 &  92 & 114 & 0.35 & 2.0 &  93 &  92 &  92 & 114 \\
\enddata

\end{deluxetable}

\clearpage

\begin{deluxetable}{cccccccccccc}
\footnotesize
\tablecaption{Detection efficiency percentage for the $\Omega _0$=0.1 set of 
parameters\label{tbl-3}}
\tablewidth{0pt}
\tablehead{
\multicolumn{6}{c}{{\it total detection efficiency}} & \multicolumn{6}{c}{
{\it Bright SHARC detection efficiency}} \\
\cline{1-6} \cline{7-12} \\
\colhead{z} & \colhead{$L_{44}$} & \colhead{1st$^a$} & \colhead{2nd$^a$} & 
\colhead{3rd$^a$} & \colhead{4th$^a$} &
\colhead{z} & \colhead{$L_{44}$} & \colhead{1st$^a$} & \colhead{2nd$^a$} & 
\colhead{3rd$^a$} & \colhead{4th$^a$}
}
\startdata
0.25 & 1.0 &  29 &  39 &  32 &  14 & 0.25 & 1.0 &   4 &   6 &   9 &  10 \\
0.30 & 1.0 &  19 &  20 &  14 &   8 & 0.30 & 1.0 &   0 &   1 &   1 &   1 \\
0.35 & 1.0 &   9 &  13 &   5 &   6 & 0.35 & 1.0 &   1 &   1 &   0 &   0 \\
0.40 & 1.0 &   7 &   4 &   1 &   2 & 0.40 & 1.0 &   0 &   0 &   0 &   0 \\
0.45 & 1.0 &   2 &   1 &   1 &   1 & 0.45 & 1.0 &   0 &   0 &   0 &   0 \\
0.25 & 2.0 &  72 &  76 &  56 &  59 & 0.25 & 2.0 &  67 &  75 &  56 &  58 \\
0.30 & 2.0 &  64 &  72 &  55 &  54 & 0.30 & 2.0 &  17 &  53 &  41 &  51 \\
0.35 & 2.0 &  44 &  53 &  43 &  30 & 0.35 & 2.0 &   3 &   4 &   5 &  18 \\
\enddata

\end{deluxetable}

\clearpage

\begin{deluxetable}{cccccccccccc}
\footnotesize
\tablecaption{Recovered luminosity percentage for the $\Omega _0$=0.1 set of 
parameters\label{tbl-3}}
\tablewidth{0pt}
\tablehead{
\multicolumn{6}{c}{total sample} & \multicolumn{6}{c}{Bright 
SHARC sample} \\
\cline{1-6} \cline{7-12} \\
\colhead{z} & \colhead{$L_{44}$} & \colhead{1st$^a$} & \colhead{2nd$^a$} & 
\colhead{3rd$^a$} & \colhead{4th$^a$} &
\colhead{z} & \colhead{$L_{44}$} & \colhead{1st$^a$} & \colhead{2nd$^a$} & 
\colhead{3rd$^a$} & \colhead{4th$^a$}
}
\startdata
0.25 & 1.0 & 109 & 105 & 108 & 118 & 0.25 & 1.0 & 147 & 141 & 125 & 119 \\
0.30 & 1.0 & 111 &  95 & 105 & 105 & 0.30 & 1.0 &   0 & 140 & 127 & 139 \\
0.35 & 1.0 & 133 & 111 & 107 & 124 & 0.35 & 1.0 & 141 & 999 &   0 &   0 \\
0.40 & 1.0 & 138 & 119 &  64 &  99 & 0.40 & 1.0 &   0 &   0 &   0 &   0 \\
0.45 & 1.0 &  75 &  81 & 192 &  99 & 0.45 & 1.0 &   0 &   0 &   0 &   0 \\
0.25 & 2.0 & 103 &  99 & 100 & 113 & 0.25 & 2.0 & 103 &  99 & 100 & 113 \\
0.30 & 2.0 &  95 &  94 &  93 & 114 & 0.30 & 2.0 &  98 &  98 &  96 & 116 \\
0.35 & 2.0 &  90 &  90 &  88 & 116 & 0.35 & 2.0 & 107 & 146 & 106 & 121 \\
\enddata

\end{deluxetable}

\clearpage

\begin{deluxetable}{cccccccccccc}
\footnotesize
\tablecaption{Detection efficiency percentage for the $\Omega _0$=0.4 + 
$\Lambda$=0.6 set of parameters\label{tbl-3}}
\tablewidth{0pt}
\tablehead{
\multicolumn{6}{c}{{\it total detection efficiency}} & \multicolumn{6}{c}{
{\it Bright SHARC detection efficiency}} \\
\cline{1-6} \cline{7-12} \\
\colhead{z} & \colhead{$L_{44}$} & \colhead{1st$^a$} & \colhead{2nd$^a$} & 
\colhead{3rd$^a$} & \colhead{4th$^a$} &
\colhead{z} & \colhead{$L_{44}$} & \colhead{1st$^a$} & \colhead{2nd$^a$} & 
\colhead{3rd$^a$} & \colhead{4th$^a$}
}
\startdata
0.25 & 1.0 &  39 &  41 &  38 &  30 & 0.25 & 1.0 &   4 &   4 &   7 &  17 \\
0.30 & 1.0 &  21 &  26 &  13 &   9 & 0.30 & 1.0 &   0 &   2 &   0 &   0 \\
0.35 & 1.0 &  10 &  10 &   9 &   9 & 0.35 & 1.0 &   0 &   0 &   0 &   1 \\
0.40 & 1.0 &   7 &   7 &   1 &   2 & 0.40 & 1.0 &   0 &   2 &   0 &   1 \\
0.45 & 1.0 &   3 &   2 &   2 &   1 & 0.45 & 1.0 &   0 &   0 &   0 &   0 \\
0.25 & 2.0 &  75 &  78 &  60 &  53 & 0.25 & 2.0 &  67 &  78 &  57 &  53 \\
0.30 & 2.0 &  63 &  73 &  54 &  40 & 0.30 & 2.0 &  23 &  40 &  30 &  39 \\
0.35 & 2.0 &  49 &  51 &  46 &  37 & 0.35 & 2.0 &   2 &   4 &   6 &  23 \\
\enddata

\end{deluxetable}

\clearpage

\begin{deluxetable}{cccccccccccc}
\footnotesize
\tablecaption{Recovered luminosity percentage for the $\Omega _0$=0.4 + 
$\Lambda$=0.6 set of parameters\label{tbl-3}}
\tablewidth{0pt}
\tablehead{
\multicolumn{6}{c}{total sample} & \multicolumn{6}{c}{Bright 
SHARC sample} \\
\cline{1-6} \cline{7-12} \\
\colhead{z} & \colhead{$L_{44}$} & \colhead{1st$^a$} & \colhead{2nd$^a$} & 
\colhead{3rd$^a$} & \colhead{4th$^a$} &
\colhead{z} & \colhead{$L_{44}$} & \colhead{1st$^a$} & \colhead{2nd$^a$} & 
\colhead{3rd$^a$} & \colhead{4th$^a$}
}
\startdata
0.25 & 1.0 & 120 & 106 & 106 & 117 & 0.25 & 1.0 & 166 & 145 & 132 & 126 \\
0.30 & 1.0 & 110 & 111 & 101 & 123 & 0.30 & 1.0 &   0 & 167 &   0 &   0 \\
0.35 & 1.0 & 126 & 133 & 139 & 137 & 0.35 & 1.0 &   0 &   0 &   0 & 999 \\
0.40 & 1.0 & 163 & 999 & 121 & 999 & 0.40 & 1.0 &   0 & 999 &   0 & 999 \\
0.45 & 1.0 & 999 & 156 & 999 & 999 & 0.45 & 1.0 &   0 &   0 &   0 &   0 \\
0.25 & 2.0 & 109 & 104 & 102 & 118 & 0.25 & 2.0 & 107 & 104 & 102 & 118 \\
0.30 & 2.0 & 106 &  96 &  97 & 114 & 0.30 & 2.0 & 118 & 101 & 103 & 114 \\
0.35 & 2.0 &  98 &  95 &  99 & 117 & 0.35 & 2.0 & 143 & 146 & 142 & 118 \\
\enddata

\end{deluxetable}

\clearpage

\begin{deluxetable}{cccccccccccc}
\footnotesize
\tablecaption{Detection efficiency percentage for the moderate
cooling-flow-like set of parameters\label{tbl-3}}
\tablewidth{0pt}
\tablehead{
\multicolumn{6}{c}{{\it total detection efficiency}} & \multicolumn{6}{c}{
{\it Bright SHARC detection efficiency}} \\
\cline{1-6} \cline{7-12} \\
\colhead{z} & \colhead{$L_{44}$} & \colhead{1st$^a$} & \colhead{2nd$^a$} & 
\colhead{3rd$^a$} & \colhead{4th$^a$} &
\colhead{z} & \colhead{$L_{44}$} & \colhead{1st$^a$} & \colhead{2nd$^a$} & 
\colhead{3rd$^a$} & \colhead{4th$^a$}
}
\startdata
0.25 & 1.0 &  31 &  48 &  24 &  34 & 0.25 & 1.0 &   1 &   8 &  12 &  25 \\
0.30 & 1.0 &  13 &  28 &   9 &  12 & 0.30 & 1.0 &   0 &   1 &   0 &   8 \\
0.35 & 1.0 &   8 &   9 &   7 &   3 & 0.35 & 1.0 &   0 &   0 &   0 &   1 \\
0.40 & 1.0 &   4 &   5 &   3 &   4 & 0.40 & 1.0 &   0 &   0 &   0 &   1 \\
0.45 & 1.0 &   2 &   4 &   1 &   1 & 0.45 & 1.0 &   0 &   0 &   0 &   0 \\
0.25 & 2.0 &  79 &  60 &  50 &  47 & 0.25 & 2.0 &  74 &  60 &  50 &  47 \\
0.30 & 2.0 &  57 &  57 &  51 &  41 & 0.30 & 2.0 &  36 &  52 &  45 &  40 \\
0.35 & 2.0 &  49 &  53 &  40 &  30 & 0.35 & 2.0 &   8 &  22 &  21 &  27 \\
\enddata

\end{deluxetable}

\clearpage

\begin{deluxetable}{cccccccccccc}
\footnotesize
\tablecaption{Recovered luminosity percentage for the moderate 
cooling-flow-like set of parameters\label{tbl-3}}
\tablewidth{0pt}
\tablehead{
\multicolumn{6}{c}{total sample} & \multicolumn{6}{c}{Bright 
SHARC sample} \\
\cline{1-6} \cline{7-12} \\
\colhead{z} & \colhead{$L_{44}$} & \colhead{1st$^a$} & \colhead{2nd$^a$} & 
\colhead{3rd$^a$} & \colhead{4th$^a$} &
\colhead{z} & \colhead{$L_{44}$} & \colhead{1st$^a$} & \colhead{2nd$^a$} & 
\colhead{3rd$^a$} & \colhead{4th$^a$}
}
\startdata
0.25 & 1.0 & 105 &  99 & 105 & 111 & 0.25 & 1.0 & 123 & 119 & 112 & 117 \\
0.30 & 1.0 & 111 &  95 & 108 & 146 & 0.30 & 1.0 &   0 & 116 &   0 & 161 \\
0.35 & 1.0 & 170 & 119 &  98 & 145 & 0.35 & 1.0 &   0 &   0 &   0 & 186 \\
0.40 & 1.0 & 107 & 124 & 106 & 153 & 0.40 & 1.0 &   0 &   0 &   0 & 999 \\
0.45 & 1.0 & 999 & 159 & 131 & 124 & 0.45 & 1.0 &   0 &   0 &   0 &   0 \\
0.25 & 2.0 & 106 & 103 & 100 & 113 & 0.25 & 2.0 & 106 & 103 & 100 & 113 \\
0.30 & 2.0 & 103 &  97 &  99 & 112 & 0.30 & 2.0 & 101 &  98 & 100 & 112 \\
0.35 & 2.0 &  96 &  92 &  95 & 116 & 0.35 & 2.0 & 115 & 100 & 101 & 117 \\
\enddata

\end{deluxetable}

\clearpage

\begin{deluxetable}{cccccccccccc}
\footnotesize
\tablecaption{Detection efficiency percentage for the major cooling-flow-like
set of parameters\label{tbl-3}}
\tablewidth{0pt}
\tablehead{
\multicolumn{6}{c}{{\it total detection efficiency}} & \multicolumn{6}{c}{
{\it Bright SHARC detection efficiency}} \\
\cline{1-6} \cline{7-12} \\
\colhead{z} & \colhead{$L_{44}$} & \colhead{1st$^a$} & \colhead{2nd$^a$} & 
\colhead{3rd$^a$} & \colhead{4th$^a$} &
\colhead{z} & \colhead{$L_{44}$} & \colhead{1st$^a$} & \colhead{2nd$^a$} & 
\colhead{3rd$^a$} & \colhead{4th$^a$}
}
\startdata
0.25 & 1.0 &  64 &  63 &  51 &  52 & 0.25 & 1.0 &  64 &  63 &  51 &  52 \\
0.30 & 1.0 &  41 &  38 &  28 &  45 & 0.30 & 1.0 &  40 &  38 &  28 &  45 \\
0.35 & 1.0 &   8 &  15 &   9 &  14 & 0.35 & 1.0 &   0 &   5 &   2 &  14 \\
0.40 & 1.0 &  83 &  57 &  60 &  52 & 0.40 & 1.0 &  49 &  33 &  35 &  43 \\
0.45 & 1.0 &  35 &  33 &  23 &  23 & 0.45 & 1.0 &  23 &  13 &  11 &  12 \\
0.25 & 2.0 &  62 &  80 &  68 &  63 & 0.25 & 2.0 &  62 &  80 &  68 &  63 \\
0.30 & 2.0 &  48 &  63 &  53 &  50 & 0.30 & 2.0 &  48 &  63 &  53 &  50 \\
0.35 & 2.0 & 100 & 100 & 100 &  97 & 0.35 & 2.0 & 100 & 100 & 100 &  97 \\
\enddata

\end{deluxetable}

\clearpage

\begin{deluxetable}{cccccccccccc}
\footnotesize
\tablecaption{Recovered luminosity percentage for the major cooling-flow-like 
set of parameters\label{tbl-3}}
\tablewidth{0pt}
\tablehead{
\multicolumn{6}{c}{total sample} & \multicolumn{6}{c}{Bright 
SHARC sample} \\
\cline{1-6} \cline{7-12} \\
\colhead{z} & \colhead{$L_{44}$} & \colhead{1st$^a$} & \colhead{2nd$^a$} & 
\colhead{3rd$^a$} & \colhead{4th$^a$} &
\colhead{z} & \colhead{$L_{44}$} & \colhead{1st$^a$} & \colhead{2nd$^a$} & 
\colhead{3rd$^a$} & \colhead{4th$^a$}
}
\startdata
0.25 & 1.0 &  78 &  76 &  76 &  86 & 0.25 & 1.0 &  78 &  76 &  76 &  86 \\
0.30 & 1.0 &  72 &  74 &  74 &  83 & 0.30 & 1.0 &  72 &  74 &  74 &  83 \\
0.35 & 1.0 &  72 &  71 &  73 &  81 & 0.35 & 1.0 &  72 &  73 &  73 &  81 \\
0.40 & 1.0 &  72 &  75 &  69 &  84 & 0.40 & 1.0 &  75 &  99 &  80 &  87 \\
0.45 & 1.0 &  94 &  77 &  72 &  84 & 0.45 & 1.0 & 128 & 127 &   1 &  96 \\
0.25 & 2.0 &  75 &  77 &  77 &  84 & 0.25 & 2.0 &  75 &  77 &  77 &  84 \\
0.30 & 2.0 &  72 &  73 &  73 &  84 & 0.30 & 2.0 &  72 &  73 &  73 &  84 \\
0.35 & 2.0 &  72 &  70 &  69 &  86 & 0.35 & 2.0 &  72 &  70 &  69 &  86 \\
\enddata

\end{deluxetable}

\clearpage

\begin{deluxetable}{cccccccccccc}
\footnotesize
\tablecaption{Detection efficiency percentage for the $\beta$=0.55
set of parameters\label{tbl-3}}
\tablewidth{0pt}
\tablehead{
\multicolumn{6}{c}{{\it total detection efficiency}} & \multicolumn{6}{c}{
{\it Bright SHARC detection efficiency}} \\
\cline{1-6} \cline{7-12} \\
\colhead{z} & \colhead{$L_{44}$} & \colhead{1st$^a$} & \colhead{2nd$^a$} & 
\colhead{3rd$^a$} & \colhead{4th$^a$} &
\colhead{z} & \colhead{$L_{44}$} & \colhead{1st$^a$} & \colhead{2nd$^a$} & 
\colhead{3rd$^a$} & \colhead{4th$^a$}
}
\startdata
0.25 & 1.0 &   2 &   5 &   1 &   2 & 0.25 & 1.0 &   0 &   2 &   0 &   0 \\
0.30 & 1.0 &   2 &   0 &   1 &   1 & 0.30 & 1.0 &   0 &   0 &   0 &   0 \\
0.35 & 1.0 &   0 &   3 &   2 &   0 & 0.35 & 1.0 &   0 &   1 &   0 &   0 \\
0.40 & 1.0 &   0 &   0 &   0 &   0 & 0.40 & 1.0 &   0 &   0 &   0 &   0 \\
0.45 & 1.0 &   0 &   1 &   0 &   0 & 0.45 & 1.0 &   0 &   0 &   0 &   0 \\
0.25 & 2.0 &   9 &  17 &   8 &  15 & 0.25 & 2.0 &   1 &   4 &   0 &   8 \\
0.30 & 2.0 &  10 &  11 &   8 &   5 & 0.30 & 2.0 &   0 &   2 &   1 &   2 \\
0.35 & 2.0 &   3 &   9 &   2 &   4 & 0.35 & 2.0 &   0 &   0 &   0 &   2 \\
\enddata

\end{deluxetable}

\clearpage

\begin{deluxetable}{cccccccccccc}
\footnotesize
\tablecaption{Recovered luminosity percentage for the $\beta$=0.55 
set of parameters\label{tbl-3}}
\tablewidth{0pt}
\tablehead{
\multicolumn{6}{c}{total sample} & \multicolumn{6}{c}{Bright 
SHARC sample} \\
\cline{1-6} \cline{7-12} \\
\colhead{z} & \colhead{$L_{44}$} & \colhead{1st$^a$} & \colhead{2nd$^a$} & 
\colhead{3rd$^a$} & \colhead{4th$^a$} &
\colhead{z} & \colhead{$L_{44}$} & \colhead{1st$^a$} & \colhead{2nd$^a$} & 
\colhead{3rd$^a$} & \colhead{4th$^a$}
}
\startdata
0.25 & 1.0 & 999 & 999 & 122 & 154 & 0.25 & 1.0 &   0 & 999 &   0 &   0 \\
0.30 & 1.0 & 999 &   0 & 104 & 999 & 0.30 & 1.0 &   0 &   0 &   0 &   0 \\
0.35 & 1.0 &   0 & 999 & 999 &   0 & 0.35 & 1.0 &   0 & 999 &   0 &   0 \\
0.40 & 1.0 &   0 &   0 &   0 &   0 & 0.40 & 1.0 &   0 &   0 &   0 &   0 \\
0.45 & 1.0 &   0 & 999 &   0 &   0 & 0.45 & 1.0 &   0 &   0 &   0 &   0 \\
0.25 & 2.0 & 114 & 102 & 106 & 112 & 0.25 & 2.0 & 999 & 114 &   0 & 123 \\
0.30 & 2.0 & 106 & 114 & 108 & 126 & 0.30 & 2.0 &   0 & 178 & 118 & 156 \\
0.35 & 2.0 & 110 & 102 & 148 & 161 & 0.35 & 2.0 &   0 &   0 &   0 & 999 \\
\enddata

\end{deluxetable}

\clearpage

\begin{deluxetable}{cccccccccccc}
\footnotesize
\tablecaption{Detection efficiency percentage for the $\beta$=0.75
set of parameters\label{tbl-3}}
\tablewidth{0pt}
\tablehead{
\multicolumn{6}{c}{{\it total detection efficiency}} & \multicolumn{6}{c}{
{\it Bright SHARC detection efficiency}} \\
\cline{1-6} \cline{7-12} \\
\colhead{z} & \colhead{$L_{44}$} & \colhead{1st$^a$} & \colhead{2nd$^a$} & 
\colhead{3rd$^a$} & \colhead{4th$^a$} &
\colhead{z} & \colhead{$L_{44}$} & \colhead{1st$^a$} & \colhead{2nd$^a$} & 
\colhead{3rd$^a$} & \colhead{4th$^a$}
}
\startdata
0.25 & 1.0 &  77 &  83 &  62 &  42 & 0.25 & 1.0 &  25 &  57 &  39 &  40 \\
0.30 & 1.0 &  56 &  72 &  43 &  36 & 0.30 & 1.0 &   3 &  12 &  10 &  31 \\
0.35 & 1.0 &  31 &  54 &  33 &  14 & 0.35 & 1.0 &   1 &   3 &   3 &   5 \\
0.40 & 1.0 &  26 &  31 &   9 &  11 & 0.40 & 1.0 &   0 &   4 &   0 &   1 \\
0.45 & 1.0 &  18 &  15 &   8 &   2 & 0.45 & 1.0 &   0 &   1 &   1 &   0 \\
0.25 & 2.0 &  79 &  91 &  79 &  48 & 0.25 & 2.0 &  79 &  91 &  79 &  48 \\
0.30 & 2.0 &  70 &  80 &  68 &  57 & 0.30 & 2.0 &  65 &  80 &  68 &  57 \\
0.35 & 2.0 &  69 &  82 &  64 &  36 & 0.35 & 2.0 &  25 &  50 &  43 &  35 \\
\enddata

\end{deluxetable}

\clearpage

\begin{deluxetable}{cccccccccccc}
\footnotesize
\tablecaption{Recovered luminosity percentage for the $\beta$=0.75 
set of parameters\label{tbl-3}}
\tablewidth{0pt}
\tablehead{
\multicolumn{6}{c}{total sample} & \multicolumn{6}{c}{Bright 
SHARC sample} \\
\cline{1-6} \cline{7-12} \\
\colhead{z} & \colhead{$L_{44}$} & \colhead{1st$^a$} & \colhead{2nd$^a$} & 
\colhead{3rd$^a$} & \colhead{4th$^a$} &
\colhead{z} & \colhead{$L_{44}$} & \colhead{1st$^a$} & \colhead{2nd$^a$} & 
\colhead{3rd$^a$} & \colhead{4th$^a$}
}
\startdata
0.25 & 1.0 & 109 & 101 & 104 & 116 & 0.25 & 1.0 & 120 & 104 & 110 & 117 \\
0.30 & 1.0 & 103 & 102 &  99 & 123 & 0.30 & 1.0 & 151 & 124 & 121 & 125 \\
0.35 & 1.0 & 111 & 103 & 102 & 136 & 0.35 & 1.0 & 999 & 147 & 161 & 151 \\
0.40 & 1.0 &  99 & 108 & 103 & 126 & 0.40 & 1.0 &   0 & 188 &   0 & 999 \\
0.45 & 1.0 & 118 & 120 & 113 & 178 & 0.45 & 1.0 &   0 & 999 & 999 &   0 \\
0.25 & 2.0 & 108 & 104 & 104 & 116 & 0.25 & 2.0 & 108 & 104 & 104 & 116 \\
0.30 & 2.0 & 102 &  95 &  95 & 116 & 0.30 & 2.0 & 100 &  95 &  95 & 116 \\
0.35 & 2.0 &  97 &  95 &  94 & 111 & 0.35 & 2.0 & 107 &  99 &  97 & 111 \\
\enddata

\end{deluxetable}

\clearpage

\begin{deluxetable}{cccccccccccc}
\footnotesize
\tablecaption{Detection efficiency percentage for the $c_r$=100 kpc
set of parameters\label{tbl-3}}
\tablewidth{0pt}
\tablehead{
\multicolumn{6}{c}{{\it total detection efficiency}} & \multicolumn{6}{c}{
{\it Bright SHARC detection efficiency}} \\
\cline{1-6} \cline{7-12} \\
\colhead{z} & \colhead{$L_{44}$} & \colhead{1st$^a$} & \colhead{2nd$^a$} & 
\colhead{3rd$^a$} & \colhead{4th$^a$} &
\colhead{z} & \colhead{$L_{44}$} & \colhead{1st$^a$} & \colhead{2nd$^a$} & 
\colhead{3rd$^a$} & \colhead{4th$^a$}
}
\startdata
0.25 & 1.0 &  77 &  89 &  71 &  21 & 0.25 & 1.0 &  25 &  41 &  35 &  21 \\
0.30 & 1.0 &  69 &  72 &  48 &  17 & 0.30 & 1.0 &   8 &  12 &   9 &  10 \\
0.35 & 1.0 &  35 &  47 &  21 &  12 & 0.35 & 1.0 &   2 &   2 &   1 &   2 \\
0.40 & 1.0 &  22 &  26 &  20 &   7 & 0.40 & 1.0 &   0 &   0 &   0 &   1 \\
0.45 & 1.0 &   6 &   6 &   8 &   3 & 0.45 & 1.0 &   0 &   0 &   0 &   0 \\
0.25 & 2.0 &  91 &  97 &  82 &  22 & 0.25 & 2.0 &  91 &  97 &  82 &  22 \\
0.30 & 2.0 &  80 &  85 &  66 &  29 & 0.30 & 2.0 &  80 &  85 &  66 &  29 \\
0.35 & 2.0 &  73 &  71 &  58 &  21 & 0.35 & 2.0 &  32 &  53 &  38 &  21 \\
\enddata

\end{deluxetable}

\clearpage

\begin{deluxetable}{cccccccccccc}
\footnotesize
\tablecaption{Recovered luminosity percentage for the $c_r$=100 kpc 
set of parameters\label{tbl-3}}
\tablewidth{0pt}
\tablehead{
\multicolumn{6}{c}{total sample} & \multicolumn{6}{c}{Bright 
SHARC sample} \\
\cline{1-6} \cline{7-12} \\
\colhead{z} & \colhead{$L_{44}$} & \colhead{1st$^a$} & \colhead{2nd$^a$} & 
\colhead{3rd$^a$} & \colhead{4th$^a$} &
\colhead{z} & \colhead{$L_{44}$} & \colhead{1st$^a$} & \colhead{2nd$^a$} & 
\colhead{3rd$^a$} & \colhead{4th$^a$}
}
\startdata
0.25 & 1.0 & 113 & 104 & 105 & 122 & 0.25 & 1.0 & 122 & 105 & 114 & 122 \\
0.30 & 1.0 & 116 & 108 & 106 & 123 & 0.30 & 1.0 & 157 & 152 & 142 & 121 \\
0.35 & 1.0 & 121 & 103 &  97 & 118 & 0.35 & 1.0 & 999 & 140 & 164 & 159 \\
0.40 & 1.0 & 124 &  97 & 111 & 168 & 0.40 & 1.0 &   0 &   0 &   0 & 999 \\
0.45 & 1.0 & 143 & 143 & 124 & 139 & 0.45 & 1.0 &   0 &   0 &   0 &   0 \\
0.25 & 2.0 & 108 & 103 & 108 & 125 & 0.25 & 2.0 & 108 & 103 & 108 & 125 \\
0.30 & 2.0 & 103 &  98 &  96 & 127 & 0.30 & 2.0 & 103 &  98 &  96 & 127 \\
0.35 & 2.0 &  99 &  94 &  94 & 125 & 0.35 & 2.0 & 105 &  96 &  98 & 125 \\
\enddata

\end{deluxetable}

\clearpage

\begin{deluxetable}{cccccccccccc}
\footnotesize
\tablecaption{Detection efficiency percentage for the $c_r$=400 kpc
set of parameters\label{tbl-3}}
\tablewidth{0pt}
\tablehead{
\multicolumn{6}{c}{{\it total detection efficiency}} & \multicolumn{6}{c}{
{\it Bright SHARC detection efficiency}} \\
\cline{1-6} \cline{7-12} \\
\colhead{z} & \colhead{$L_{44}$} & \colhead{1st$^a$} & \colhead{2nd$^a$} & 
\colhead{3rd$^a$} & \colhead{4th$^a$} &
\colhead{z} & \colhead{$L_{44}$} & \colhead{1st$^a$} & \colhead{2nd$^a$} & 
\colhead{3rd$^a$} & \colhead{4th$^a$}
}
\startdata
0.25 & 1.0 &  13 &   9 &   4 &   5 & 0.25 & 1.0 &   0 &   3 &   0 &   1 \\
0.30 & 1.0 &   8 &   7 &   3 &   2 & 0.30 & 1.0 &   0 &   0 &   0 &   0 \\
0.35 & 1.0 &   7 &   4 &   4 &   2 & 0.35 & 1.0 &   0 &   0 &   0 &   0 \\
0.40 & 1.0 &   2 &   1 &   4 &   3 & 0.40 & 1.0 &   0 &   0 &   0 &   1 \\
0.45 & 1.0 &   2 &   0 &   2 &   1 & 0.45 & 1.0 &   0 &   0 &   0 &   0 \\
0.25 & 2.0 &  27 &  32 &  18 &  24 & 0.25 & 2.0 &  19 &  31 &  14 &  23 \\
0.30 & 2.0 &  35 &  25 &  19 &  13 & 0.30 & 2.0 &   9 &  12 &   7 &  10 \\
0.35 & 2.0 &  14 &  23 &  14 &  18 & 0.35 & 2.0 &   3 &   3 &   5 &  11 \\
\enddata

\end{deluxetable}

\clearpage

\begin{deluxetable}{cccccccccccc}
\footnotesize
\tablecaption{Recovered luminosity percentage for the $c_r$=400 kpc 
set of parameters\label{tbl-3}}
\tablewidth{0pt}
\tablehead{
\multicolumn{6}{c}{total sample} & \multicolumn{6}{c}{Bright 
SHARC sample} \\
\cline{1-6} \cline{7-12} \\
\colhead{z} & \colhead{$L_{44}$} & \colhead{1st$^a$} & \colhead{2nd$^a$} & 
\colhead{3rd$^a$} & \colhead{4th$^a$} &
\colhead{z} & \colhead{$L_{44}$} & \colhead{1st$^a$} & \colhead{2nd$^a$} & 
\colhead{3rd$^a$} & \colhead{4th$^a$}
}
\startdata
0.25 & 1.0 &  98 & 101 &  99 & 104 & 0.25 & 1.0 &   0 & 113 &   0 &  90 \\
0.30 & 1.0 & 115 &  97 & 115 & 105 & 0.30 & 1.0 &   0 &   0 &   0 &   0 \\
0.35 & 1.0 & 120 &  77 & 113 &  87 & 0.35 & 1.0 &   0 &   0 &   0 &   0 \\
0.40 & 1.0 & 113 &  85 & 110 & 197 & 0.40 & 1.0 &   0 &   0 &   0 & 999 \\
0.45 & 1.0 &  99 &   0 & 121 & 192 & 0.45 & 1.0 &   0 &   0 &   0 &   0 \\
0.25 & 2.0 &  96 &  91 &  93 &  98 & 0.25 & 2.0 &  94 &  91 &  94 &  99 \\
0.30 & 2.0 &  90 &  83 &  97 &  97 & 0.30 & 2.0 &  94 &  87 & 123 & 101 \\
0.35 & 2.0 &  87 &  82 &  91 & 114 & 0.35 & 2.0 & 116 & 110 & 103 & 126 \\
\enddata

\end{deluxetable}

\clearpage

\begin{deluxetable}{cccccccccccc}
\footnotesize
\tablecaption{Detection efficiency percentage for the ellipticity equal to 0.01
set of parameters\label{tbl-3}}
\tablewidth{0pt}
\tablehead{
\multicolumn{6}{c}{{\it total detection efficiency}} & \multicolumn{6}{c}{
{\it Bright SHARC detection efficiency}} \\
\cline{1-6} \cline{7-12} \\
\colhead{z} & \colhead{$L_{44}$} & \colhead{1st$^a$} & \colhead{2nd$^a$} & 
\colhead{3rd$^a$} & \colhead{4th$^a$} &
\colhead{z} & \colhead{$L_{44}$} & \colhead{1st$^a$} & \colhead{2nd$^a$} & 
\colhead{3rd$^a$} & \colhead{4th$^a$}
}
\startdata
0.25 & 1.0 &  39 &  46 &  31 &  29 & 0.25 & 1.0 &   7 &  10 &   8 &  21 \\
0.30 & 1.0 &  25 &  33 &  17 &  16 & 0.30 & 1.0 &   2 &   1 &   2 &   3 \\
0.35 & 1.0 &  15 &  20 &  17 &  13 & 0.35 & 1.0 &   0 &   0 &   0 &   0 \\
0.40 & 1.0 &   6 &  16 &  11 &   5 & 0.40 & 1.0 &   0 &   1 &   0 &   0 \\
0.45 & 1.0 &   7 &  11 &   5 &   1 & 0.45 & 1.0 &   0 &   1 &   0 &   0 \\
0.25 & 2.0 &  63 &  74 &  68 &  44 & 0.25 & 2.0 &  59 &  74 &  68 &  44 \\
0.30 & 2.0 &  67 &  75 &  52 &  46 & 0.30 & 2.0 &  42 &  66 &  50 &  46 \\
0.35 & 2.0 &  59 &  75 &  42 &  34 & 0.35 & 2.0 &   9 &  27 &  18 &  29 \\
\enddata

\end{deluxetable}

\clearpage

\begin{deluxetable}{cccccccccccc}
\footnotesize
\tablecaption{Recovered luminosity percentage for the ellipticity equal to 0.01 
set of parameters\label{tbl-3}}
\tablewidth{0pt}
\tablehead{
\multicolumn{6}{c}{total sample} & \multicolumn{6}{c}{Bright 
SHARC sample} \\
\cline{1-6} \cline{7-12} \\
\colhead{z} & \colhead{$L_{44}$} & \colhead{1st$^a$} & \colhead{2nd$^a$} & 
\colhead{3rd$^a$} & \colhead{4th$^a$} &
\colhead{z} & \colhead{$L_{44}$} & \colhead{1st$^a$} & \colhead{2nd$^a$} & 
\colhead{3rd$^a$} & \colhead{4th$^a$}
}
\startdata
0.25 & 1.0 & 116 & 101 & 104 & 123 & 0.25 & 1.0 & 133 & 106 & 111 & 128 \\
0.30 & 1.0 & 112 &  99 & 114 & 115 & 0.30 & 1.0 & 196 & 137 & 176 & 139 \\
0.35 & 1.0 & 132 &  92 &  89 & 120 & 0.35 & 1.0 &   0 &   0 &   0 &   0 \\
0.40 & 1.0 & 110 & 120 & 122 & 139 & 0.40 & 1.0 &   0 & 200 &   0 &   0 \\
0.45 & 1.0 & 115 & 161 & 129 & 999 & 0.45 & 1.0 &   0 & 999 &   0 &   0 \\
0.25 & 2.0 & 110 & 101 & 103 & 114 & 0.25 & 2.0 & 109 & 101 & 103 & 114 \\
0.30 & 2.0 & 100 &  96 &  97 & 115 & 0.30 & 2.0 &  97 &  97 &  97 & 115 \\
0.35 & 2.0 &  98 &  91 &  94 & 114 & 0.35 & 2.0 & 109 &  99 & 101 & 117 \\
\enddata

\end{deluxetable}

\clearpage

\begin{deluxetable}{cccccccccccc}
\footnotesize
\tablecaption{Detection efficiency percentage for the ellipticity equal to 0.30
set of parameters\label{tbl-3}}
\tablewidth{0pt}
\tablehead{
\multicolumn{6}{c}{{\it total detection efficiency}} & \multicolumn{6}{c}{
{\it Bright SHARC detection efficiency}} \\
\cline{1-6} \cline{7-12} \\
\colhead{z} & \colhead{$L_{44}$} & \colhead{1st$^a$} & \colhead{2nd$^a$} & 
\colhead{3rd$^a$} & \colhead{4th$^a$} &
\colhead{z} & \colhead{$L_{44}$} & \colhead{1st$^a$} & \colhead{2nd$^a$} & 
\colhead{3rd$^a$} & \colhead{4th$^a$}
}
\startdata
0.25 & 1.0 &  34 &  44 &  26 &  28 & 0.25 & 1.0 &   5 &  12 &   8 &  25 \\
0.30 & 1.0 &  24 &  32 &  20 &  27 & 0.30 & 1.0 &   4 &   2 &   1 &  12 \\
0.35 & 1.0 &  15 &  14 &  11 &   8 & 0.35 & 1.0 &   0 &   0 &   0 &   0 \\
0.40 & 1.0 &   9 &  11 &   5 &   3 & 0.40 & 1.0 &   2 &   0 &   0 &   0 \\
0.45 & 1.0 &   6 &   8 &   5 &   2 & 0.45 & 1.0 &   1 &   2 &   0 &   0 \\
0.25 & 2.0 &  62 &  66 &  60 &  52 & 0.25 & 2.0 &  59 &  66 &  60 &  52 \\
0.30 & 2.0 &  63 &  66 &  57 &  48 & 0.30 & 2.0 &  35 &  62 &  53 &  46 \\
0.35 & 2.0 &  47 &  52 &  39 &  33 & 0.35 & 2.0 &   8 &  21 &  14 &  29 \\
\enddata

\end{deluxetable}

\clearpage

\begin{deluxetable}{cccccccccccc}
\footnotesize
\tablecaption{Recovered luminosity percentage for the ellipticity equal to 0.30 
set of parameters\label{tbl-3}}
\tablewidth{0pt}
\tablehead{
\multicolumn{6}{c}{total sample} & \multicolumn{6}{c}{Bright 
SHARC sample} \\
\cline{1-6} \cline{7-12} \\
\colhead{z} & \colhead{$L_{44}$} & \colhead{1st$^a$} & \colhead{2nd$^a$} & 
\colhead{3rd$^a$} & \colhead{4th$^a$} &
\colhead{z} & \colhead{$L_{44}$} & \colhead{1st$^a$} & \colhead{2nd$^a$} & 
\colhead{3rd$^a$} & \colhead{4th$^a$}
}
\startdata
0.25 & 1.0 & 109 & 100 & 106 & 117 & 0.25 & 1.0 & 127 & 119 & 118 & 119 \\
0.30 & 1.0 & 117 & 102 &  98 & 119 & 0.30 & 1.0 & 183 & 149 & 125 & 129 \\
0.35 & 1.0 & 115 &  92 & 105 & 120 & 0.35 & 1.0 &   0 &   0 &   0 &   0 \\
0.40 & 1.0 & 144 & 111 & 126 & 161 & 0.40 & 1.0 & 999 &   0 &   0 &   0 \\
0.45 & 1.0 & 999 & 177 & 132 & 176 & 0.45 & 1.0 & 999 & 999 &   0 &   0 \\
0.25 & 2.0 & 106 & 100 & 100 & 114 & 0.25 & 2.0 & 104 & 100 & 100 & 114 \\
0.30 & 2.0 & 103 &  94 &  95 & 112 & 0.30 & 2.0 & 105 &  94 &  96 & 113 \\
0.35 & 2.0 &  97 &  92 &  93 & 107 & 0.35 & 2.0 & 109 & 101 & 101 & 107 \\
\enddata

\end{deluxetable}

\clearpage

\begin{deluxetable}{cccccccccccc}
\footnotesize
\tablecaption{Detection efficiency percentage for the standard set of 
parameters and the type 2 pointings. \label{tbl-3}}
\tablewidth{0pt}
\tablehead{
\multicolumn{6}{c}{{\it total detection efficiency}} & \multicolumn{6}{c}{
{\it Bright SHARC detection efficiency}} \\
\cline{1-6} \cline{7-12} \\
\colhead{z} & \colhead{$L_{44}$} & \colhead{1st$^a$} & \colhead{2nd$^a$} & 
\colhead{3rd$^a$} & \colhead{4th$^a$} &
\colhead{z} & \colhead{$L_{44}$} & \colhead{1st$^a$} & \colhead{2nd$^a$} & 
\colhead{3rd$^a$} & \colhead{4th$^a$}
}
\startdata
0.25 & 1.0 &  34 &  35 &  29 &  25 & 0.25 & 1.0 &   6 &  11 &   6 &  21 \\
0.30 & 1.0 &  16 &  36 &  20 &  13 & 0.30 & 1.0 &   1 &   4 &   0 &   6 \\
0.35 & 1.0 &  10 &  18 &   9 &  16 & 0.35 & 1.0 &   0 &   1 &   0 &   2 \\
0.40 & 1.0 &   7 &  10 &   2 &   4 & 0.40 & 1.0 &   0 &   1 &   0 &   0 \\
0.45 & 1.0 &   3 &   8 &   2 &   2 & 0.45 & 1.0 &   0 &   0 &   0 &   0 \\
0.25 & 2.0 &  69 &  69 &  56 &  52 & 0.25 & 2.0 &  65 &  69 &  56 &  52 \\
0.30 & 2.0 &  63 &  74 &  54 &  49 & 0.30 & 2.0 &  37 &  68 &  51 &  49 \\
0.35 & 2.0 &  63 &  71 &  44 &  38 & 0.35 & 2.0 &   7 &  14 &  18 &  34 \\
\enddata

\end{deluxetable}

\clearpage

\begin{deluxetable}{cccccccccccc}
\footnotesize
\tablecaption{Recovered luminosity percentage for the standard set of 
parameters and the type 2 pointings. \label{tbl-3}}
\tablewidth{0pt}
\tablehead{
\multicolumn{6}{c}{total sample} & \multicolumn{6}{c}{Bright 
SHARC sample} \\
\cline{1-6} \cline{7-12} \\
\colhead{z} & \colhead{$L_{44}$} & \colhead{1st$^a$} & \colhead{2nd$^a$} & 
\colhead{3rd$^a$} & \colhead{4th$^a$} &
\colhead{z} & \colhead{$L_{44}$} & \colhead{1st$^a$} & \colhead{2nd$^a$} & 
\colhead{3rd$^a$} & \colhead{4th$^a$}
}
\startdata
0.25 & 1.0 & 102 & 108 & 107 & 111 & 0.25 & 1.0 & 119 & 123 & 123 & 113 \\
0.30 & 1.0 & 105 & 104 & 103 & 133 & 0.30 & 1.0 & 183 & 156 &   0 & 159 \\
0.35 & 1.0 &  98 & 104 &  98 & 121 & 0.35 & 1.0 &   0 & 150 &   0 & 155 \\
0.40 & 1.0 & 102 & 122 & 150 & 115 & 0.40 & 1.0 &   0 & 999 &   0 &   0 \\
0.45 & 1.0 & 175 & 132 & 105 & 195 & 0.45 & 1.0 &   0 &   0 &   0 &   0 \\
0.25 & 2.0 & 103 & 104 & 101 & 116 & 0.25 & 2.0 & 102 & 104 & 101 & 116 \\
0.30 & 2.0 &  98 &  95 & 101 & 116 & 0.30 & 2.0 & 102 &  95 & 101 & 116 \\
0.35 & 2.0 &  91 &  93 &  94 & 116 & 0.35 & 2.0 & 101 & 110 & 102 & 118 \\
\enddata

\end{deluxetable}

\clearpage

\begin{deluxetable}{cccccccccccc}
\footnotesize
\tablecaption{Detection efficiency percentage for the standard set of 
parameters and the type 3 pointings. \label{tbl-3}}
\tablewidth{0pt}
\tablehead{
\multicolumn{6}{c}{{\it total detection efficiency}} & \multicolumn{6}{c}{
{\it Bright SHARC detection efficiency}} \\
\cline{1-6} \cline{7-12} \\
\colhead{z} & \colhead{$L_{44}$} & \colhead{1st$^a$} & \colhead{2nd$^a$} & 
\colhead{3rd$^a$} & \colhead{4th$^a$} &
\colhead{z} & \colhead{$L_{44}$} & \colhead{1st$^a$} & \colhead{2nd$^a$} & 
\colhead{3rd$^a$} & \colhead{4th$^a$}
}
\startdata
0.25 & 1.0 &  39 &  49 &  40 &  29 & 0.25 & 1.0 &   4 &  14 &  12 &  24 \\
0.30 & 1.0 &  28 &  32 &  14 &  18 & 0.30 & 1.0 &   2 &   3 &   2 &   4 \\
0.35 & 1.0 &  10 &  22 &  14 &  10 & 0.35 & 1.0 &   0 &   4 &   0 &   1 \\
0.40 & 1.0 &   6 &  10 &   5 &   1 & 0.40 & 1.0 &   0 &   0 &   0 &   1 \\
0.45 & 1.0 &   3 &   4 &   5 &   0 & 0.45 & 1.0 &   0 &   0 &   0 &   0 \\
0.25 & 2.0 &  62 &  71 &  58 &  52 & 0.25 & 2.0 &  61 &  71 &  58 &  52 \\
0.30 & 2.0 &  69 &  67 &  55 &  56 & 0.30 & 2.0 &  37 &  58 &  50 &  56 \\
0.35 & 2.0 &  50 &  53 &  38 &  32 & 0.35 & 2.0 &  11 &  21 &  13 &  30 \\
\enddata

\end{deluxetable}

\clearpage

\begin{deluxetable}{cccccccccccc}
\footnotesize
\tablecaption{Recovered luminosity percentage for the standard set of 
parameters and the type 3 pointings. \label{tbl-3}}
\tablewidth{0pt}
\tablehead{
\multicolumn{6}{c}{total sample} & \multicolumn{6}{c}{Bright 
SHARC sample} \\
\cline{1-6} \cline{7-12} \\
\colhead{z} & \colhead{$L_{44}$} & \colhead{1st$^a$} & \colhead{2nd$^a$} & 
\colhead{3rd$^a$} & \colhead{4th$^a$} &
\colhead{z} & \colhead{$L_{44}$} & \colhead{1st$^a$} & \colhead{2nd$^a$} & 
\colhead{3rd$^a$} & \colhead{4th$^a$}
}
\startdata
0.25 & 1.0 & 105 & 103 & 107 & 111 & 0.25 & 1.0 & 121 & 119 & 119 & 114 \\
0.30 & 1.0 & 111 & 103 & 109 & 120 & 0.30 & 1.0 & 999 & 133 & 140 & 142 \\
0.35 & 1.0 & 106 & 137 & 117 & 129 & 0.35 & 1.0 &   0 & 999 &   0 & 166 \\
0.40 & 1.0 & 118 & 110 & 128 & 158 & 0.40 & 1.0 &   0 &   0 &   0 & 158 \\
0.45 & 1.0 & 132 & 134 & 162 &   0 & 0.45 & 1.0 &   0 &   0 &   0 &   0 \\
0.25 & 2.0 & 105 & 103 & 102 & 112 & 0.25 & 2.0 & 104 & 103 & 102 & 112 \\
0.30 & 2.0 &  94 &  98 &  97 & 115 & 0.30 & 2.0 &  95 &  99 &  97 & 115 \\
0.35 & 2.0 &  96 & 100 &  94 & 117 & 0.35 & 2.0 & 105 & 111 & 104 & 119 \\
\enddata

\end{deluxetable}

\clearpage

\begin{deluxetable}{cccccccccccc}
\footnotesize
\tablecaption{Detection efficiency percentage for the standard set of 
parameters and the type 6 pointings. \label{tbl-3}}
\tablewidth{0pt}
\tablehead{
\multicolumn{6}{c}{{\it total detection efficiency}} & \multicolumn{6}{c}{
{\it Bright SHARC detection efficiency}} \\
\cline{1-6} \cline{7-12} \\
\colhead{z} & \colhead{$L_{44}$} & \colhead{1st$^a$} & \colhead{2nd$^a$} & 
\colhead{3rd$^a$} & \colhead{4th$^a$} &
\colhead{z} & \colhead{$L_{44}$} & \colhead{1st$^a$} & \colhead{2nd$^a$} & 
\colhead{3rd$^a$} & \colhead{4th$^a$}
}
\startdata
0.25 & 1.0 &  35 &  43 &  23 &  28 & 0.25 & 1.0 &   4 &  17 &   9 &  22 \\
0.30 & 1.0 &  31 &  27 &  28 &  14 & 0.30 & 1.0 &   1 &   2 &   5 &   4 \\
0.35 & 1.0 &  12 &  14 &   6 &   7 & 0.35 & 1.0 &   0 &   5 &   0 &   2 \\
0.40 & 1.0 &   6 &   9 &   1 &   4 & 0.40 & 1.0 &   0 &   0 &   0 &   3 \\
0.45 & 1.0 &   4 &   5 &   1 &   2 & 0.45 & 1.0 &   0 &   4 &   0 &   1 \\
0.25 & 2.0 &  58 &  78 &  60 &  55 & 0.25 & 2.0 &  54 &  78 &  60 &  55 \\
0.30 & 2.0 &  66 &  74 &  55 &  47 & 0.30 & 2.0 &  46 &  65 &  53 &  47 \\
0.35 & 2.0 &  60 &  65 &  44 &  34 & 0.35 & 2.0 &   9 &  24 &  19 &  30 \\
\enddata

\end{deluxetable}

\clearpage

\begin{deluxetable}{cccccccccccc}
\footnotesize
\tablecaption{Recovered luminosity percentage for the standard set of 
parameters and the type 6 pointings. \label{tbl-3}}
\tablewidth{0pt}
\tablehead{
\multicolumn{6}{c}{total sample} & \multicolumn{6}{c}{Bright 
SHARC sample} \\
\cline{1-6} \cline{7-12} \\
\colhead{z} & \colhead{$L_{44}$} & \colhead{1st$^a$} & \colhead{2nd$^a$} & 
\colhead{3rd$^a$} & \colhead{4th$^a$} &
\colhead{z} & \colhead{$L_{44}$} & \colhead{1st$^a$} & \colhead{2nd$^a$} & 
\colhead{3rd$^a$} & \colhead{4th$^a$}
}
\startdata
0.25 & 1.0 & 116 & 103 &  96 & 123 & 0.25 & 1.0 & 121 & 111 & 105 & 127 \\
0.30 & 1.0 & 124 & 102 & 105 & 117 & 0.30 & 1.0 & 192 & 999 & 120 & 153 \\
0.35 & 1.0 & 129 & 113 &  95 & 148 & 0.35 & 1.0 &   0 & 146 &   0 & 182 \\
0.40 & 1.0 & 127 & 110 & 148 & 999 & 0.40 & 1.0 &   0 &   0 &   0 & 999 \\
0.45 & 1.0 & 999 & 999 & 101 & 999 & 0.45 & 1.0 &   0 & 999 &   0 & 999 \\
0.25 & 2.0 & 105 & 102 & 103 & 116 & 0.25 & 2.0 & 105 & 102 & 103 & 116 \\
0.30 & 2.0 & 108 &  94 &  95 & 111 & 0.30 & 2.0 & 110 &  94 &  96 & 111 \\
0.35 & 2.0 & 101 &  93 & 100 & 123 & 0.35 & 2.0 & 120 & 100 & 115 & 125 \\
\enddata

\end{deluxetable}

\clearpage

\begin{deluxetable}{cccccccccccc}
\footnotesize
\tablecaption{Detection efficiency percentage for the standard set of 
parameters and the type 7 pointings. \label{tbl-3}}
\tablewidth{0pt}
\tablehead{
\multicolumn{6}{c}{{\it total detection efficiency}} & \multicolumn{6}{c}{
{\it Bright SHARC detection efficiency}} \\
\cline{1-6} \cline{7-12} \\
\colhead{z} & \colhead{$L_{44}$} & \colhead{1st$^a$} & \colhead{2nd$^a$} & 
\colhead{3rd$^a$} & \colhead{4th$^a$} &
\colhead{z} & \colhead{$L_{44}$} & \colhead{1st$^a$} & \colhead{2nd$^a$} & 
\colhead{3rd$^a$} & \colhead{4th$^a$}
}
\startdata
0.25 & 1.0 &  36 &  39 &  24 &  19 & 0.25 & 1.0 &   5 &  11 &   5 &  16 \\
0.30 & 1.0 &  26 &  29 &  14 &  17 & 0.30 & 1.0 &   4 &   4 &   5 &   5 \\
0.35 & 1.0 &   6 &  16 &   4 &   7 & 0.35 & 1.0 &   0 &   0 &   1 &   3 \\
0.40 & 1.0 &   6 &   4 &   4 &   1 & 0.40 & 1.0 &   0 &   0 &   1 &   0 \\
0.45 & 1.0 &   5 &   1 &   1 &   0 & 0.45 & 1.0 &   0 &   0 &   0 &   0 \\
0.25 & 2.0 &  71 &  71 &  61 &  47 & 0.25 & 2.0 &  67 &  71 &  61 &  47 \\
0.30 & 2.0 &  71 &  68 &  57 &  42 & 0.30 & 2.0 &  45 &  58 &  52 &  42 \\
0.35 & 2.0 &  59 &  50 &  40 &  35 & 0.35 & 2.0 &  10 &  17 &  15 &  33 \\
\enddata

\end{deluxetable}

\clearpage

\begin{deluxetable}{cccccccccccc}
\footnotesize
\tablecaption{Recovered luminosity percentage for the standard set of 
parameters and the type 7 pointings. \label{tbl-3}}
\tablewidth{0pt}
\tablehead{
\multicolumn{6}{c}{total sample} & \multicolumn{6}{c}{Bright 
SHARC sample} \\
\cline{1-6} \cline{7-12} \\
\colhead{z} & \colhead{$L_{44}$} & \colhead{1st$^a$} & \colhead{2nd$^a$} & 
\colhead{3rd$^a$} & \colhead{4th$^a$} &
\colhead{z} & \colhead{$L_{44}$} & \colhead{1st$^a$} & \colhead{2nd$^a$} & 
\colhead{3rd$^a$} & \colhead{4th$^a$}
}
\startdata
0.25 & 1.0 & 115 & 100 & 104 & 120 & 0.25 & 1.0 & 145 & 112 & 119 & 121 \\
0.30 & 1.0 & 129 &  99 & 115 & 126 & 0.30 & 1.0 & 181 & 114 & 148 & 160 \\
0.35 & 1.0 & 170 & 102 & 102 & 145 & 0.35 & 1.0 &   0 &   0 & 166 & 186 \\
0.40 & 1.0 & 123 & 144 & 167 & 148 & 0.40 & 1.0 &   0 &   0 & 999 &   0 \\
0.45 & 1.0 & 999 & 116 & 144 &   0 & 0.45 & 1.0 &   0 &   0 &   0 &   0 \\
0.25 & 2.0 & 106 &  99 & 101 & 116 & 0.25 & 2.0 & 105 &  99 & 101 & 116 \\
0.30 & 2.0 & 103 &  94 &  97 & 112 & 0.30 & 2.0 & 102 &  95 &  97 & 112 \\
0.35 & 2.0 &  98 &  92 &  93 & 120 & 0.35 & 2.0 & 116 & 104 &  98 & 121 \\
\enddata

\end{deluxetable}

\clearpage

\begin{deluxetable}{cccccccccccc}
\footnotesize
\tablecaption{Detection efficiency percentage for the standard set of 
parameters and the type 8 pointings. \label{tbl-3}}
\tablewidth{0pt}
\tablehead{
\multicolumn{6}{c}{{\it total detection efficiency}} & \multicolumn{6}{c}{
{\it Bright SHARC detection efficiency}} \\
\cline{1-6} \cline{7-12} \\
\colhead{z} & \colhead{$L_{44}$} & \colhead{1st$^a$} & \colhead{2nd$^a$} & 
\colhead{3rd$^a$} & \colhead{4th$^a$} &
\colhead{z} & \colhead{$L_{44}$} & \colhead{1st$^a$} & \colhead{2nd$^a$} & 
\colhead{3rd$^a$} & \colhead{4th$^a$}
}
\startdata
0.25 & 1.0 &  34 &  47 &  33 &  31 & 0.25 & 1.0 &   5 &  14 &   8 &  24 \\
0.30 & 1.0 &  22 &  31 &  24 &  17 & 0.30 & 1.0 &   3 &   2 &   0 &   4 \\
0.35 & 1.0 &  10 &  16 &   9 &   9 & 0.35 & 1.0 &   0 &   2 &   0 &   2 \\
0.40 & 1.0 &   5 &  13 &   3 &   1 & 0.40 & 1.0 &   0 &   0 &   0 &   0 \\
0.45 & 1.0 &   6 &   4 &   2 &   2 & 0.45 & 1.0 &   0 &   0 &   0 &   0 \\
0.25 & 2.0 &  56 &  81 &  54 &  51 & 0.25 & 2.0 &  54 &  80 &  54 &  51 \\
0.30 & 2.0 &  58 &  80 &  65 &  54 & 0.30 & 2.0 &  45 &  75 &  61 &  53 \\
0.35 & 2.0 &  57 &  57 &  55 &  52 & 0.35 & 2.0 &  17 &  12 &  25 &  46 \\
\enddata

\end{deluxetable}

\clearpage

\begin{deluxetable}{cccccccccccc}
\footnotesize
\tablecaption{Recovered luminosity percentage for the standard set of 
parameters and the type 8 pointings. \label{tbl-3}}
\tablewidth{0pt}
\tablehead{
\multicolumn{6}{c}{total sample} & \multicolumn{6}{c}{Bright 
SHARC sample} \\
\cline{1-6} \cline{7-12} \\
\colhead{z} & \colhead{$L_{44}$} & \colhead{1st$^a$} & \colhead{2nd$^a$} & 
\colhead{3rd$^a$} & \colhead{4th$^a$} &
\colhead{z} & \colhead{$L_{44}$} & \colhead{1st$^a$} & \colhead{2nd$^a$} & 
\colhead{3rd$^a$} & \colhead{4th$^a$}
}
\startdata
0.25 & 1.0 & 145 & 101 & 102 & 117 & 0.25 & 1.0 & 999 & 112 & 119 & 120 \\
0.30 & 1.0 & 120 & 103 &  97 & 112 & 0.30 & 1.0 & 173 & 135 &   0 & 121 \\
0.35 & 1.0 & 114 & 108 &  97 & 152 & 0.35 & 1.0 &   0 & 191 &   0 & 999 \\
0.40 & 1.0 & 118 & 118 & 112 & 116 & 0.40 & 1.0 &   0 &   0 &   0 &   0 \\
0.45 & 1.0 & 150 & 105 & 166 & 117 & 0.45 & 1.0 &   0 &   0 &   0 &   0 \\
0.25 & 2.0 & 107 & 101 & 104 & 115 & 0.25 & 2.0 & 106 & 101 & 104 & 115 \\
0.30 & 2.0 & 102 &  98 &  95 & 113 & 0.30 & 2.0 & 102 &  99 &  96 & 114 \\
0.35 & 2.0 & 101 &  95 &  90 & 114 & 0.35 & 2.0 & 107 & 116 &  98 & 115 \\
\enddata

\end{deluxetable}

\clearpage

\begin{deluxetable}{cccccccccccc}
\footnotesize
\tablecaption{Detection efficiency percentage for the standard set of 
parameters and the type 9 pointings. \label{tbl-3}}
\tablewidth{0pt}
\tablehead{
\multicolumn{6}{c}{{\it total detection efficiency}} & \multicolumn{6}{c}{
{\it Bright SHARC detection efficiency}} \\
\cline{1-6} \cline{7-12} \\
\colhead{z} & \colhead{$L_{44}$} & \colhead{1st$^a$} & \colhead{2nd$^a$} & 
\colhead{3rd$^a$} & \colhead{4th$^a$} &
\colhead{z} & \colhead{$L_{44}$} & \colhead{1st$^a$} & \colhead{2nd$^a$} & 
\colhead{3rd$^a$} & \colhead{4th$^a$}
}
\startdata
0.25 & 1.0 &  45 &  47 &  37 &  32 & 0.25 & 1.0 &  14 &  20 &  17 &  28 \\
0.30 & 1.0 &  32 &  36 &  28 &  15 & 0.30 & 1.0 &   0 &   4 &   4 &   9 \\
0.35 & 1.0 &  14 &  22 &  14 &   8 & 0.35 & 1.0 &   1 &   4 &   0 &   2 \\
0.40 & 1.0 &  15 &  19 &   8 &   6 & 0.40 & 1.0 &   0 &   1 &   0 &   3 \\
0.45 & 1.0 &   7 &  10 &   7 &   6 & 0.45 & 1.0 &   0 &   0 &   1 &   2 \\
0.25 & 2.0 &  86 &  91 &  77 &  57 & 0.25 & 2.0 &  86 &  91 &  77 &  57 \\
0.30 & 2.0 &  86 &  85 &  78 &  46 & 0.30 & 2.0 &  79 &  83 &  78 &  46 \\
0.35 & 2.0 &  78 &  86 &  64 &  47 & 0.35 & 2.0 &  27 &  33 &  26 &  44 \\
\enddata

\end{deluxetable}

\clearpage

\begin{deluxetable}{cccccccccccc}
\footnotesize
\tablecaption{Recovered luminosity percentage for the standard set of 
parameters and the type 9 pointings. \label{tbl-3}}
\tablewidth{0pt}
\tablehead{
\multicolumn{6}{c}{total sample} & \multicolumn{6}{c}{Bright 
SHARC sample} \\
\cline{1-6} \cline{7-12} \\
\colhead{z} & \colhead{$L_{44}$} & \colhead{1st$^a$} & \colhead{2nd$^a$} & 
\colhead{3rd$^a$} & \colhead{4th$^a$} &
\colhead{z} & \colhead{$L_{44}$} & \colhead{1st$^a$} & \colhead{2nd$^a$} & 
\colhead{3rd$^a$} & \colhead{4th$^a$}
}
\startdata
0.25 & 1.0 & 109 & 105 & 106 & 130 & 0.25 & 1.0 & 116 & 120 & 112 & 132 \\
0.30 & 1.0 &  98 & 103 & 110 & 119 & 0.30 & 1.0 &   0 & 136 & 176 & 136 \\
0.35 & 1.0 & 106 & 125 & 117 & 153 & 0.35 & 1.0 & 191 & 999 &   0 & 999 \\
0.40 & 1.0 & 117 & 113 & 128 & 186 & 0.40 & 1.0 &   0 & 999 &   0 & 999 \\
0.45 & 1.0 & 124 & 109 & 999 & 184 & 0.45 & 1.0 &   0 &   0 & 999 & 999 \\
0.25 & 2.0 & 103 & 105 & 104 & 122 & 0.25 & 2.0 & 103 & 105 & 104 & 122 \\
0.30 & 2.0 &  98 & 100 &  98 & 115 & 0.30 & 2.0 &  99 & 100 &  98 & 115 \\
0.35 & 2.0 &  94 &  97 &  97 & 120 & 0.35 & 2.0 & 101 & 108 & 111 & 121 \\
\enddata

\end{deluxetable}

\clearpage

\begin{deluxetable}{cccccccccccc}
\footnotesize
\tablecaption{Detection efficiency percentage for the standard set of 
parameters and the pointings with point sources. \label{tbl-3}}
\tablewidth{0pt}
\tablehead{
\multicolumn{6}{c}{{\it total detection efficiency}} & \multicolumn{6}{c}{
{\it Bright SHARC detection efficiency}} \\
\cline{1-6} \cline{7-12} \\
\colhead{z} & \colhead{$L_{44}$} & \colhead{1st$^a$} & \colhead{2nd$^a$} & 
\colhead{3rd$^a$} & \colhead{4th$^a$} &
\colhead{z} & \colhead{$L_{44}$} & \colhead{1st$^a$} & \colhead{2nd$^a$} & 
\colhead{3rd$^a$} & \colhead{4th$^a$}
}
\startdata
0.25 & 1.0 &  40 &  43 &  28 &  31 & 0.25 & 1.0 &   5 &  13 &   5 &  17 \\
0.30 & 1.0 &  36 &  34 &  21 &  15 & 0.30 & 1.0 &   0 &   3 &   1 &   3 \\
0.35 & 1.0 &  15 &  21 &  16 &   8 & 0.35 & 1.0 &   0 &   0 &   2 &   2 \\
0.40 & 1.0 &   3 &  15 &   6 &   3 & 0.40 & 1.0 &   0 &   0 &   1 &   0 \\
0.45 & 1.0 &   4 &   9 &   4 &   2 & 0.45 & 1.0 &   0 &   0 &   1 &   0 \\
0.25 & 2.0 &  73 &  74 &  48 &  51 & 0.25 & 2.0 &  72 &  74 &  48 &  51 \\
0.30 & 2.0 &  72 &  64 &  60 &  53 & 0.30 & 2.0 &  49 &  60 &  55 &  52 \\
0.35 & 2.0 &  62 &  67 &  61 &  43 & 0.35 & 2.0 &  12 &  24 &  30 &  41 \\
\enddata

\end{deluxetable}%

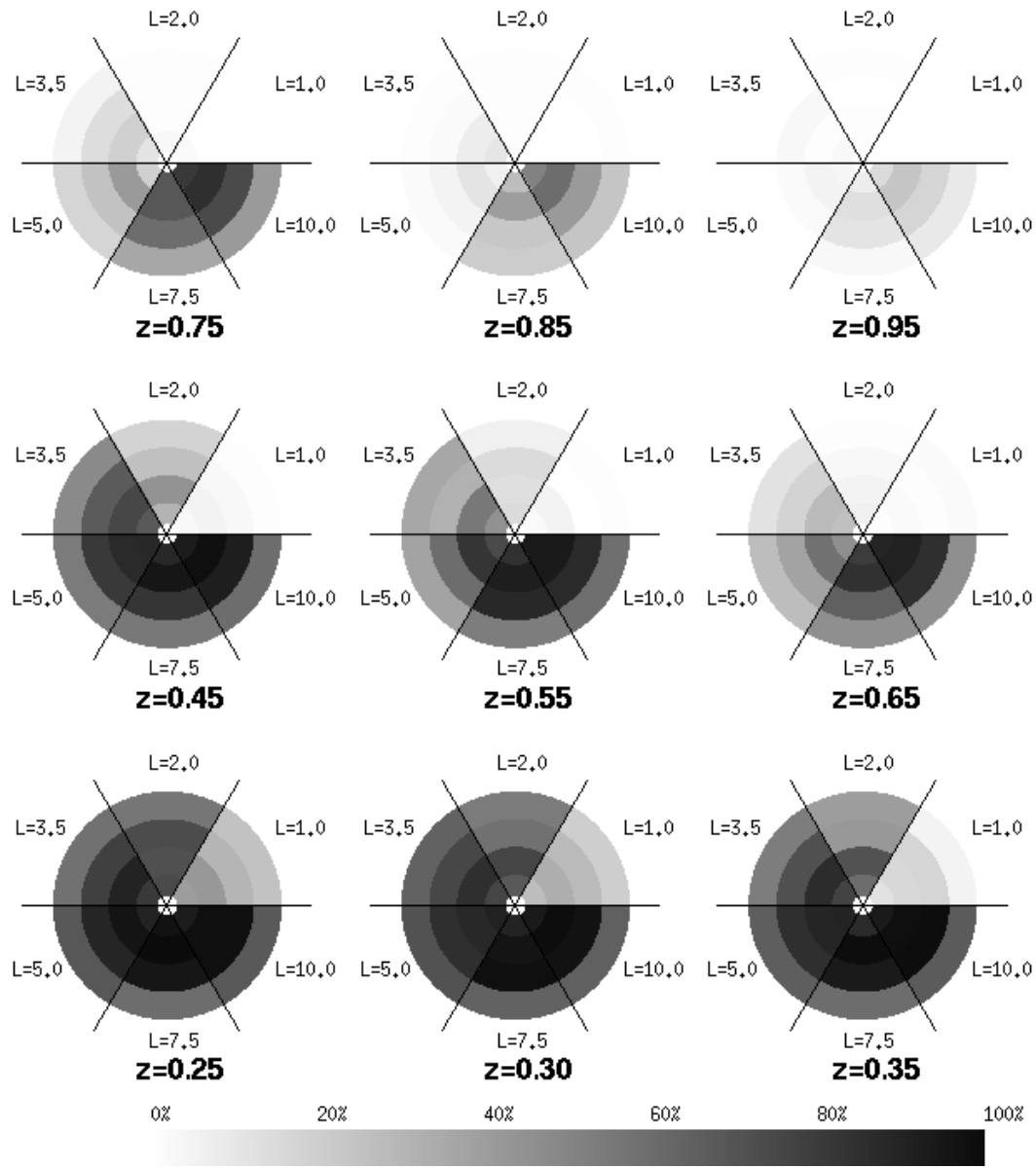
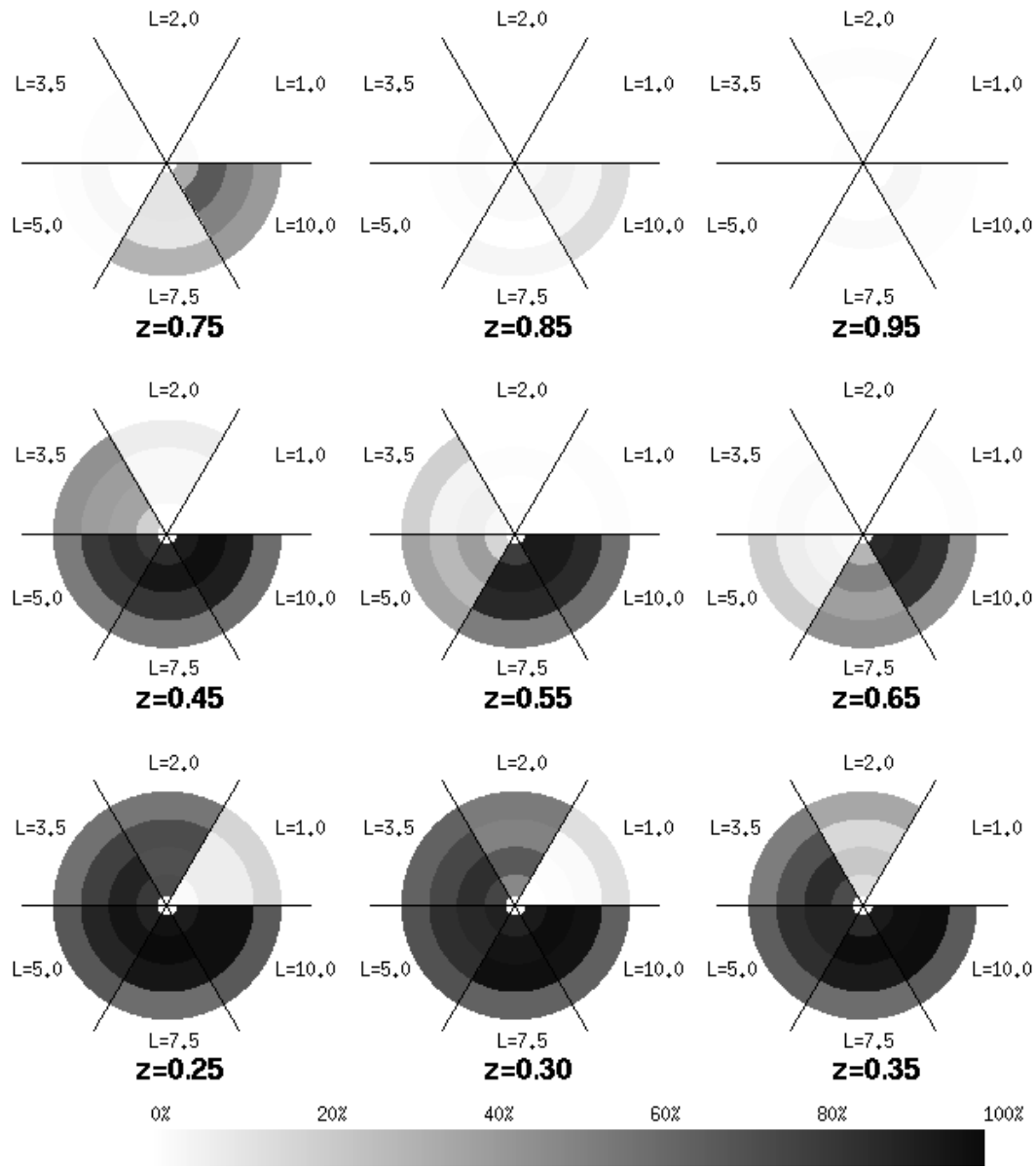
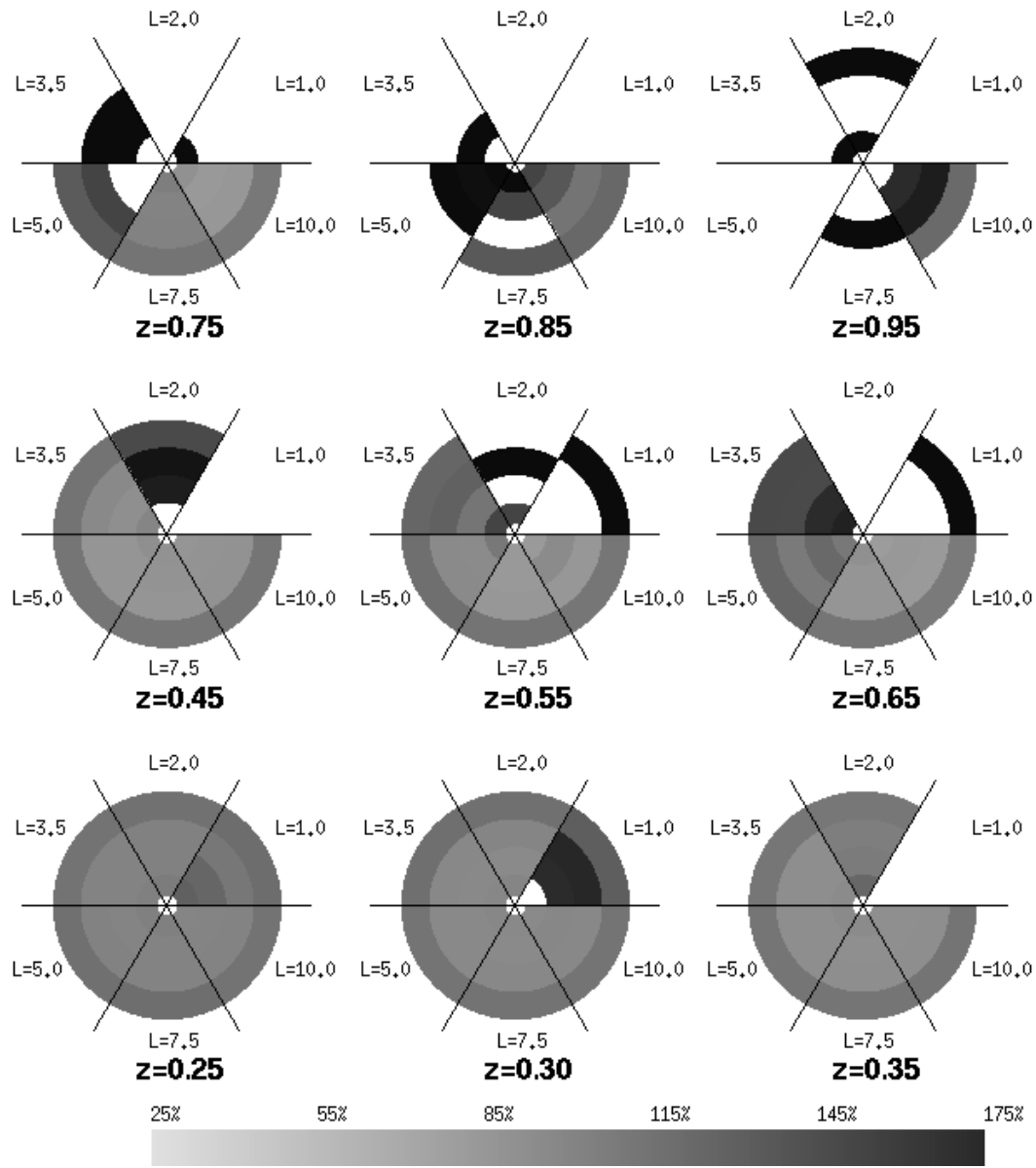
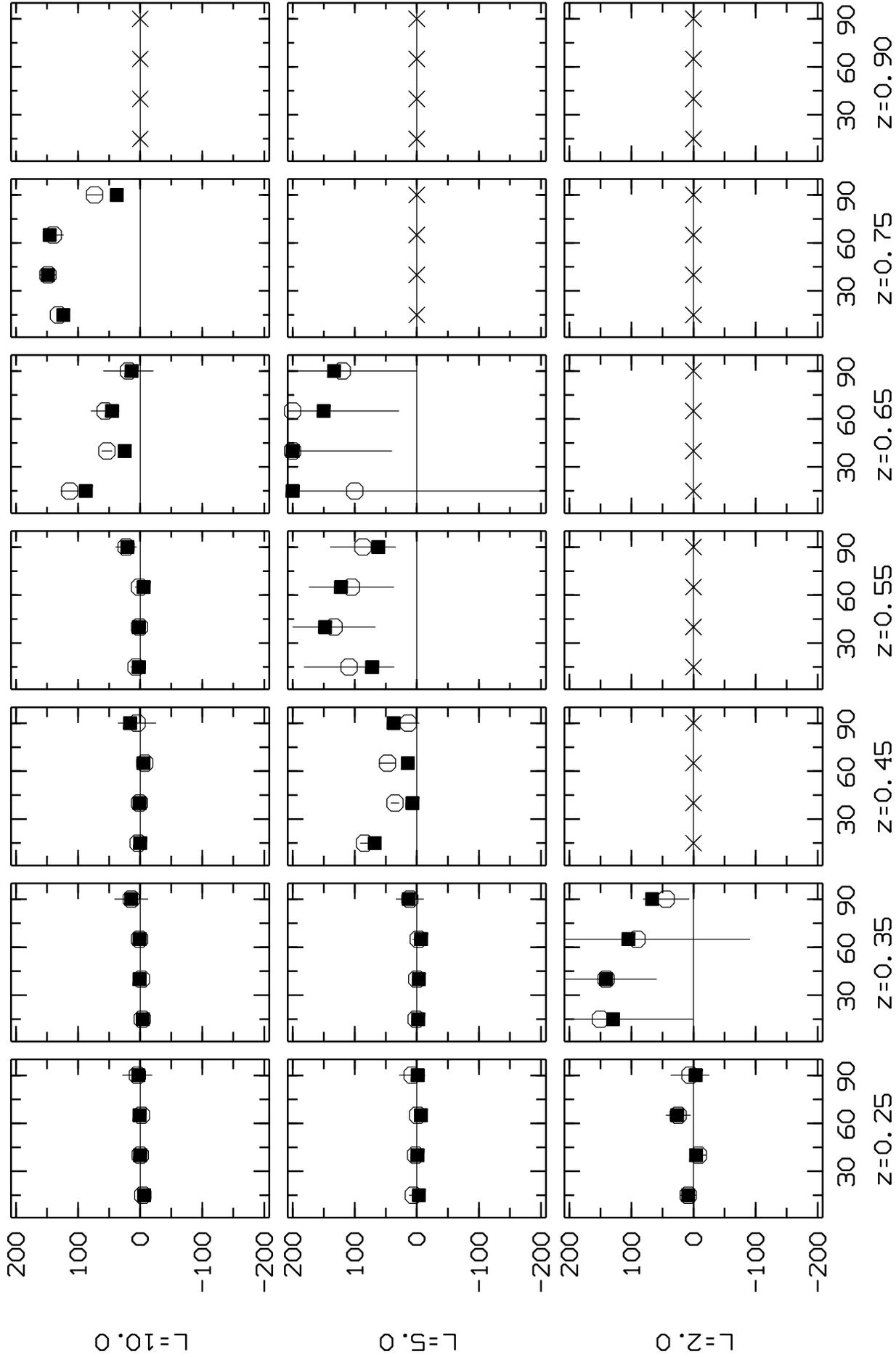
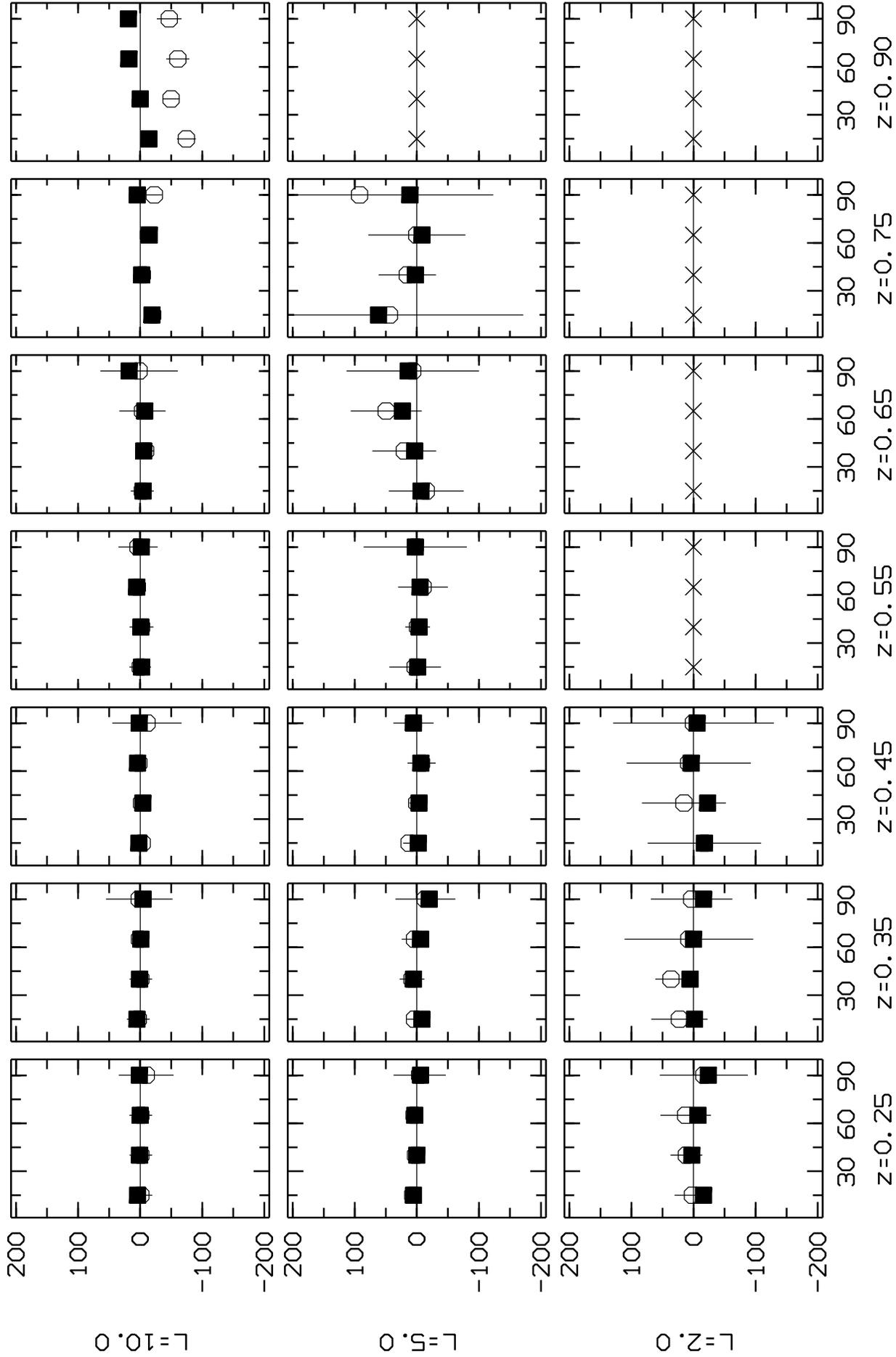
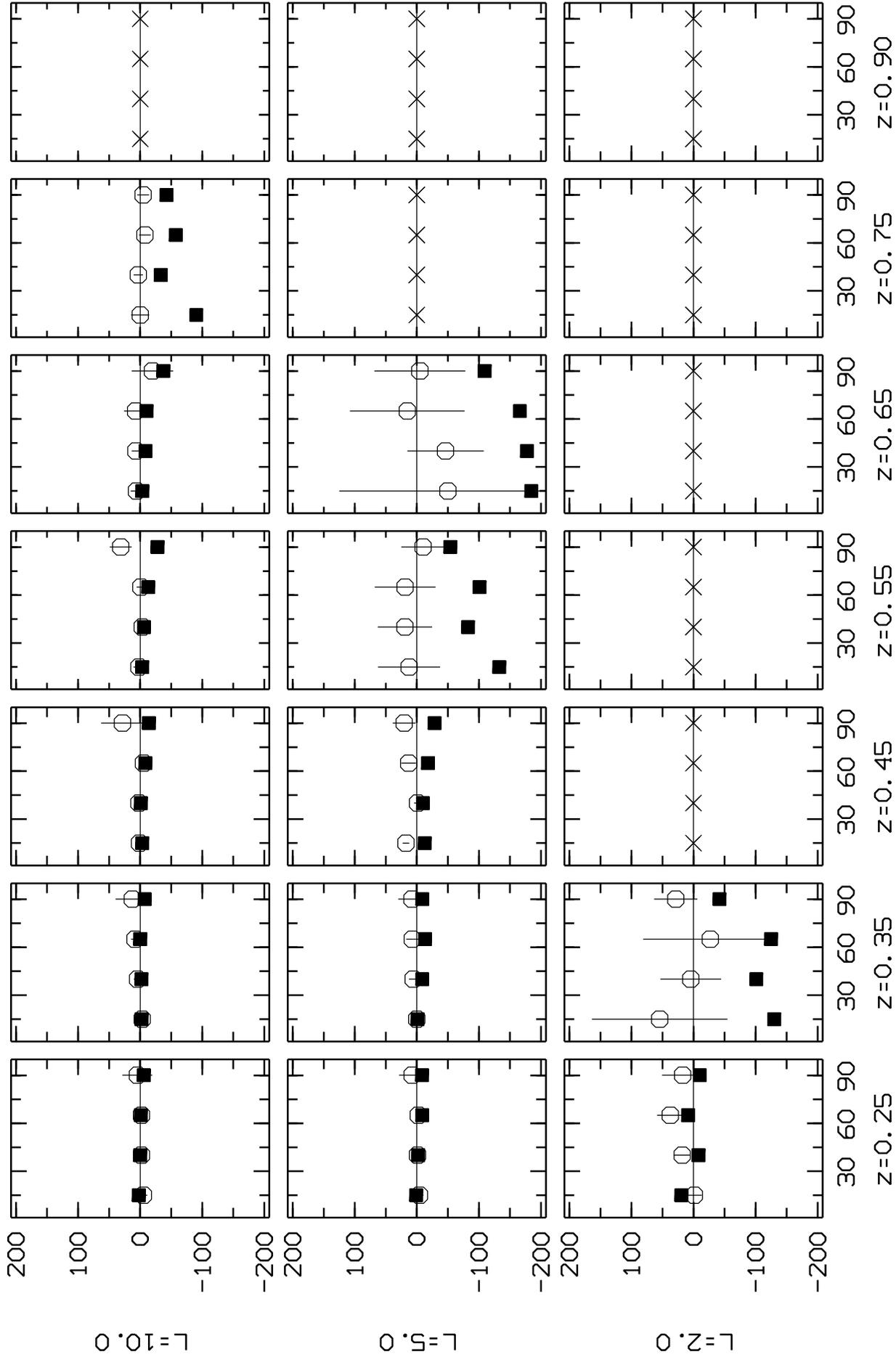
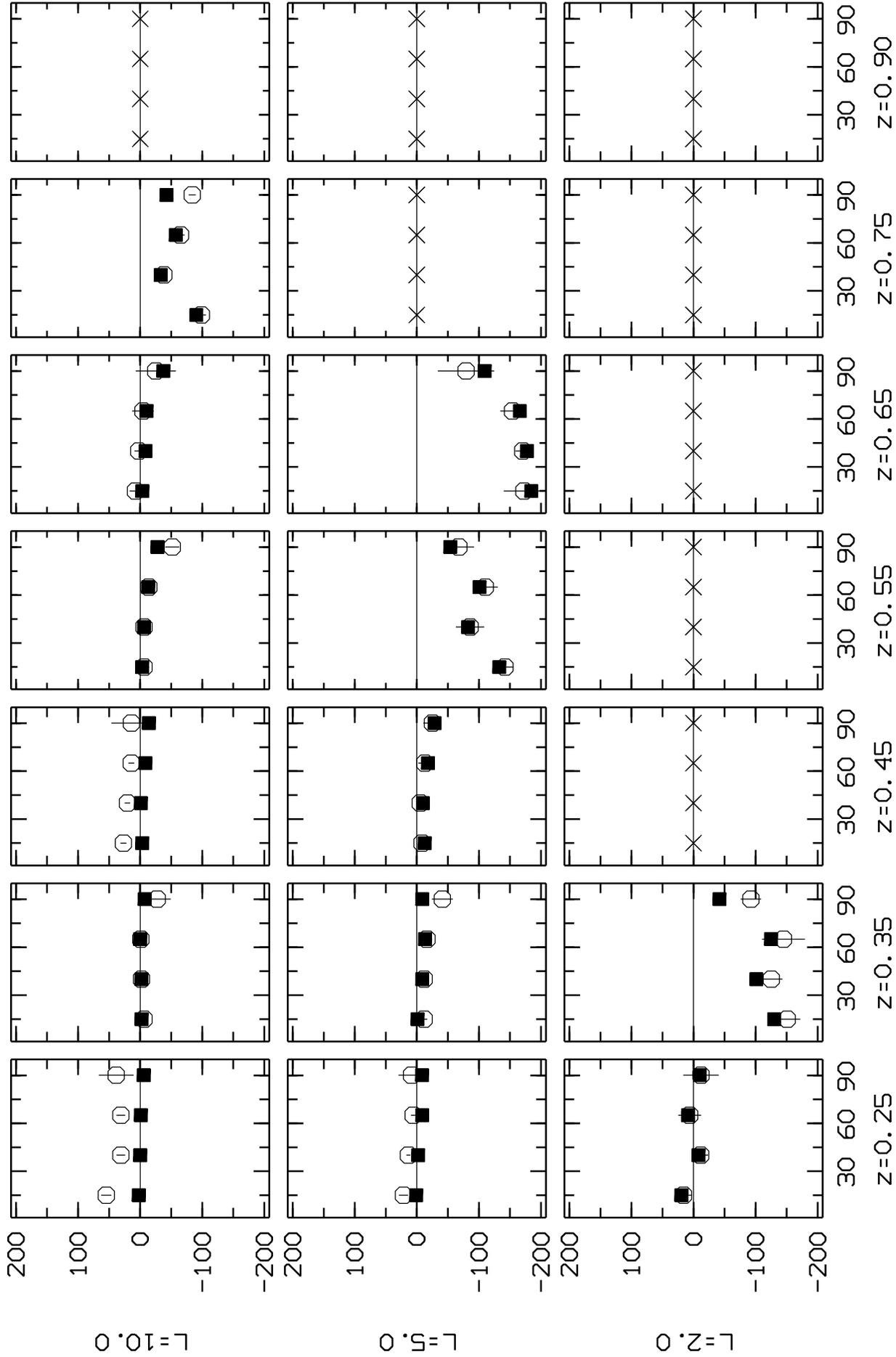
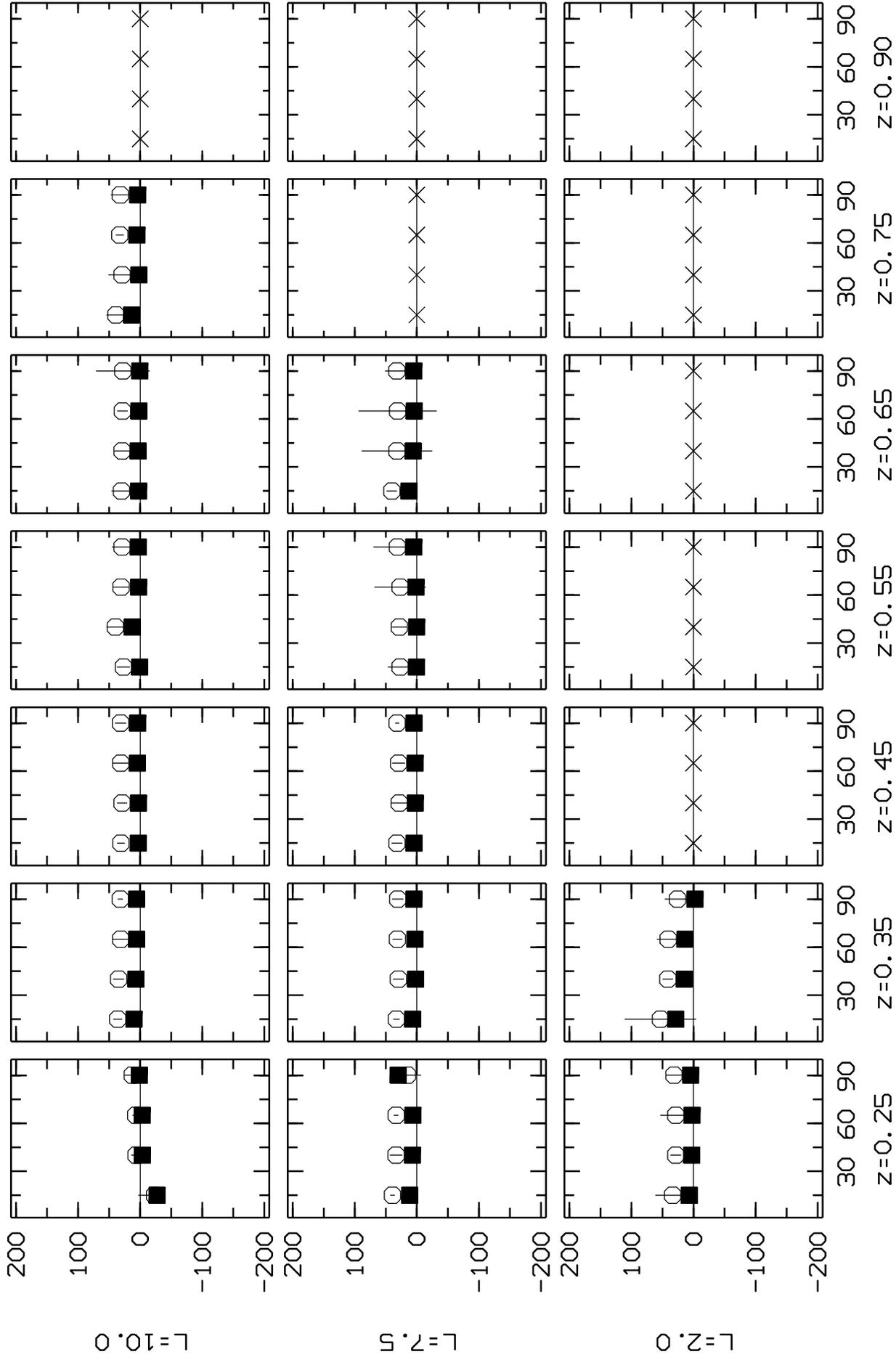
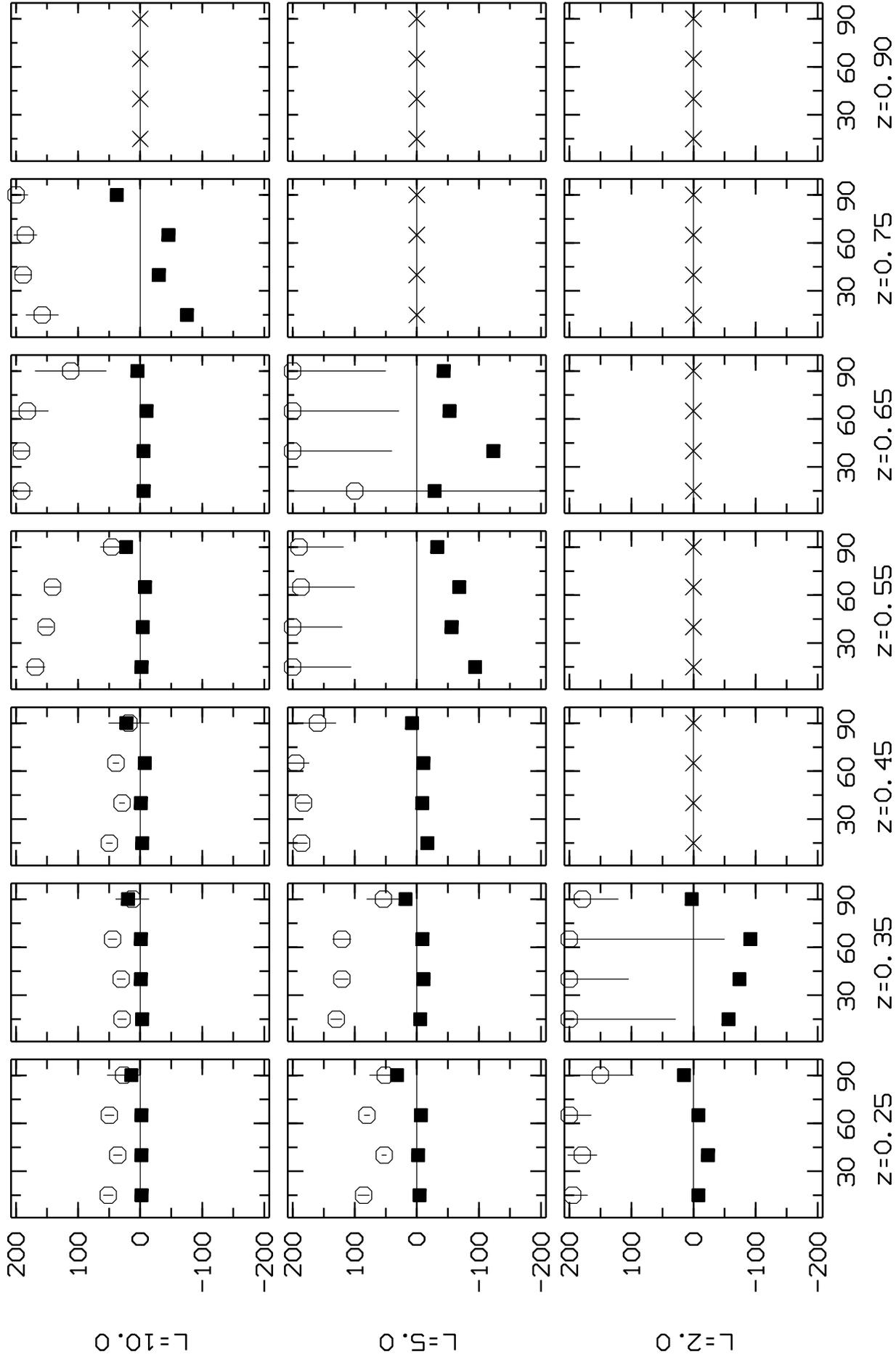
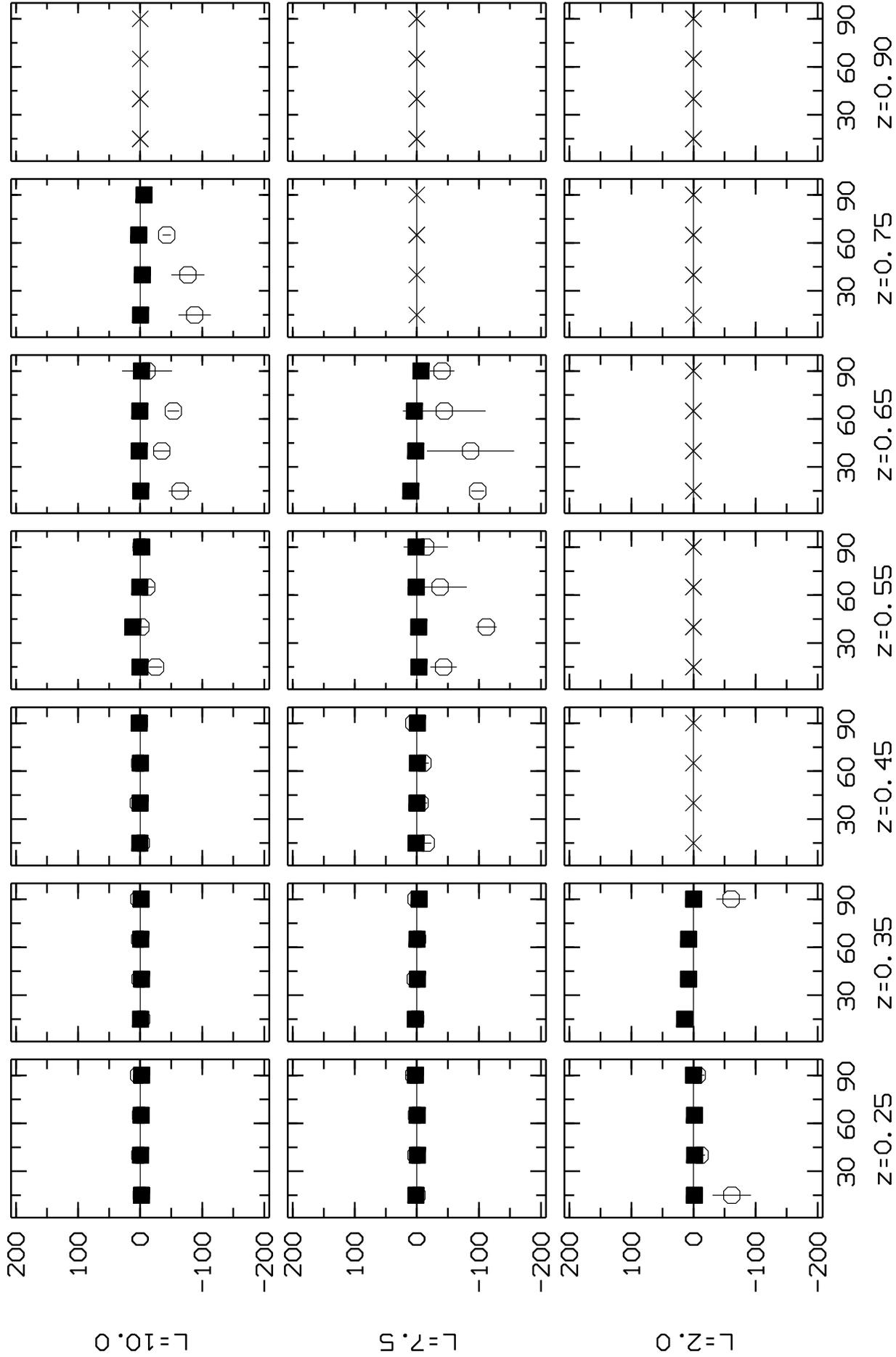
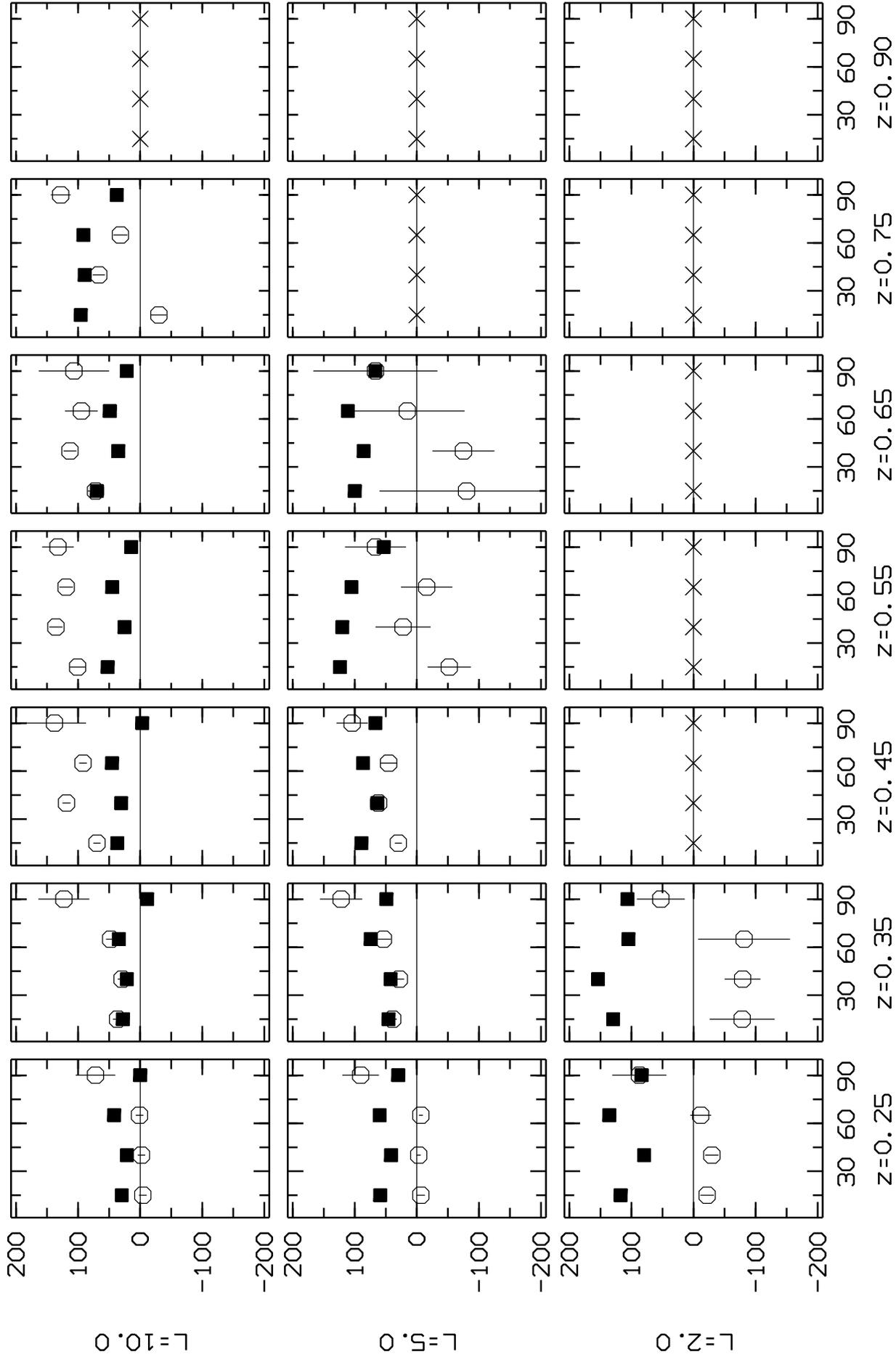
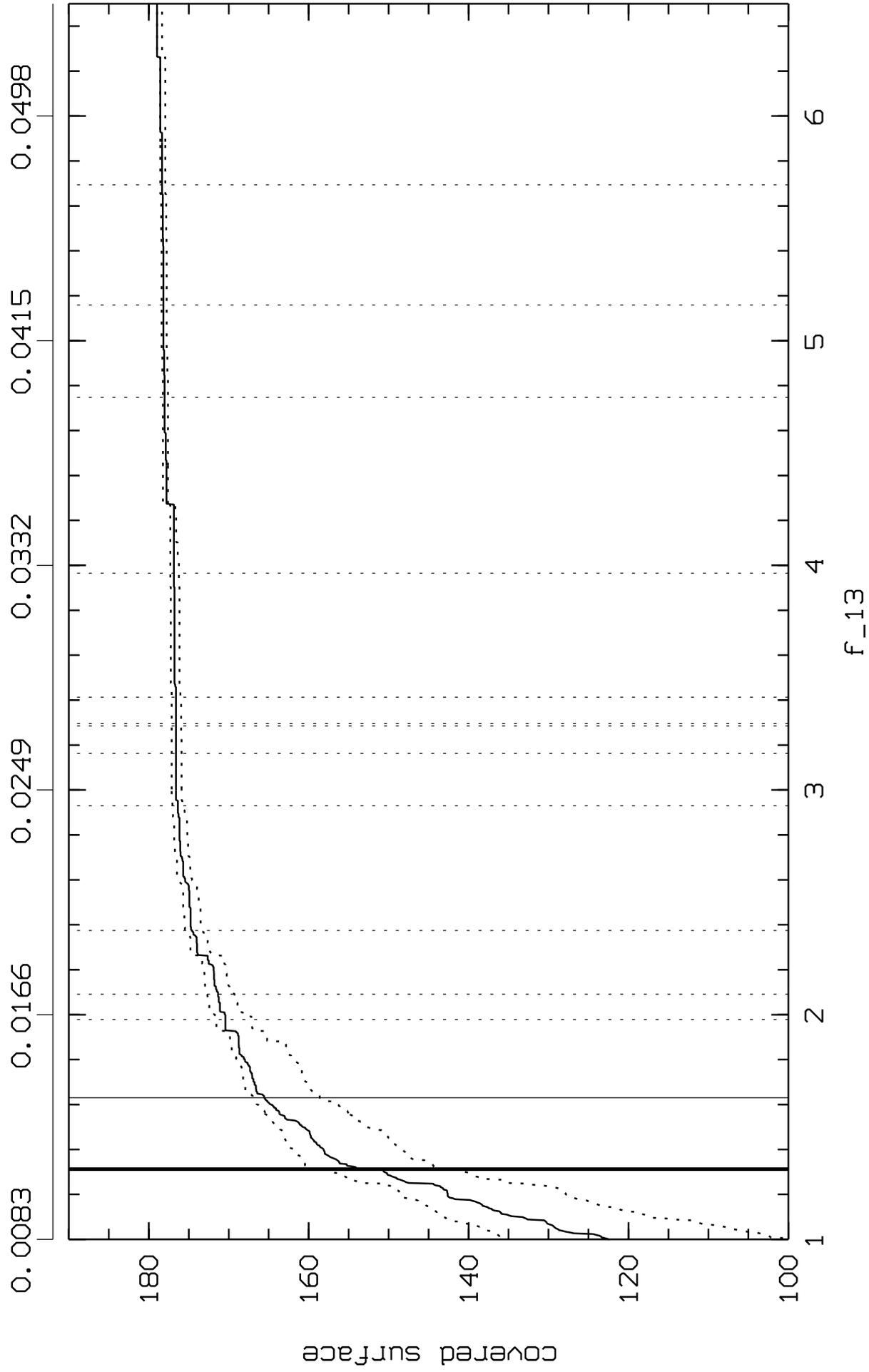
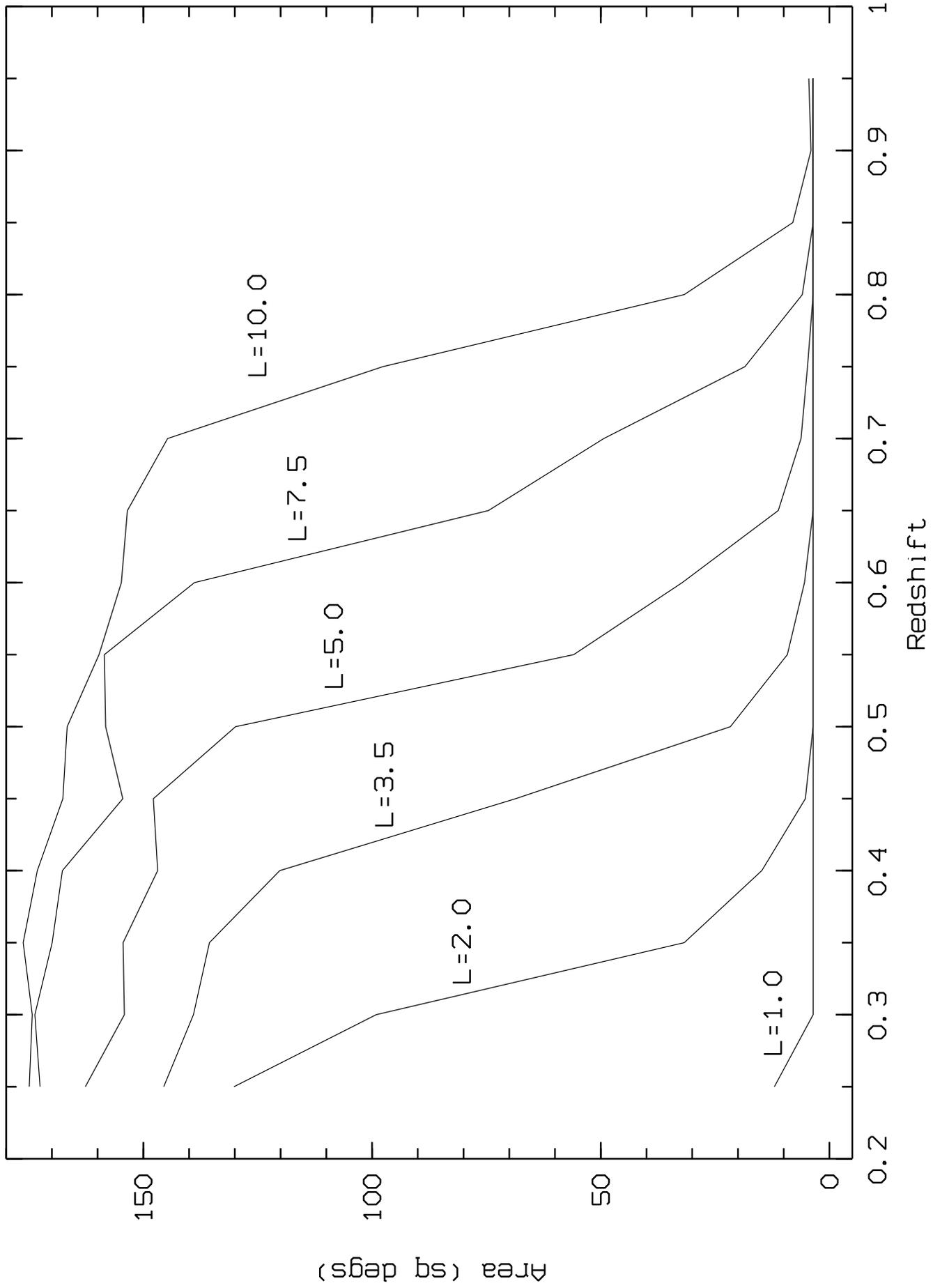
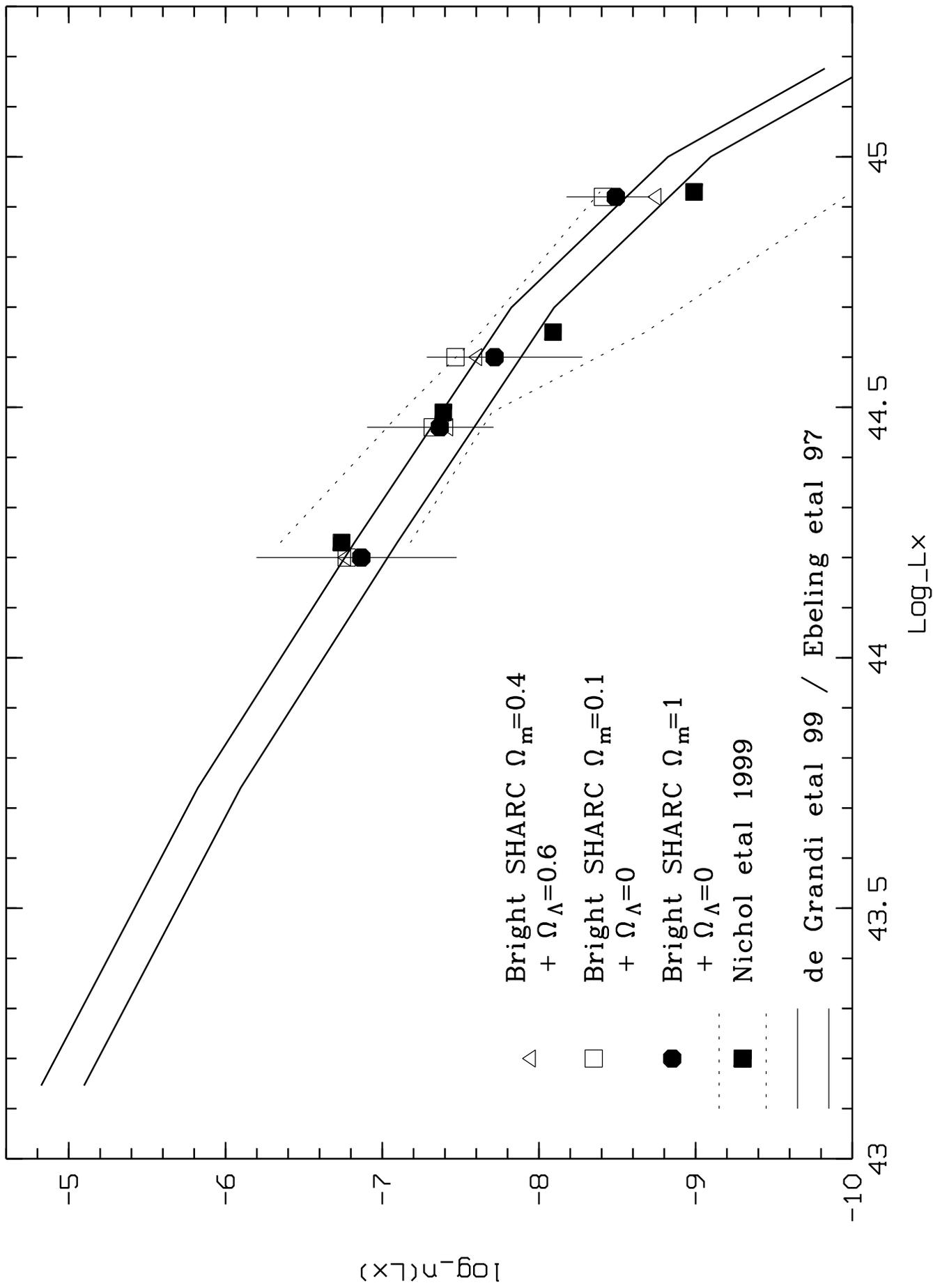
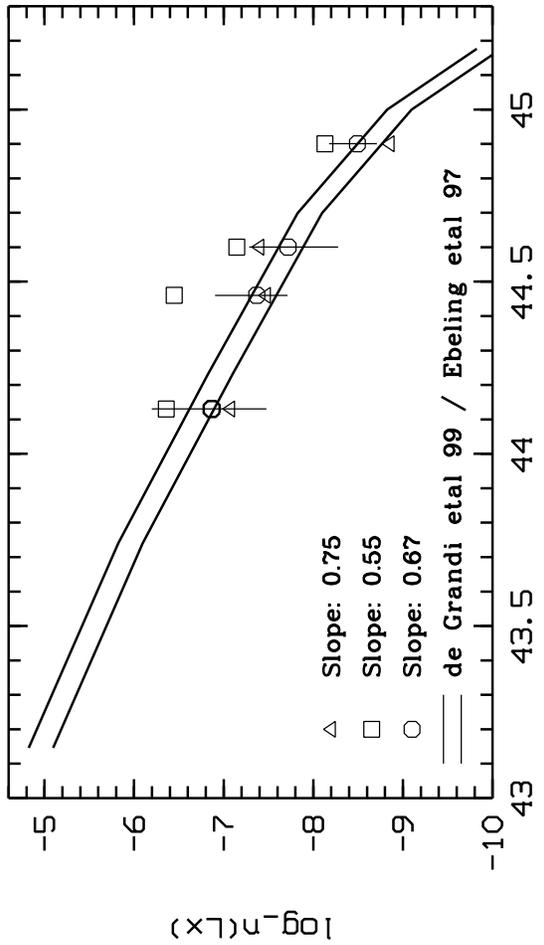
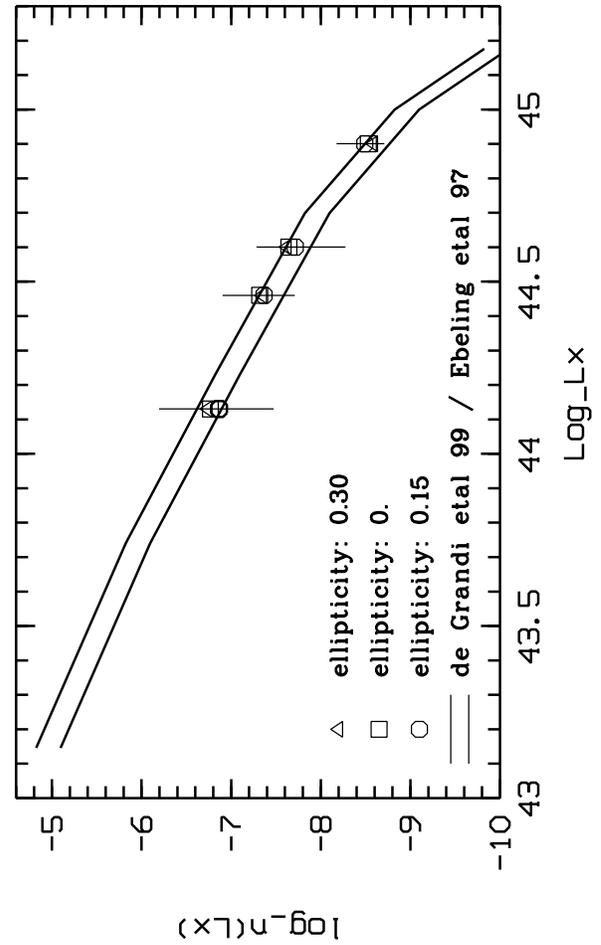
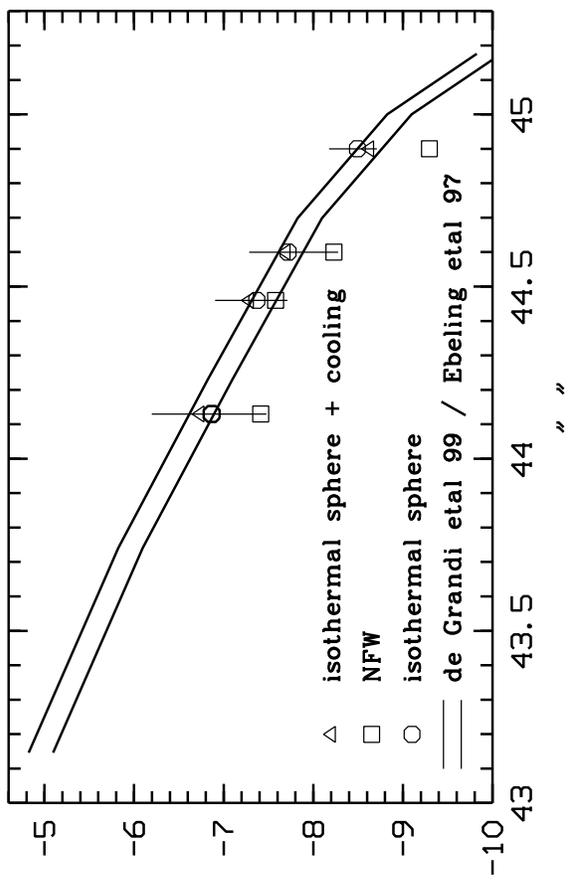
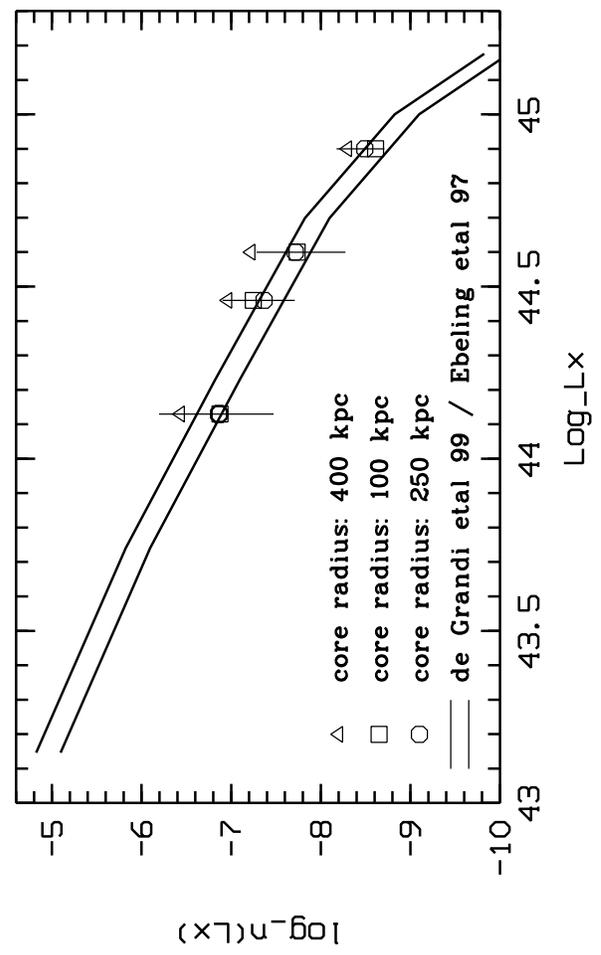
\end{document}